\documentclass[]{aastex63}

\newcommand{\ilam}{erg cm$^{-2}$ s$^{-1}$ sr$^{-1}$ \AA$^{-1}$}

\submitjournal{ApJ}
\usepackage{romannum}
\usepackage{wrapfig}
\usepackage{fancyvrb}
\VerbatimFootnotes
\usepackage{enumitem}                   
\usepackage{bm}
\shorttitle{RADYN Flare Broadening Models}
\shortauthors{Kowalski et al.}
\usepackage{amsmath}
\usepackage{lineno}

\begin{document}

\title{The Atmospheric Response to High Nonthermal Electron Beam Fluxes in Solar Flares. II. Hydrogen Broadening Predictions 
for Solar Flare Observations with the Daniel K. Inouye Solar Telescope }

\correspondingauthor{Adam F Kowalski}
\email{adam.f.kowalski@colorado.edu}

\author[0000-0001-7458-1176]{Adam F. Kowalski}
\affiliation{National Solar Observatory, University of Colorado Boulder, 3665 Discovery Drive, Boulder, CO 80303, USA}
\affiliation{Department of Astrophysical and Planetary Sciences, University of Colorado, Boulder, 2000 Colorado Ave, CO 80305, USA}
\affiliation{Laboratory for Atmospheric and Space Physics, University of Colorado Boulder, 3665 Discovery Drive, Boulder, CO 80303, USA.}

\author[0000-0003-4227-6809]{Joel C. Allred}
\affiliation{NASA Goddard Space Flight Center, Solar Physics Laboratory, Code 671, Greenbelt, MD 20771, USA}

\author[0000-0001-9218-3139]{Mats Carlsson}
\affiliation{Institute of Theoretical Astrophysics, University
              of Oslo, P.O. Box 1029 Blindern, 0315 Oslo, Norway}
              \affiliation{Rosseland Centre for Solar Physics, University of Oslo, P.O. Box 1029 Blindern, 0315 Oslo, Norway}

\author[0000-0001-5316-914X]{Graham S. Kerr}
\affiliation{Catholic University of America at NASA Goddard Space Flight Center, Solar Physics Laboratory, Code 671, Greenbelt, MD 20771, USA}

\author[0000-0001-9873-0121]{Pier-Emmanuel Tremblay}
\affiliation{Department of Physics, University of Warwick, Coventry CV4 7AL, UK}

\author[0000-0002-1297-9485]{Kosuke Namekata}
\affiliation{Department of Astronomy, Kyoto University, Sakyo, Kyoto, Japan}
\affiliation{Astronomical Observatory, Kyoto University, Sakyo, Kyoto, Japan}
\affiliation{ALMA Project, NAOJ, NINS, Osawa, Mitaka, Tokyo, Japan}

\author[0000-0003-2760-2311]{David Kuridze}
\affiliation{Department of Physics, Aberystwyth University, Ceredigion, SY23 3BZ, UK}
\affiliation{Astrophysics Research Centre, School of Mathematics and Physics, Queen’s University Belfast, Belfast BT7 1NN, UK}
\affiliation{Abastumani Astrophysical Observatory, Mount Kanobili, 0301, Abastumani, Georgia}

\author[0000-0002-2554-1351]{Han Uitenbroek}
\affiliation{National Solar Observatory, University of Colorado Boulder, 3665 Discovery Drive, Boulder, CO 80303, USA}

\begin{abstract}
Red-shifted components of chromospheric emission lines in the hard X-ray impulsive phase of solar flares have recently been studied through their 30~s evolution with the high resolution of IRIS.  Radiative-hydrodynamic flare models show that these redshifts are generally reproduced by electron-beam generated chromospheric condensations.  The models produce large ambient electron densities, and the pressure broadening of hydrogen Balmer series should be readily detected in observations.  To accurately interpret upcoming spectral data of flares with the DKIST, we incorporate non-ideal, non-adiabatic line broadening profiles of hydrogen into the RADYN code.  These improvements allow time-dependent predictions for the extreme Balmer line wing enhancements in solar flares.   We study two chromospheric condensation models, which cover a range of electron beam fluxes ($1-5 \times 10^{11}$ erg s$^{-1}$ cm$^{-2}$) and  ambient electron densities ($1 - 60 \times 10^{13}$ cm$^{-3}$) in the flare chromosphere.  Both models produce broadening and redshift variations within 10~s of the onset of beam heating.    In the chromospheric condensations, there is enhanced spectral broadening due to large optical depths at H$\alpha$, H$\beta$, and H$\gamma$, while the much lower optical depth of the Balmer series H12$-$H16 provides a translucent window into the smaller electron densities in the beam-heated layers below the condensation.   The wavelength ranges of typical DKIST/ViSP spectra of solar flares will be sufficient to test the predictions of extreme hydrogen wing broadening and accurately constrain large densities in chromospheric condensations.

\end{abstract}

\keywords{}

\section{Introduction} \label{sec:intro}
Sunspot Cycle 25 has begun, and predictions for the severity of space weather range from one of the weakest maxima on record\footnote{\url{https://www.weather.gov/news/201509-solar-cycle}}, to one of the strongest \citep{McIntosh2020}.  Solar flares are one such manifestation of magnetic activity and space weather that will increase in regularity.  We are continuing to learn much about the solar fares that occurred in Sunspot Cycle 24, due in part to missions such as RHESSI, Hinode, the Goode Solar Telescope, and the Interface Region Imaging Spectrograph (IRIS) \citep{Melrose2018, Kerr2020A, Kerr2020B, DePontieu2021}.  In particular, power-law (hereafter, ``beam'') electrons with order-of-magnitude greater inferred heating fluxes than were typically reported in the past \citep[e.g.,][]{Neidig1994} have become commonplace \citep{Neidig1993, Krucker2011, Milligan2014, Alaoui2017, Graham2020}.  
Over the next cycle maximum, the National Science Foundation's Daniel K. Inouye Solar Telescope (DKIST) will provide solar flare optical and near-infrared spectra at the highest-ever spatial, temporal, and spectral resolution \citep{Rimmele2020, Rast2021}.  As a result, there will be a growing interest in the diagnostic potential of the hydrogen Balmer and Paschen series in solar flares, especially in bright flare kernels, where the high fluxes in power-law electrons occur.  Hydrogen lines are among the brightest and broadest emission lines in solar flares. They are thought to form a crucial connection between the enhanced pressures in the tenuous flare corona through line center intensity and the heating in lower layers through the line wing broadening \citep{Canfield1984, Canfield1987, Gayley1991, HF94, Namekata2020}.

A model paradigm connecting the upper and lower chromospheric heating in flares was recently developed in \citet{Kowalski2015}, \citet{Kowalski2017Mar29} (hereafter, Paper $\textsc{i}$), and \citet{KowalskiAllred2018} using radiative-hydrdodynamic (RHD) simulations of the response to high-flux electron beam heating calculated with the RADYN code \citep{Carlsson1992B, Carlsson1995, Carlsson1997, Abbett1998, Carlsson2002, Allred2015}.   In these models, redshifted emission line components originate from a downflowing $T \approx 10^4$ K chromospheric condensation with ambient charge density that attains values that are factors of $\gtrsim 3-100$ larger than inferred from uniform slab, static modeling of solar flare spectra in the past \citep[e.g.,][]{Svestka1967b, Neidig1983, Donati1985}.  The lower chromosphere below the condensation is heated by the highest energy electrons in the beam.  Compared to the condensation, these layers attain lower electron densities by factors of $\approx 10$ or more but extend over much larger path lengths.  Since they are far less dynamic than the condensation and produce chromospheric line intensity around the rest wavelength, these layers are referred to as the stationary chromospheric flare layers.  Chromospheric condensations\footnote{We distinguish between chromospheric condensations that are transient over timescales of less than one minute (and are typically reported to occur earlier in the flare in the hard X-ray impulsive phase), and persistent redshifts observed over tens of minutes into the gradual decay phase \citep[see][and references therein]{Lacatus2017, Reep2020}.  The long-duration redshifts are typically reported in transition region lines and may be due to the cooling of evaporated material, or the so-called ``coronal rain'' in the ``post-flare'' loops.  For lack of a better word in the English language, we follow previous works and use ``condensation'' to refer to the fact that the state changes from a nearly fully ionized plasma at early times to a nearly fully neutral gas at late times.} can be produced by beam heating and thermal heat flux originating from the coronal apex \citep[][Paper I]{Fisher1989, Ashfield2021}.  Chromospheric condensations have long been thought to be important in explaining many solar flare spectral lines \citep{Livshits1981, Ichimoto1984, Gan1993, Gan1994, Falchi1997, Abbett1999, Libbrecht2019}, but only recently have they been calculated through their full evolution  in radiative-hydrodynamic (RHD) models using high electron beam fluxes, as inferred from modern measurements of impulsive phase hard X-ray radiation.  These RHD models self-consistently account for the heating in the deeper layers (stationary chromospheric flare layers and the upper photosphere) as well.

  \citet{Graham2020} recently presented stunning spectral line observations of a $10^{32}$ erg solar flare (SOL2014-09-10T17:45) observed by IRIS.  They showed that the chromospheric emission lines of Fe $\textsc{i}$, Fe $\textsc{ii}$, Si $\textsc{ii}$, C $\textsc{i}$, and Mg $\textsc{ii}$ exhibit a common temporal evolution: 
 a highly redshifted satellite component that becomes less redshifted and brighter over time as it merges with the emission at the rest wavelength, resulting in an apparently single, broad line.  
 The RADYN models of the evolution of the stationary chromospheric flare layers and chromospheric condensation  were found to be largely consistent in spectro-temporal comparisons  with the observations.  One glaring discrepancy is that the modeled redshifted satellite component becomes far too bright relative to the rest-wavelength emission line component at late times in the beam heating.   Also, the duration of the red satellite component is a factor of three too short compared to the observed development.    Paper $\textsc{i}$ found that a larger-than-observed relative brightness in the red-shifted component in the RHD model of the SOL2014-03-29T17:48 flare could be explained by averaging over exposure times during the first 8~s.  However, \citet{Graham2020} demonstrated that such temporal averaging of the model spectra could not abate this problem when compared to the SOL2014-09-10T17:45 flare observations with much higher time resolution.

 Are the electron beam-generated chromospheric condensation models too dense, thus resulting in far too much emergent intensity in the predicted red-shifted emission line component?  This presents a conundrum: a lower ambient electron density in the condensation model in \citet{Graham2020} would also produce less near-ultraviolet (NUV) continuum brightness, which was found to be consistent with the IRIS observations.  A better understanding of the maximum densities achieved in chromospheric condensations would facilitate progress in establishing fundamental physical links between RHD in the lower atmosphere and particle acceleration physics in the highly magnetized corona \citep{Liu2009, Rubio2015, Allred2020, Arnold2021}.

With the Visible Spectropolarimeter \citep[ViSP;][de Wijn et al. 2022 in prep]{Nelson2010,dewijn2012,Rimmele2020, Rast2021} on the DKIST, spectral measurements of the optical hydrogen lines with wavelength coverage into the far wings and high spatial resolution will provide new estimates on the ambient charge density and optical depth in chromospheric condensations in bright flare kernels.  Due to a high sensitivity to pressure broadening, the H $\textsc{i}$ emission line properties are (potentially) much better probes of the ambient charged particle densities in flare chromospheres than the measured brightness and broadening properties of Fe $\textsc{ii}$, Mg $\textsc{ii}$, and Ca $\textsc{ii}$.  The physical interpretation of the emission line broadening from these ions is much more affected by free parameters in stellar atmospheric modeling, such as nonthermal broadening due to micro-turbulence and macro-turbulence \citep{Rubio2017, Kowalski2017Mar29,Zhu2019}. Further, semi-empirical enhancements to damping constants are required to account for the apparent limitations of applying standard electronic collisional broadening theory to models of non-hydrogenic lines in stellar atmosphere conditions \citep{Zhu2019}.  For proper modeling of the resonance lines of Mg $\textsc{ii}$ and Ca $\textsc{ii}$, the partial frequency redistribution of opacity in dynamic atmospheres must also be considered \citep{Hawley2007, Rubio2017, Kerr2019A, Kerr2019B, Zhu2019}.   In the model of Paper $\textsc{i}$ \citep[see also][]{Kowalski2019IRIS}, the variation in the red wing intensity relative to the rest-wavelength intensity in Fe $\textsc{ii}\ \lambda 2814$, Fe $\textsc{ii}\ \lambda 2832$, and Mg $\textsc{ii}\ \lambda 2791.6$ was argued to vary with the optical depth of the line.  As hydrogen lines are much more optically thick than these lines, they are expected to be formed over a narrower height range.  Thus, the hydrogen line broadening from the brightest flare kernels is expected to be less affected by strong atmospheric velocity gradients in semi-empirical models of resonance lines \citep{Rubio2017} and RHD models of Fe $\textsc{ii}$ lines \citep{Kowalski2017Mar29, Graham2020}.

Here we present DKIST/ViSP model spectra of hydrogen Balmer lines from the RADYN flare code with a critical update to the broadening by ambient charge density, which is referred to as electronic and ionic ``electric pressure broadening''.   This update was deployed in several recent RADYN flare models published in \citet{Zhu2019}, \cite{Graham2020}, \cite{Kuridze2020}, and \cite{Namekata2020} (see discussions therein).  In Section \ref{sec:background}, we introduce the broadening theory for hydrogen lines and summarize the various formulations that have been used in RADYN flare models in the past.  In Section \ref{sec:implementation}, we describe the method of implementing the new hydrogen profiles in RADYN.   In this section, we also summarize the hydrodynamics of the two chromospheric condensation models that are used in the broadening analysis and ViSP predictions, which are presented in Section \ref{sec:results}.
In Section \ref{sec:discussion}, we discuss the connection to models and observations of M dwarf flares gleaned by the new line broadening treatment, and we speculate on several areas of improvement in models of solar and stellar flare chromospheric condensations.   In Section \ref{sec:conclusions}, we summarize our findings and conclusions.

\section{The Broadening of Hydrogen Lines in Stellar Atmospheres} \label{sec:background}
Electric fields split the degenerate energy levels of hydrogen atoms, Rydberg atoms, and hydrogen-like ions by shifts that are directly proportional to the field magnitude \citep[$\propto E^1$; e.g.,][]{Condon1963, Gallagher2006, Goldman2006}, which is commonly known as the linear Stark, or Stark-Lo Surdo\footnote{This effect was discovered nearly simultaneously in experiments conducted by Johannes Stark and Antonino Lo Surdo in 1913 and 1914.  The relevant history is thoroughly summarized in \cite{Leone2004} and \cite{Longair2013}.  See \citet{Kleppner1981} for an excellent high-level overview.}, effect. Electric microfields from ambient charges in partially ionized plasmas lead to broadening of spectral lines, which is well-established as the source of prominent hydrogen absorption wings in main sequence A stars and white dwarfs.  This is sometimes called pressure, or collisional, broadening and is thought to be an important effect in the interpretation of symmetric broadening of solar and stellar flare hydrogen lines as well \citep{Svestka1962,Svestka1963, Svestka1965, Svestka1967, Cram1982, Neidig1983, Worden1984, Canfield1987, JohnsKrull1997, HP91, Allred2006, Paulson2006, Namekata2020}.

The densities of ambient protons ($n_p$) and electrons ($n_e$) affect the broadening of radiative transitions differently  \citep[for modern reviews, see][]{Gigosos2014, Hubeny2014}.  The slow-moving ambient protons and ions produce quasi-static microfields, meaning that the microfield changes slowly compared to the timescale over which radiation is emitted or absorbed \citep{Baranger1962, Griem1974, HM88, Barklem2016}.  The ionic microfield splits the energy levels of hydrogen into sub-states given by quantum numbers $(n,m_l,q)$.  The microfield magnitude probability distribution is described by a Hooper distribution, which is essentially a Holtsmark distribution that accurately accounts for Debye screening and plasma correlations \citep{Nayfonov1999}.  The perturbations from ambient electrons are more complicated \citep{Baranger1958, Vidal1970}.  The electron collisions produce quasi-static broadening at large detunings ($|\Delta \nu| = |\nu - \nu_o|$; in this paper, we use ``detuning''  interchangeably to refer to the wavelength, frequency, angular frequency distance from the rest value, which is denoted by the naught subscript).  At smaller detunings, the large thermal speeds of ambient electrons lead to damping, which is given by a Lorentzian/impact line profile.  This damping includes lifetime broadening through collisional depopulation \citep{Bransden, BohmVitense1989, HM88, Kunze2009} and cumulative phase shifts \citep[e.g.,][]{Foley1946, Cooper1966, Jefferies1968, Griem1974} that can be comparable in magnitude to the ionic microfield energy level splitting \citep[eg.,][]{Griem1959}.  
Transient electron collisions with hydrogen can be non-adiabatic within a large range of impact parameters, which means that there are broadening effects due to transitions among the microfield-split states within the same principal quantum number $n$  \citep[e.g.,][]{Baranger1958b, Kolb1958, Smith1969}.  Thus, quantum calculations are necessary. 

There are two widely used theoretical frameworks that include accurate calculations of the broadening due to ambient electrons from their quasi-static (far wing) through their  impact (near wing and line center) limits self-consistently with the ionic microfield splitting.  These are the model microfield method \citep{MMM1, MMM2, Seidel1977, Stehle1993, Stehle1994, Stehle1999} and the ``unified theory'' \citep[][hereafter, VCS]{Vidal1970, Vidal1971, Vidal1973};  see \citet{Barklem2000}, \citet{Tremblay2009}, \citet{Hubeny2014}, and \citet{Barklem2016} for reviews. 
 The VCS unified theory is used in models of hot star main sequence atmospheres \citep{Kurucz1979}\footnote{\url{http://kurucz.harvard.edu/grids/gridp00/}}. \citet{Tremblay2009} (hereafter, TB09) extended the VCS profile calculations to hydrogen line series members with higher upper levels ($n_j$) and recalculated their profiles to account for non-ideal gas effects (level dissolution) due to large ionic microfield splittings \citep{Seaton1990} and inelastic collisions with electrons \citep[][hereafter, HM88]{HM88}.  Here, we implement the updated VCS broadening profile calculations from TB09 into the time-dependent RADYN flare code.  Following \citet{Kowalski2017Broadening}, we use the nomenclature ``TB09$+$HM88''  to refer to these line profile functions, $\phi$, which have been widely adopted into modern-day models of hot star atmospheres \citep{Tremblay2011,Bohlin2014,TLusty,Bohlin2020} and high-density, laboratory experiments  \citep{Falcon2015, Gomez2016}.

 \subsection{Previous implementation of Hydrogen broadening in RADYN}

\begin{figure}
\gridline{\fig{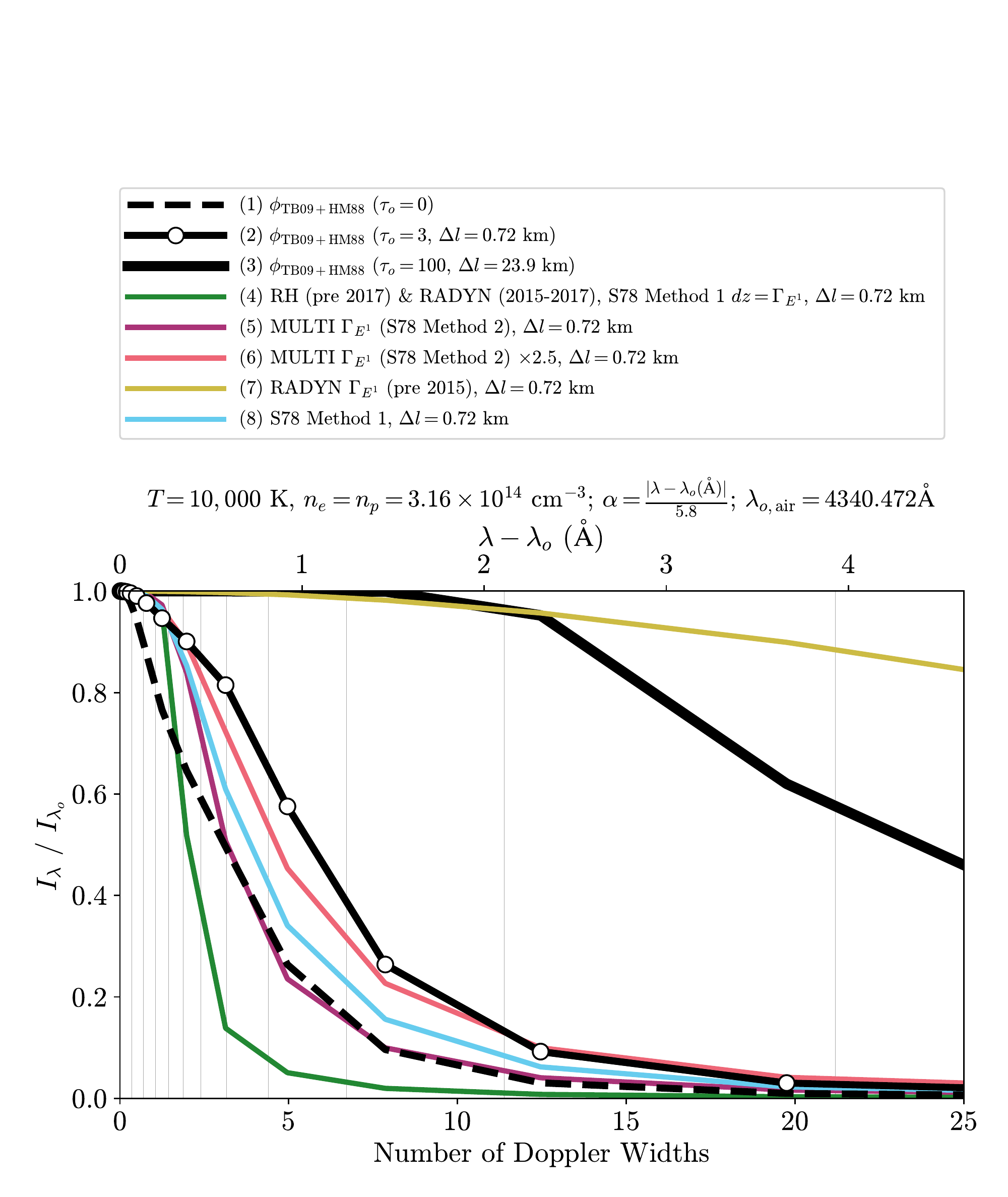}{0.45\textwidth}{(a)}}
\gridline{
          \fig{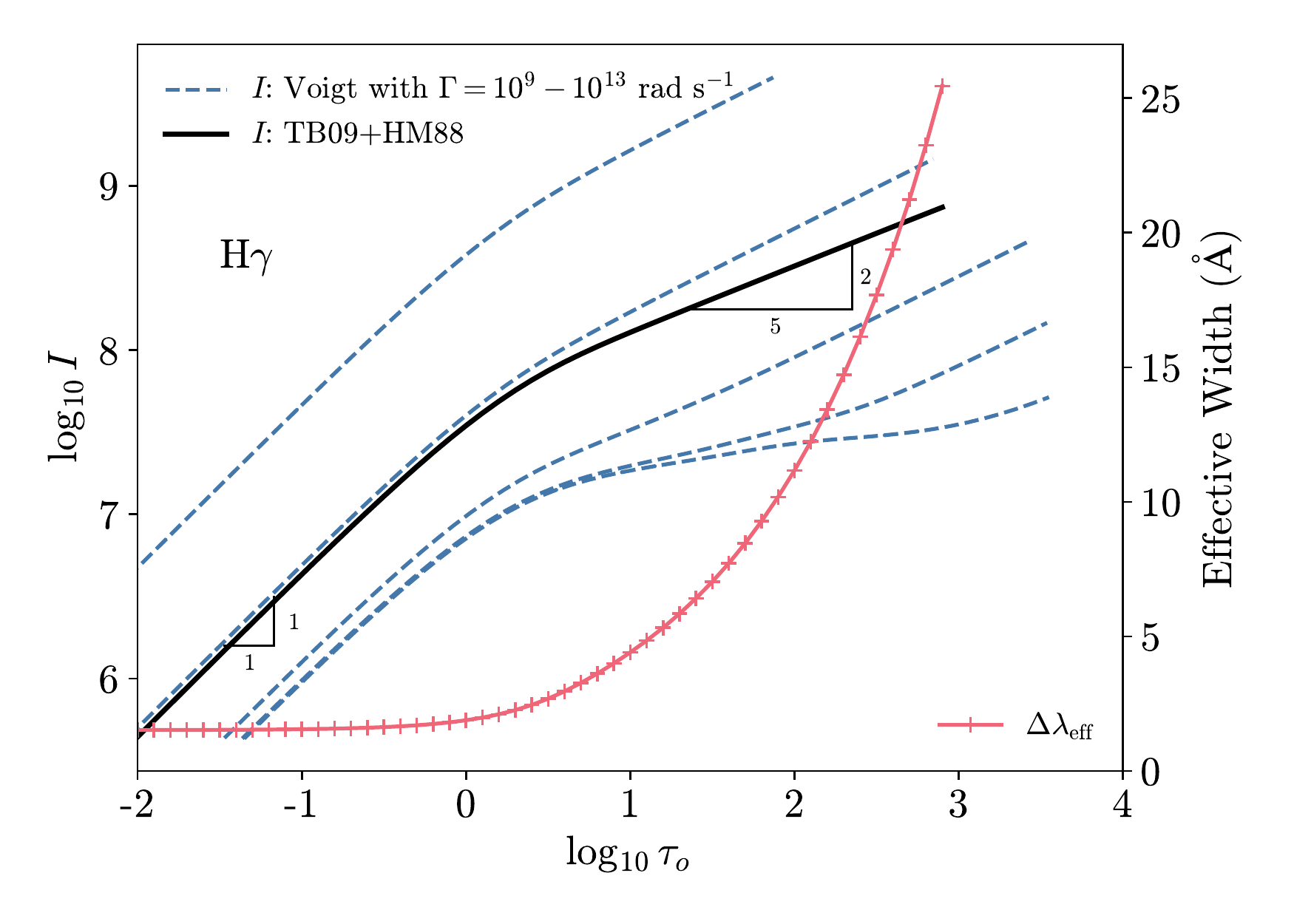}{0.45\textwidth}{(b)}
          \fig{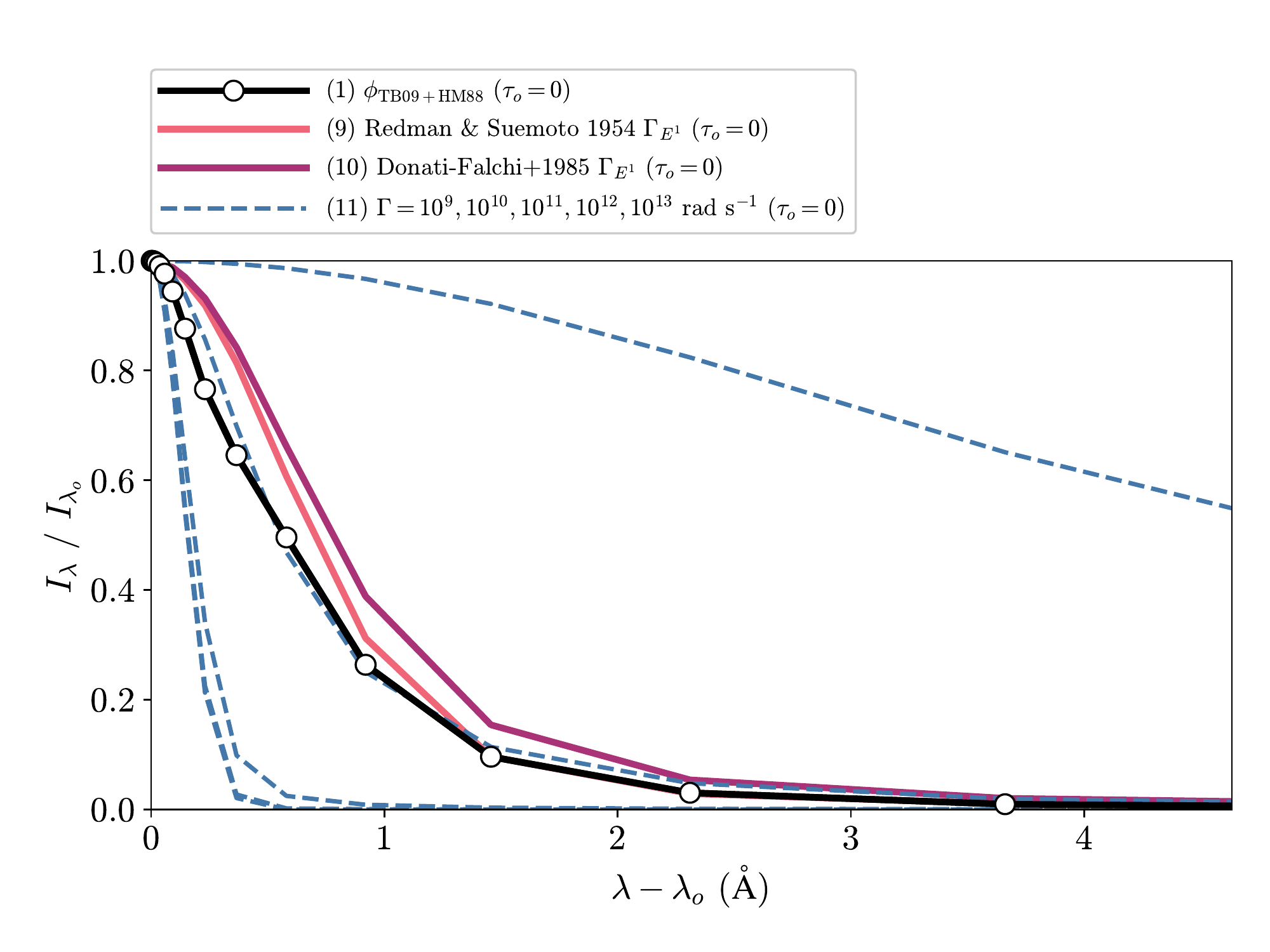}{0.45\textwidth}{(c)}}
\caption{(a) Emergent intensity spectra of hydrogen Balmer $\gamma$ (H$\gamma$) calculated in LTE, summarizing the long history of pressure broadening employed in model hydrogen spectra of solar flares.   All profiles include thermal Doppler broadening for $T=10,000$ K. All profiles in this figure are calculated at the detuning grid of the original TB09$+$HM88 profiles, indicated by circles in spectrum (2).  These are interpolated to the detuning grid in the RADYN models (see text), which is indicated for H$\gamma$ by the vertical grey lines.  For those calculations with a $\Gamma_{E^{1}}$ in the legend, a Voigt profile is used as $\phi$.  The black curves show the spectra using the TB09$+$HM88 profiles, which are now incorporated into detailed calculations of hydrogen  in RADYN.  For the spectra with solid linestyles (except for \textbf{(3)}), a path length ($\Delta l$) of 0.72 km is used in the calculations.  Over the maximum extent of the slab, $\tau=1$ occurs at 4.4 Doppler widths ($\Delta \lambda_{\rm{Dopp}} = 0.19$ \AA) for the calculation of spectrum \textbf{(2)}, which has an effective width of 2.6 \AA, assuming symmetry about the rest wavelength.  In spectrum \textbf{(1)}, the perturubing electrons are in the impact limit at $|\lambda - \lambda_o| \lesssim 20$ \AA\ \citep{Mihalas1978}.  (b) Curve-of-growth for the TB09$+$HM88 H$\gamma$ line at $n_e = 3.16 \times 10^{14}$ cm$^{-3}$ and $T=10,000$ K. The TB09$+$HM88 effective width is calculated and shown on the right axis.  In the Holtsmark wing regime at $\tau_o \gtrsim 5$, the effective width and wavelength-integrated line intensity (left-axis) increase as $\tau_o^{\frac{2}{5}}$.  For reference, we show the curve-of-growths for Voigt profiles with a range of damping parameters, $a=0.0027, 0.027, 0.27, 2.7, $ and $27$ with $\Delta \nu_{\rm{Dopp}} = 2.95 \times 10^{10}$ Hz.  (c) Several more broadening schemes that have been implemented in older studies of solar flare Balmer lines.   For reference, we show Voigt profiles with the same range of damping constants as in panel (b); note that the profiles with the two lowest damping constants overlap.
\label{fig:summary}}
\end{figure}

RADYN and other synthesis codes used in solar and stellar physics have implemented several schemes for the broadening of hydrogen lines.  These schemes are summarized and compared to the  TB09$+$HM88 profiles in Figure \ref{fig:summary} for spectra of the hydrogen Balmer $\gamma$ transition ($n_{i}=2 \rightarrow n_{j} = 5$; $\lambda_{o} = 4340.472$ \AA\ in air), hereafter ``H$\gamma$''.   The peak-normalized emergent intensity, $I_{\lambda}$/$I_{\lambda, \rm{max}}$, spectra from a slab with uniform gas density, $\rho=10^{-9}$ g cm$^{-3}$, and temperature, $T=10^4$ K, over a physical depth range, $\Delta l$, are calculated with the bound-bound opacity in local thermodynamic equilibrium (LTE).  

Optical depth effects are critical in accurate inferences of electron densities from hydrogen line spectra in flares \citep[e.g.,][]{Svestka1962, JohnsKrull1997, Namekata2020}.
The emergent intensity differences in the wings of the spectra in Figure \ref{fig:summary}(a)  demonstrate the broadening enhancements that result from optical depth variations in general curve-of-growth analyses of hydrogen lines \citep[e.g.,][]{Sutton1978, BohmVitense1989}.  These broadening effects are related to what is sometimes described as ``self-absorption'' over the line profile \citep[e.g.,][]{Kunze2009}.    Wing enhancements occur when the emergent intensity profile at wavelengths around $\lambda_o$ experience strong amounts of ``opacity broadening'', which is defined for optically thick, non-hydrogenic lines formed over a height-dependent non-LTE source function in the quiet Sun in \citet{Rathore2015A}. The optical depth through the maximum extent of the slab at $\lambda_o$ is indicated for each calculation in Figure \ref{fig:summary}(a);  the emergent intensities for the $\tau_o =0$ calculations are optically thin and are directly proportional to the line profile function $\phi$.   The intensity spectra plotted as black curves \textbf{(1, 2, 3)} are the TB09$+$HM88 profiles for a range of values of $\tau_o = 0, 3, $ and $100$, which are calculated by varying $\Delta l$.  These simplified profile calculations demonstrate that the optically thin wings become enhanced relative to the saturated line core as the optical depth (path length) increases.

A curve-of-growth of H$\gamma$ at this density and temperature is shown in Figure \ref{fig:summary}(b) compared to Voigt profiles with a range of generic damping constants $\Gamma$;  note that the term ``curve-of-growth'' here is not related to typical applications in abundance analyses of non-hydrogenic photospheric lines.   In the quasi-static wing regime of the curve-of-growth, log$_{10}$ ($I$) v.\ log$_{10}$ $(\tau_o)$ (where $I$ is the wavelength-integrated intensity), the slope is $\approx \frac{2}{5}$, as expected for H$\gamma$ at these densities where non-ideal effects are not important.  The linear and quasi-static slopes are indicated in Figure \ref{fig:summary}(b).   The three representative optical depths ($\tau_o = 0, 3, 100$) in Figure \ref{fig:summary}(a) respectively span the linear, saturated, and quasi-static wing regimes in the curve-of-growth of this line.  In flare atmospheres, the Balmer-emitting atmospheric regions exhibit large optical depths over small path lengths ($\approx 0.5-20$ km) with heterogeneous densities and temperatures, which rapidly evolve on short timescales ($\approx 1$~s).

Prior to 2015, Voigt profiles with electric pressure broadening damping constants $\Gamma_{E^1}$ have been used in RADYN to model hydrogen lines, according 
to the expression
\begin{equation}
\Gamma_{E^1} = G_S \times n_e.
\end{equation}

\noindent The $G_S$ values are specified in the atomic input files and have typically varied among modeling applications.  Several representative values
that have been chosen for flare modeling with RADYN in the past are $G_S =  3.836 \times 10^{-2}$ for Balmer $\beta$ (H$\beta$), $G_S = 6.715 \times 10^{-2}$ for H$\gamma$, $G_S = 3.212 \times 10^{-4}$ for Paschen $\alpha$.  The value of  $\Gamma_{E^1}$ for Balmer $\alpha$ (H$\alpha$) is given by $4.737 \times 10^{-7} \times n_{\rm{HI}, n=1}$.
The prescription for H$\gamma$ is shown in Figure \ref{fig:summary}(a) as spectrum \textbf{(7)}, which is clearly far too broad at this electron density ($n_e = 3.16 \times 10^{14}$ cm$^{-3}$) in comparison to the TB09$+$HM88 spectrum \textbf{(2)}.  
  
  In most RADYN flare models presented in the literature from 2015-2017, the value of $dz$ from Eq. 17 in \citet{Sutton1978} (hereafter, S78) was equated to $\Gamma_{E^1}$ in the Voigt profiles in RADYN \citep[see the description in Section 5.2 of][]{Allred2015}.  This implementation is used in the calculation of spectrum \textbf{(4)} in Figure \ref{fig:summary}(a), in the solar  flare models in Paper I, and in the M dwarf flare models of \citet{Kowalski2015, Kowalski2016}.  This simplification was done for direct comparisons to spectra calculated by the standard version of the RH code \citep{Uitenbroek2001} and was a temporary placeholder in RADYN, as discussed in these papers. It has long been recognized that the S78 profiles result in significantly narrower lines for transitions with high $n_j$ compared to proper theoretical treatments \citep{JohnsKrull1997, Kowalski2017Broadening}.  

The radiative transfer codes MULTI \citep{MULTI1, MULTI2} and MULTI3D \citep{MULTI3D} are also widely used in the solar and stellar atmospheric modeling community.  The MULTI code multiplies the expression for $dz$ in S78 by 0.425 $\times 4\pi$ to obtain $\Gamma_{E^1}$ for the electron damping constant.  This is equivalent to Method \#2 in S78 with an additional $4\pi$ factor that converts from the S78 form of $\gamma$ (which is half of a Lorentzian full-width-at-half-maximum (FWHM) in units of s$^{-1}$) to the proper form of $\Gamma$ (the FWHM of a Lorentzian in units of rad s$^{-1}$) that is added into the typical Voigt damping parameter $a = \frac{\Gamma}{4 \pi \Delta \nu_{\rm{Dopp}}}$ \citep[e.g.,][]{BohmVitense1989, Rutten2003}.   MULTI's implementation has been used in the solar flare models of \citet{HF94} and is shown in Figure \ref{fig:summary}(a) as spectrum \textbf{(5)} with  $\Gamma_{E^1}=3.1 \times 10^{11}$ rad s$^{-1}$.  For this density and path length, the spectrum is nearly half as broad as the TB09$+$HM88 spectrum \textbf{(2)}. For the infrared hydrogen lines ($n_{i} \ge 4$), MULTI adds additional damping from ambient proton and ion densities, and the resulting spectra well match the quiet Sun absorption profiles \citep{Carlsson1992}.  

Spectrum \textbf{(6)} is the result of multiplying the value of $\Gamma_{E^1}$ in MULTI's calculation by a factor of 2.5, which was determined by equating the effective width to that of the TB09$+$HM88 spectrum \textbf{(2)}.  The effective width is 

\begin{equation} \label{eq:effective_width}
\Delta \lambda_{\rm{eff}} = \int_{\lambda_1}^{\lambda_2} \frac{I_{\lambda}(\lambda) - I_{\rm{cont}}(\lambda)}{I_{\lambda}(\lambda_{\rm{max}})- I_{\lambda,\rm{cont}}(\lambda_{\rm{max}})  } d\lambda 
\end{equation}

\noindent where $\lambda_{\rm{max}}$ is the wavelength of maximum emergent intensity over the emission line, $I_{\lambda,\rm{cont}}$ is the local continuum intensity that is interpolated to the emission line wavelengths,  and $\lambda_1$ and $\lambda_2$ define the limits of the emission line\footnote{For a Lorentzian and a Gaussian profile with the same FWHM, the effective width for a Gaussian is 5\% greater than the FWHM, while the effective width for a Lorentzian is 50\% greater. The effective width is a full width across a spectral line, and it is sometimes referred to as an equivalent width in light curve analysis \citep{Aschwanden1995}.}. The units of $I_\lambda (\lambda) $ are those of specific intensity (\ilam).  The effective width will be used to compare the broadening magnitudes for model spectra in Section \ref{sec:results}. Spectrum \textbf{(6)} highlights the different shapes of Lorentzian and Holtsmark wings after the respective saturation regimes in the curve-of-growth (Figure \ref{fig:summary}b).  
The unified theory profiles can indeed be accurately represented with a Lorentzian profile that has a detuning-dependent damping \citep[][and for a succinct review see Section 8.5 pp. 255-257 of \citealt{Hubeny2014}]{Vidal1970, Griem1974, Cooper1989}.  However, these approximations do not include the improvements implemented in TB09.  For very low electron densities ($n_e < 10^{12}$ cm$^{-3}$), Lorentzian profiles with detuning-independent damping have been used for the H$\beta$ line \citep{Stehle1983, Stehle1984, Stehle1988}.  For the electron density in Figure \ref{fig:summary}, however, spectrum \textbf{(1)} has a slope of $-\frac{5}{2}$ in log$_{10} \, I_{\lambda}$ vs log$_{10} \, |\lambda - \lambda_o|$ space
from detunings of  $\approx 1.4$ \AA\ to $60$ \AA. S78 proposed a different analytic profile function (their Method \# 1) that approaches a Holtsmark power-law ($\propto |\lambda - \lambda_o|^{-5/2}$) in the wings; this is convolved with Doppler broadening and is shown as spectrum \textbf{(8)}, which is clearly not sufficiently accurate in comparison to the TB09+HM88 spectrum \textbf{(2)}.

To facilitate comparison with older solar flare analyses we show two additional methods compared to spectrum $\textbf{(1)}$ for $\tau_o=0$ in Figure \ref{fig:summary}(c). 
\citet{vanDien1949} parameterized the microfield splitting, which was adopted in Voigt profiles as the ion damping in solar spectral analysis in \citet{Redman1954} and \citet{Svestka1963}.  This approach is discussed in \citet{JohnsKrull1997} and is shown as spectrum \textbf{(9)}.  This is close to the TB09$+$HM88 line profile function in this wavelength range.  \citet{Svestka1965} and \citet{Svestka1967} later added detuning-dependent electron damping from the modified impact theory \citep{Griem1960, Griem1962} to the Voigt profiles  in solar flare spectral analyses.    The models of high$-n_j$ Balmer lines in solar flares in \citet{Donati1985} used Griem's electron damping half, half-widths \citep[see also][]{deFeiter1964} to normalize the Voigt profiles with effective electron damping constants, which were added to estimates of the ion broadening from \citet{Griem1960}.  This method is applied to H$\gamma$ and is shown as spectrum \textbf{(10)}.  At $|\lambda - \lambda_o| \gtrsim 5$ \AA, the Holtsmark wings of the TB09$+$HM88 spectrum \textbf{(1)} diverge from the Lorentzian power-law in spectra \textbf{(9-10)}. The various implementations of Sutton's formulae, Griem's modified electron damping constants, and van Dien's microfield broadening parameterizations have lead to large systematic errors by factors of $\approx 3-10$ in inferences of the electron densities from solar and stellar flare spectra \citep{Svestka1967, JohnsKrull1997, Kowalski2017Broadening}.  Generic Voigt profiles with $\tau_o=0$ and a range of damping constants are shown for comparison in Figure \ref{fig:summary}(c).  Their curve-of-growth variations in Figure \ref{fig:summary}(b) further demonstrate that carefully chosen damping parameters may show an approximate match in intensity spectra at $\tau_o=0$ but are increasingly inaccurate at larger values of $\tau_o$.

The broadening differences in Figure \ref{fig:summary}(a)-(b) are far more severe for higher $n_j$ Balmer lines, such as H10 and H14 (\emph{cf} Fig.\ 1 of \citealt{Kowalski2017Broadening} and Fig.\ 7 of \citealt{JohnsKrull1997}), but the differences are far less severe for  H$\alpha$.  Thus it is desirable to implement a self-consistent theory of the broadening of all hydrogen Balmer lines into the far wings \citep[e.g., Figure 8 of][]{Vidal1971} for large ambient densities, which are expected to be generated from Coulomb heating of the solar chromosphere by high fluxes of nonthermal electrons during flares.

\section{New Hydrogen Broadening Profiles in RADYN} \label{sec:implementation}

In our modifications to the RADYN code, we interpolate the grid of the TB09$+$HM88 line profile functions, $\phi_{\alpha}(\alpha, n_e, T|n_{\rm{i}} \rightarrow n_{\rm{j}})$.  These profiles have been convolved with thermal Doppler broadening outside of RADYN.  The grid of $n_e$ ranges from $10^{10}$ to $10^{18}$ cm$^{-3}$, and the temperature grid ranges from $2500$ K to $160,000$ K.  The detuning parameter $\alpha$ is $(\lambda - \lambda_{o}) / F_{\rm{Norm}}$, where $\lambda$ is in units of $\rm{\AA}$,  $F_{\rm{Norm}} = 1.25 \times10^{-9} Z_p n_e^{2/3}$ in \emph{cgs} units (1 statvolt cm$^{-1} = 30,000$ V m$^{-1}$), and $Z_p=1$.  $F_{\rm{Norm}}$ is known as the normal field strength (sometimes referred to as $F_o$), and it is assumed that all ionic charge is distributed through the plasma as protons (see HM88) giving $n_e = n_p$.   In the TB09$+$HM88 line profiles, there is no nonthermal Doppler broadening (microturbulent velocity), which is 2 km s$^{-1}$ in RADYN by default for non-hydrogen lines that are calculated in detail.   The self-broadening (resonance and van der Waals broadening due to H-H collisions) from \citet{Barklem2000} and natural damping for the Balmer lines are a factor of $\approx 10^{-4}$ of the TB09$+$HM88 broadening at the densities, temperatures, and ionization fractions expected in flare chromospheres (e.g., Figure \ref{fig:summary}); these damping contributions are excluded in the new RADYN flare simulations.

RADYN uses a 6-level hydrogen atom including the H II ionization stage in its detailed calculations.  Thus, only TB09$+$HM88 profiles for Ly $\gamma$ and Ly $\delta$, H$\alpha$, H$\beta$, and H$\gamma$, Paschen (Pa) $\alpha$ and Pa $\beta$, and Brackett (Br) $\alpha$ are included.  We do not change the Ly $\alpha$ and Ly $\beta$ profiles from previous implementations in RADYN; these two lines are least affected by the choice of pressure broadening scheme, but they will be investigated in detail \citep[e.g., following][]{Cooper1989, Kerr2019A} in future work.
We use bilinear interpolation\footnote{We experimented with interpolation of log$_{10}\: \phi_{\alpha} (\alpha)$ on a grid of (log$_{10}\: n_e$, log$_{10}\: T$) but found no differences. In the density regime of $n_e = 3\times 10^{12}$ to $10^{16}$ cm$^{-3}$, the $\phi_{\alpha}(\alpha)$ curves overlap within each low $n_j$ transition.  Thus, uncertainties in the interpolation are minimized in $\alpha$ space.  This would not be true if instead log$_{10}$ $\phi_\lambda (\lambda)$ were interpolated.} of log$_{10}\: \phi_{\alpha} (\alpha,n_e, T)$ over the grid of ($n_e, T$).  Beyond the grid ranges, we experimented with several methods.  We found that setting the hydrogen profiles to the profiles at the nearest edge of the grid prevented RADYN 
from stalling with small time-steps in the relaxation of the starting atmosphere in the transition region at $T = 160,000$ K, where the neutral hydrogen density is only $10^{-10}$ to $10^{-12}$ of the total population density.
By adapting the profiles at the edges of our line profile grid at locations in the atmosphere that exceed the range in one or both variables (electron density and/or temperature), we ensure smooth derivatives in the population densities even where the population densities are relatively small but not quite small enough that they are ignored in the error terms in the convergence.

The line profile function $\phi_{\alpha}(\alpha)$ is the conditional transition probability density, which is symmetric and unit-normalized from $\alpha = -\infty$ to $+\infty$ \citep{Vidal1973}.  A change of variables is done according to 

\begin{equation}
  \phi_{\alpha}(\alpha) |d\alpha|  = \phi_{\nu}(\nu) |d\nu| \\
   \Rightarrow \phi_{\nu}(\nu) = \phi_{\alpha}(\alpha)\frac{c}{F_{\rm{Norm}} \nu^2}  
\end{equation}

\noindent and $\phi_{\nu}(\nu)$ has units of probability per unit frequency.  The NLTE bound-bound opacity \citep{Mihalas1978} for $n_{i} \rightarrow n_{j}$ is given by 

\begin{equation}\label{eq:bbopacity}
\chi(\nu,  n_{i} \rightarrow n_{j}) [\mathrm{cm}^{-1}] = \frac{\pi e^2}{m_e c} f_{\rm{osc}}  \left( n_{i} - \frac{g_{i}}{g_{j}} n_{j} \right) \phi_{\nu}(\nu|n_i \rightarrow n_j)
\end{equation}

\noindent where $n_{i}$ and $n_{j}$ are the non-equilibrium population densities for the lower and upper levels, respectively, of the transition. 
The opacity is shifted by the gas velocity at each depth point, and complete frequency redistribution (i.e., the line absorption profile function, $\phi_{\nu}$, is equal to the line emission profile function, $\psi_{\nu}$) over each transition is assumed in RADYN.  The TB09$+$HM88 profiles are linearly interpolated to the observer-frame frequency grids, which are set \emph{a priori} for each line in RADYN. 
These calculations are implemented in one new routine, \verb|phi_tb09.f|. 

The Balmer H$\gamma$, H$\beta$, and H$\alpha$ profiles are the only hydrogen lines that are analyzed in detail in this study.  The H$\gamma$ and H$\beta$ lines are calculated on a frequency grid with 31 points (the sampling near $\lambda_o$ on the red side of H$\gamma$ is indicated by vertical thin grey lines in Figure \ref{fig:summary}a), whereas  the H$\alpha$ profile is calculated at 51 frequency points.   Due to the different wavelength sampling in the wings, we find that trapezoidal integrations of the  line profile functions systematically result in 5\% larger integrations (e.g., Eq.\ \ref{eq:effective_width}) of H$\gamma$ compared to H$\alpha$.  These interline systematic errors were evaluated by shifting the TB09$+$HM88 line profile functions on their original detuning grids (e.g, circle symbols in Figure \ref{fig:summary}) by $-45$ km s$^{-1}$, linearly interpolating  to RADYN's frequency grid as in  phi\_tb09.f, and comparing to other interpolation schemes, such as a four-point, third-order polynomial interpolation scheme in log$_{10} \, \phi_{\alpha}$ vs.\ log$_{10} \, \alpha$ space, on a very fine frequency grid.  To estimate systematic effects on the emergent line intensity, we use a solver of the equation of radiative transfer \citep[by the Feautrier method as described in][]{MULTI1} with several line profile interpolation schemes\footnote{Now provided in RADYN's Python and IDL analysis tools.} on coarse frequency grids.  Such tests show systematic $1-5$\% variations of the integrated emergent intensity for a wide range of flare atmospheres in our models. The RADYN interpolations systematically overpredict the wings by this percentage range in comparison to more sophisticated detuning interpolation schemes on a coarse grid.  However,  RADYN well reproduces the Holtsmark limit in the power-law in the optically thin wings of the emergent intensity spectra.   

To our knowledge, the modifications to RADYN described above encompass the first time that an accurate theory of hydrogen Balmer line broadening due to charged particle perturbations has been included in a time-dependent modeling code that is widely used by the solar and stellar atmospheres communities.  Our method dramatically reduces the intractable systematic errors of one hundred percent and larger in many previous time-dependent models (\emph{cf.} Figure \ref{fig:summary}a).  The new levels of systematics discussed above are tractable, and, moreover, are far smaller than the trends that we discuss in detail in Section \ref{sec:results}.  If finer frequency grid sampling is required, emergent intensity spectra can be calculated with the RH code as discussed in \citet{Kowalski2017Broadening} and in Section \ref{sec:lz} here.

\vspace{10mm}
\subsection{Solar flare simulations} \label{sec:rhd}

The RADYN flare simulations that are used for hydrogen Balmer line spectra analysis (Section \ref{sec:results}) are described and analyzed in  \citet{Graham2020}, \citet{Kuridze2020}, and 
Paper $\textsc{i}$.  These simulations employ collisional heating due to a nonthermal electron beam using a Fokker-Planck solver \citep{Allred2015, Allred2020}.  The model IDs are listed in Table \ref{table:models} with the injected beam energy flux density (F11 or 5F11), power-law index ($\delta$), low-energy cutoff ($E_c$), the injected pitch-angle distribution for the beam electrons ($\mu_o$), and the duration of the beam energy injection. The initial atmosphere from Paper I was relaxed to hydrostatic equilibrium with the TB09$+$HM88 profiles.  This initial atmosphere is representative of an active region plage with a transition region at a mass column density of log$_{10} \: m = - 4.9$ g cm$^{-2}$ \citep{Metcalf1990, HF94, Abbett1998}.   The 5F11 models of Paper $\textsc{i}$ have been re-calculated with the TB09$+$HM88 profiles, and the F11 models in \citet{Graham2020} and \citet{Kuridze2020} already include the new profiles. We refer the reader to these works for further details about the model setup.
Other updates to the RADYN flare code that have been made since \citet{Allred2015} are outside the scope of this paper to describe in full. These updates do not impact the results in this study, and all models in Table \ref{table:models} were calculated with the same input files and same version of the RADYN source code.

\begin{deluxetable}{lcccccc}
\tabletypesize{\scriptsize}
\tablewidth{0pt}
\tablecaption{Electron Beam Solar Flare Models with TB09$+$HM88 Hydrogen Profiles}
\tablehead{
\colhead{model ID} &
\colhead{Beam Flux [erg s$^{-1}$ cm$^{-2}$] } & \colhead{$\delta$} & \colhead{$E_c$ [keV]} & \colhead{$\mu_o$} & \colhead{Beam Duration [s]} & \colhead{Comment} }
\startdata 
c15s-5F11-25-4.2 & $5 \times 10^{11}$ (5F11) & 4.2 & 25 & 0.1 & 15 & ``Extended heating'' 5F11 in Paper I.\\
c20s-F11-25-4 & $10^{11}$ (F11) & 4 & 25 & 0.1 & 20 & Published in \citet{Kuridze2020}. \\
c20s-F11-15-5 & $10^{11}$ (F11) & 5 & 15 & 0.1 & 20 & Published in \citet{Graham2020}. \\
\enddata
\tablecomments{ The c20s-F11-25-4 was also analyzed in Paper I, \citet{Kuridze2015}, and \citet{Kuridze2016} with the old implementation of $\Gamma_{E^1}$ (see text).  The value of $\mu_o$ refers to $\sigma_{\mu} / \sqrt{2}$, which is the $1/e$ half-width of the injected beam pitch angle distribution, measured with respect to the magnetic field \citep[see][]{Allred2015}.  The pitch angle distribution is assumed to be a Gaussian beam in $\mu$-space in the forward hemisphere. Note that an isotropic distribution in the forward hemisphere was used in Paper I for the 5F11 model;  only relatively minor differences in heating are expected between forward-hemisphere collimation and forward-hemisphere isotropy \citep{Allred2015}. }
\end{deluxetable}\label{table:models}

\begin{figure}
\begin{center}
\includegraphics[scale=0.8]{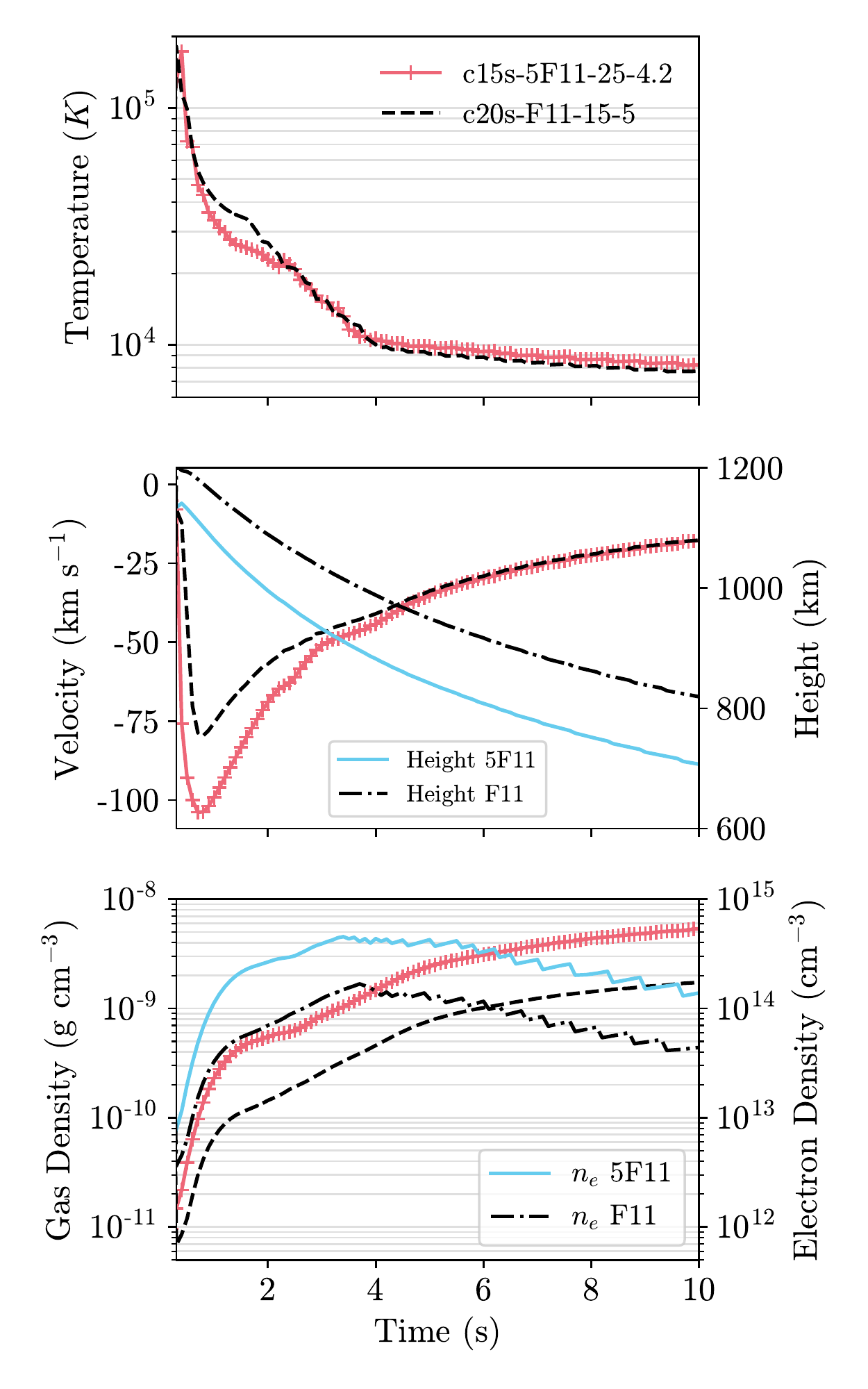}
\caption{Evolution  ($\Delta t = 0.1$~s) of the atmospheric variables at the location of maximum gas density in the chromospheric condensation in the 5F11-25-4.2 and F11-15-5 models.  The time range is shown for $t = 0.3 - 10$~s.  The maximum gas density in the condensation ($\rho_{\rm{max}}$) is shown in the bottom panel (left axis).  The locations of maximum densities in the respective condensations are shown in the middle panel, right axis. \label{fig:ccevol} }
\end{center}
\end{figure}

The three models in Table \ref{table:models} were chosen because they sample a range of chromospheric condensation densities and beam parameters.   
We focus the analysis of the broadening differences over the time interval of $t=0-10$s in the c15s-5F11-25-4.2 (hereafter, 5F11-25-4.2) and the c20s-F11-15-5 (hereafter, F11-15-5)
models.  The evolution of the thermodynamic properties at the maximum gas mass density ($\rho_{\rm{max}}$) in the chromospheric condensations  in these two models is summarized in Figure \ref{fig:ccevol}, following the analyses in Paper I and \citet{Kowalski2015}.  The height locations (middle panel, right axis) of the maximum densities descend as the condensations accrue mass, radiatively cool, and decelerate. The temporal maximum of the  electron densities at $\rho_{\rm{max}}(t)$ are $1.5 \times 10^{14}$ cm$^{-3}$ and $4.5 \times 10^{14}$ cm$^{-3}$ in these two models, respectively.  Remarkably, the maximum downflow velocities are consistent with the analytic relationship of \citet{Fisher1989}, who predicts (\emph{cf.} their Eq 34b) $-86$ km s$^{-1}$ and $-100$ km s$^{-1}$ for the F11 and and 5F11 beams, respectively.  Though the F11 beam has a factor of five less energy flux, its lower low-energy cutoff ($E_c = 15$ keV) and softer ($\delta = 5$) power-law produces a runaway temperature increase higher up in the chromosphere at lower ambient density, thus resulting in a similar dynamical evolution.  
The F11 model with a higher low-energy cutoff and harder beam (c20s-F11-25-4, hereafter F11-25-4) develops a condensation after  $t \approx 10$~s of heating. However, the condensation cools to only $T\approx$35,000 K by the end of the beam heating at $t=20$~s, producing the enhancement in the violet peak of H$\alpha$ spectrum as described in \citet{Kuridze2015}.

As discussed in Paper I and \citet{Graham2020}, the flaring layers below the condensation consist of a radiatively backwarmed photosphere and lower chromospheric flare layers that are directly heated by the highest energy beam electrons.  In this paper, we refer to the arc distance $z$ along the semi-circular flare loop from $\tau_{500} = 1$ as the ``height''.  The chromospheric flaring layers from $z \approx 400$ km up to the heights of the bottom of the condensation are referred to as the ``stationary chromospheric flare layers'', following the terminology\footnote{There are very small upflows in these layers, $<1$ km s$^{-1}$ by 10~s in the 5F11 model, as expected from the thermal pressure gradient formed in these layers in the flare.} used in previous works.  Since negative velocities in RADYN models correspond to flows toward the bottom of the loop, negative velocities represent spectral redshifts and positive velocities represent spectral blueshifts throughout this paper. 

\section{Modeling Analysis and Results} \label{sec:results}
From the models in Table \ref{table:models}, we first present revised predictions of H$\alpha$ spectra that have been previously analyzed using broadening method \textbf{(4)} in Figure \ref{fig:summary} (Section \ref{sec:mar29}).  In Section \ref{sec:mar29}, the updated 5F11-25-4.2 emergent intensity spectra are presented for $\mu=0.77$, which is the viewing angle for this model that was analyzed in Paper I.  All other emergent intensity spectra are presented for $\mu=0.95$.  
In Section \ref{sec:visp}, we analyze the high-time resolution evolution of the broadening of the H$\gamma$ line in the two models (5F11-25-4.2 and F11-15-5) in Figure \ref{fig:ccevol}  with 
prompt chromospheric condensations that develop and cool rapidly.  We further focus our analysis on the H$\gamma$ line for the 5F11 model:  we discuss its formation properties using the Eddington-Barbier relation, the emergent spectral broadening that results from large optical depths and electron densities, and the degree to which this line is formed near LTE.  In Section \ref{sec:dec}, we compare the broadening and wavelength-integrated intensity of H$\gamma$ to the other low-$n_j$ Balmer lines and to Ca II K in both models.  
In Section \ref{sec:lz}, we complement these prediction to new calculations of the extremely broad hydrogen lines at the Balmer limit ($\lambda = 3646$ \AA) using the RH code \citep{Uitenbroek2001}, which reveals a new diagnostic of the electron densities below the optically thick layers in the condensation where the H$\alpha$, H$\beta$, and H$\gamma$ lines are formed.

\subsection{Revised Predictions for Fabry-Perot Spectral Imaging Observations of H$\alpha$} \label{sec:mar29}
Fabry-Perot Imaging Spectrographs are some of the most unique instruments in solar physics.
The IBIS instrument on the Dunn Solar Telescope and the CRISP instrument on the Swedish Solar Telescope provide
 line scans of H$\alpha$, Ca $\textsc{ii}$, Na $\textsc{i}$, and He $\textsc{i}$ at high spatial resolution in solar flares \citep[e.g.,][]{Kuridze2015, Kleint2015, Libbrecht2019}.  The pre-filters employed on Fabry-Perot 
spectrographs have typically sampled out to $\lambda_o \pm 1.5$ \AA\ or $\pm 1.7$ \AA\ and do not cover the broad wing profiles in flares.  This has led to incomplete coverage in these otherwise stunning flare observations.

\begin{figure}
\gridline{\fig{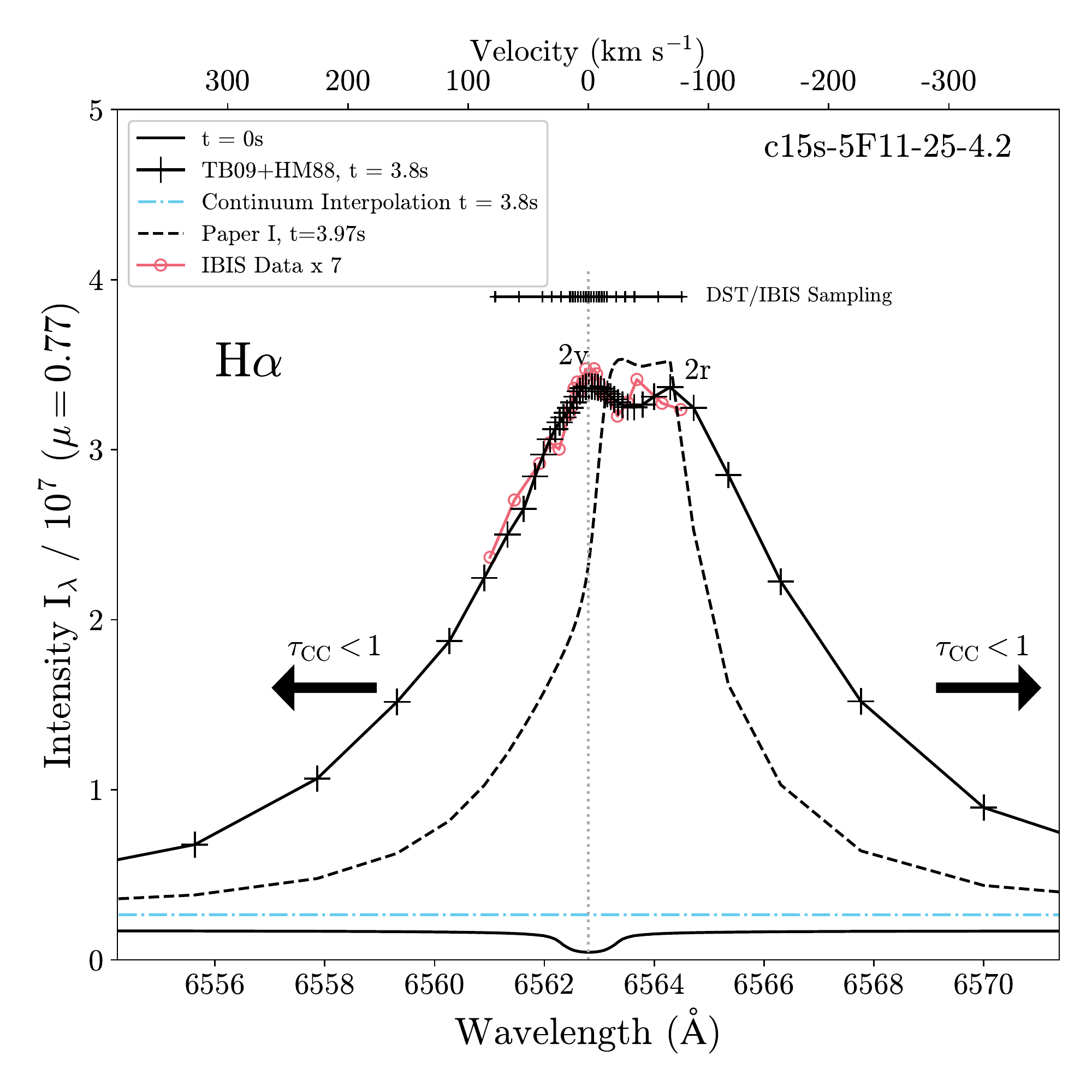}{0.4\textwidth}{(a)}
          \fig{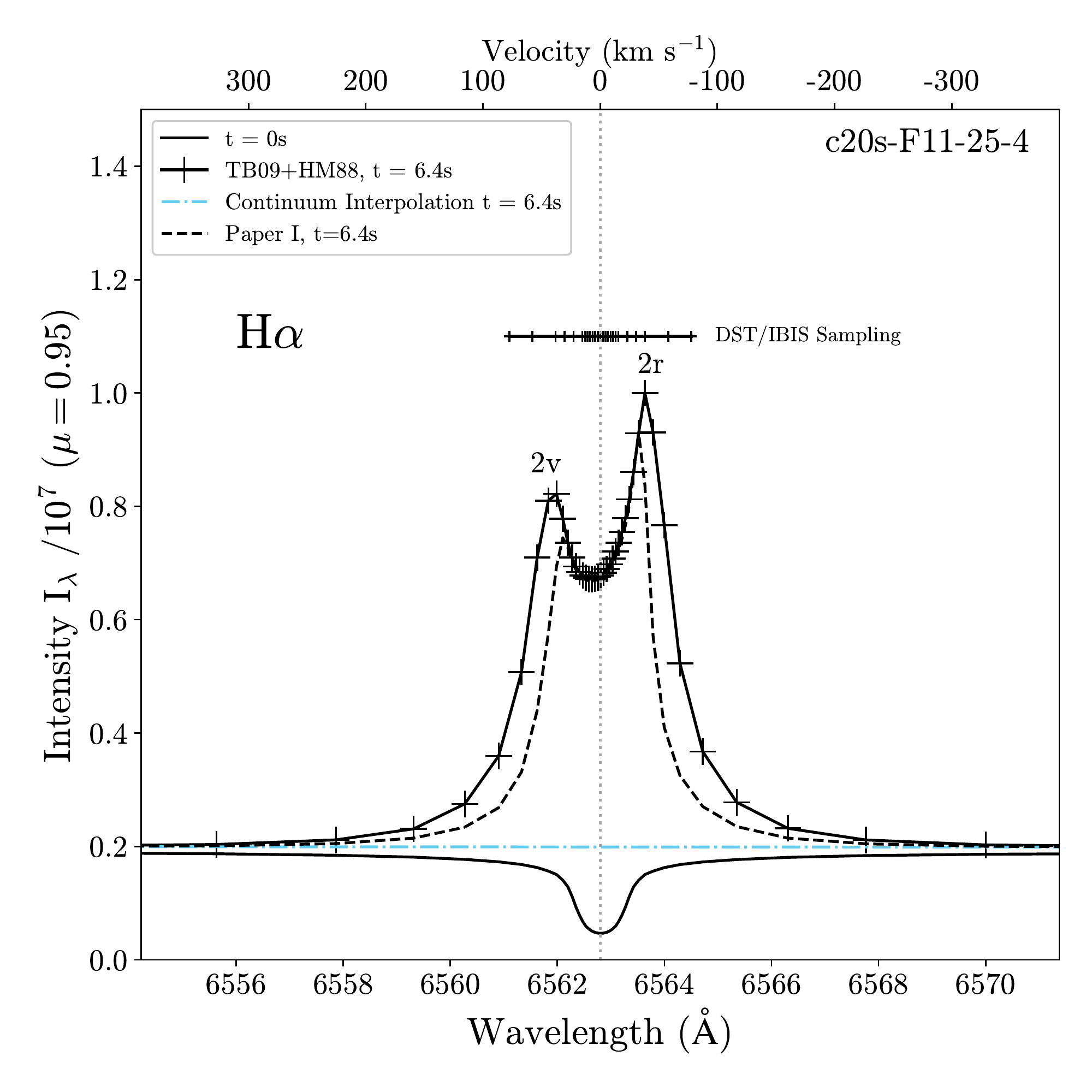}{0.4\textwidth}{(b)}
          }
\caption{Model H$\alpha$ flare spectra using the TB09$+$HM88 profiles in RADYN. These are improvements on recent works (dashed lines), which used the profiles corresponding to the broadening implementation of spectrum \textbf{(4)} in Figure \ref{fig:summary}(a). These two snapshots were chosen as the most similar in time to the snapshots that were analyzed in Figure 8 of Paper I (panel (a) here) and in Figure 7 in \citet{Kuridze2015} (panel (b) here).  In this figure, the same viewing angles ($\mu$) are presented as  in these two previous works.  The flare continuum emergent intensity is interpolated across the line to demonstrate that in the higher beam flux model, the wings of the H$\alpha$ line extend beyond this wavelength range ($\Delta \lambda = 17.15$ \AA, chosen as a representative wavelength range of a DKIST/ViSP window). The arrows in the left panel indicate the wavelengths at which the chromospheric condensation becomes optically thin over the TB09$+$HM88 spectrum (see text).
\label{fig:ibis_redux}}
\end{figure}

\citet{Rubio2016} compared RADYN flare models to H$\alpha$ and Ca $\textsc{ii}$ spectra from IBIS.  The observed H$\alpha$ spectra exhibit rather flat profiles, which slope gently to the red.  Paper I discusses the H$\alpha$ profiles from the dense chromospheric condensation in RADYN models with a higher beam flux of 5F11.  The re-calculated 5F11 model H$\alpha$ spectrum from Paper I (5F11-25-4.2) at $t=3.8$~s is shown in Figure \ref{fig:ibis_redux}(a), compared to the intensity profile at the most similar state of temperature, density, and velocity in the $t=3.97$~s snapshot that is analyzed in Paper I.  The TB09$+$HM88 broadening greatly enhances the wings compared to the hydrogen broadening prescription (\textbf{(4)} in Figure \ref{fig:summary}) in Paper I, resulting in an increase in the effective width by over 100\%.  Notably, the entire line profile is shifted to the red by $\approx -40$ km s$^{-1}$.  The IBIS wavelength sampling and the observed profile at 17:46:13 from \citet{Rubio2016} (\emph{cf} their Figure 2b) are overplotted.  While the new model ostensibly explains the shape of the H$\alpha$ profile over the limited wavelength range of IBIS, the intensity is $\approx 7$x too bright, which could be attributed to a modest filling factor.   In the model comparison in Figure \ref{fig:ibis_redux}(a), we have not taken into account the irregular wavelength step cadence over 18~s or the effects from variable seeing \citep{Rubio2016}.

\citet{Kuridze2015} and \citet{Kuridze2016} analyzed the evolution of H$\alpha$ profiles in response to lower flux beam heating in the c20s-F11-25-4 model, which is also discussed in Paper I.  \citet{Kuridze2015} found a similarity in the evolution of the relative brightness of the peaks to the red (2r; following standard labeling of quiet Sun emission line profiles of near-ultraviolet resonance lines) and violet (2v) of the central reversal of the H$\alpha$ model in comparison to CRISP data.  In Figure \ref{fig:ibis_redux}(b), we show the model with the TB09$+$HM88 hydrogen broadening profiles. Though the broadening of H$\alpha$ is larger in the new calculation, the asymmetries in the 2v and 2r features of H$\alpha$ follow a nearly identical evolution as in \citet{Kuridze2015}.  Notably, the 5F11 model profile of H$\alpha$ also exhibits broad 2v and 2r peaks due to opacity broadening and a central reversal due to NLTE effects, which are discussed in detail in Section \ref{sec:ebapprox}.

\begin{figure}
\begin{center}
\includegraphics[scale=0.4]{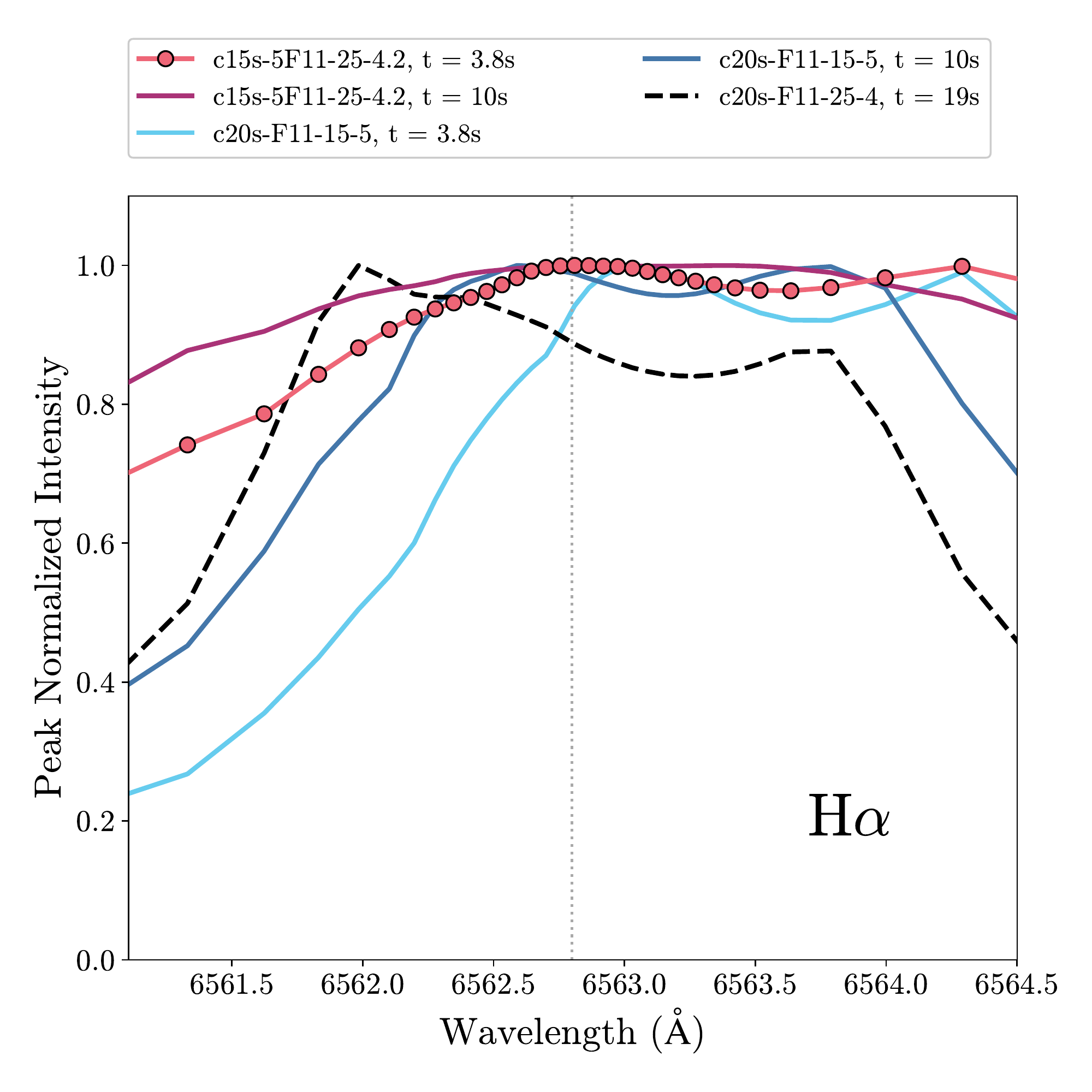}
\caption{Representative H$\alpha$ flare spectra from the three models in Table \ref{table:models} normalized to the peak intensity.  The wavelength range ($\Delta \lambda = 3.4$ \AA) corresponds to a typical setup with the IBIS instrument on the Dunn Solar Telescope. In the near-blue wing, there are discernible differences in the slopes between the F11-15-5 (increasing to red side) and 5F11 (nearly flat across this limited wavelength range). The F11-25-4 model is shown at 19~s when the condensation begins to have an effect on the hydrogen line spectra \citep[see][]{Kuridze2015}.
\label{fig:ibis_all}}
\end{center}
\end{figure}

From all models in Table \ref{table:models}, representative predictions for H$\alpha$ within a typical wavelength range of IBIS are shown in Figure \ref{fig:ibis_all}.  These spectra have been normalized to the peak intensity to facilitate comparison of the spectral slopes.  A wide range of behaviors among the line broadening for different electron densities are evident.  Even for the narrowband wavelength sampling in Figure \ref{fig:ibis_all}, the slope in the blue wing varies as a function of the flare heating.  However, spectra that extend farther into the wings are essential for robust physical interpretation of the broadest and most highly redshifted, 5F11 profiles produced in these models.   The extended wavelength ranges of Figure \ref{fig:ibis_redux}(a) and Figure \ref{fig:ibis_redux}(b)  are shown for a representative configuration with the ViSP using its online instrument performance calculator\footnote{\url{https://nso.edu/telescopes/dkist/instruments/visp/}}.

\subsection{Predictions for the Time-Evolution of H$\gamma$ in ViSP Observations} \label{sec:visp}
With the DKIST/ViSP, it will be possible to observe several hydrogen lines simultaneously with coverage of their far wings.  The large number of possible line-window combinations with the ViSP beckons for novel observational strategies to reveal critical missing microphysical processes in current models of solar flares (e.g., Section \ref{sec:discussion}). H$\alpha$ has traditionally been the most popular line for solar flare studies, but it will not always be possible to include in configurations with the ViSP when combined with other instruments on the DKIST.  The H$\gamma$ line will be readily observable with the ViSP and is the highest-$n_j$ Balmer line calculated in RADYN.  Thus we focus the model analysis on the spectra of this emission line.  The interpretation of symmetric broadening of higher $-n_j$ Balmer lines at high spatial and temporal resolution is generally less obfuscated compared to H$\alpha$ for several reasons.   Even at low electron densities around $3\times 10^{13}$ cm$^{-3}$, the H$\gamma$ line is less optically thick by a factor of $\approx 30$ (estimated  from LTE at $\lambda_o$), it experiences larger pressure broadening from ambient charged particles by a factor of six (measured in the wings at $\lambda_o+5$\AA), and the relative contributions from self-broadening are less by a factor of three \citep[measured at $\lambda_o + 5$\AA;][]{Barklem2000}.     While thermal Doppler broadening at $T=10,000 - 20,000$ K affects the line opacity significantly for both lines within $\pm1$ \AA\ of their respective values of $\lambda_o$ (at this particular electron density), the broadening of the H$\gamma$ line profile function is $6-8$x more sensitive to electron density variations around $n_e = 3 \times 10^{13}$ cm$^{-3}$.
  Also, a large wavelength span across the H$\gamma$ profile is more likely to be formed in conditions that are closer to LTE in dense chromospheric condensations.  These advantages are discussed further below.

Dramatic changes occur in the brightness, broadening, and Doppler shift of the low$-n_j$ Balmer lines within the first 10~s of heating in the 5F11-25-4.2 and F11-15-5 models. Figure \ref{fig:time_evol} shows the time-evolution of the emergent H$\gamma$ intensity spectra ($\mu=0.95$) for the 5F11-25-4.2 (a) and F11-15-5 (b) models over a wavelength range that is possible with the ViSP.  
Though the 5F11-25-4.2 exhibits profiles that are $\approx 2$x  brighter and broader than the lower flux model, there are some remarkable similarities.    Throughout the paper, we quantify the Doppler shift of broadened and symmetric profiles using the mean wavelength, $\lambda_{\rm{cen}}$, which is calculated from the continuum-subtracted, emergent intensity flare spectra. For H$\gamma$, $\lambda_{\rm{cen}}$ typically corresponds to $\lambda_{\rm{max}}$.   At first ($t = 1$~s), relatively narrow, nearly symmetric emission lines form at $\lambda_{\rm{cen}} \approx \lambda_{o} - 6$ km s$^{-1}$, where $\lambda_o$ is the rest wavelength. In the 5F11 model, the far red wing within $-100$ km s$^{-1}$ of $\lambda_{o}$ is enhanced over the blue wing.  In the F11 model, a similar faint redshifted emission component develops at $t \approx 2$~s at a small displacement from $\lambda_{o}$.   These prompt, red-shifted components are due to the condensations cooling to $T \approx 25,000 -35,000$ K but not yet accruing enough density to become very optically thick over a broad wavelength range.  The red shift in the 5F11 shows up before the similar feature in the F11 because the 5F11 condensation cools slightly quicker and its electron density is already near $10^{14}$ cm$^{-3}$ by $t=1$~s (Figure \ref{fig:ccevol}).  The optical depth corresponding to $\approx 1$ at $\lambda_o$ occurs in the deeper stationary flare layers at these early times and thus the emergent intensity spectra are mostly symmetric around $\lambda_o$ within the first few seconds.   As the condensations cool further and accrue more mass, they become optically thick within an increasingly large spectral range around the rest wavelength in the frame of the condensation.  Extremely broad wings form in the spectra, and the entire profiles become more symmetric and redshifted at $\lambda_{\rm{cen}} \approx -40$ km s$^{-1}$ (Figure \ref{fig:time_evol}a-b).  These spectral changes occur over only $\Delta t \approx 2$~s in each model.  Over the next $\Delta t \approx 6$~s, the maximum emergent intensities over the spectra then become fainter and less red shifted at $\approx -20$ km s$^{-1}$, while the symmetric broadening further increases. At smaller values of $\mu$, the wings are slightly broader while the bulk redshifts are slightly smaller due to line-of-sight projection effects.

 Following \citet{Carlsson1997}, we calculate the contribution function to the emergent intensity for the H$\gamma$ and H$\alpha$ spectra.  The contribution functions are  analyzed at all time steps in all models.  Here, we summarize the most relevant points for the H$\gamma$ formation from $t=3.8$~s until $t=10$~s in the 5F11 model.  The contribution functions to the emergent intensity ($\mu=0.95$) at $t=0$~s and $t=3.8$~s are shown in Figure \ref{fig:ci_hgamma_5F11}(a)-(b).   The chromospheric condensation is evident as the $\Delta z \approx 25$ km, downflowing region with a line-of-sight (L.o.S.) gas velocity of $v \approx -5$ to $v \approx -50$ km s$^{-1}$, at locations where the contribution function across the line is brightest.  The $\tau=1$ contours indicate that the broad wings start to become optically thin in the condensation beyond $\lambda_{\rm{cen}} \pm 1.5$ \AA.
 
We also calculate the normalized cumulative contribution function (Paper I) and the optical depth at the bottom of the chromospheric condensation, $\tau_{\rm{CC}}$.  Following Paper I,
 the normalized cumulative contribution function is calculated with limits of integration as $z=10,000$ km (the top of the model atmosphere) and $z=400$ km (the bottom of the chromospheric flare layers).  The contours corresponding to 10\% and 90\% in Figure \ref{fig:ci_hgamma_5F11}(b) indicate where the majority of the emergent flare intensity originates in the chromosphere. Only a  small fraction of the emergent wing intensity over this wavelength range is formed below the condensation in the stationary flare chromospheric layers ($z \approx 400 - 865$ km).
 At  $\lambda_{\rm{cen}} \approx \lambda_o - 40 $ km s$^{-1} $, the emergent intensity is formed in a very narrow height range around $\tau=1$ at the top of the condensation ($T\approx 15,400$ K), and $\tau_{\rm{CC}}$ attains large values ($\tau_{\rm{CC}} \approx 35$) at the bottom of the condensation where $T \approx 9500$ K.  At wavelengths beyond the range of Figure \ref{fig:ci_hgamma_5F11}(b), the value of $\tau_{\rm{CC}}$ for the H$\gamma$ line decreases to $\tau_{\rm{CC}} \ll 1$ of the bound-free continuum, and the 90\% contour extends as deep as $z \approx 700$ km. 
 
 H$\alpha$ (not shown) is far more optically thick  in the condensations than H$\gamma$.  At $t=3.8$~s in the 5F11 model, the formation height at $\lambda_{\rm{cen}}$ is higher in the atmosphere at hotter temperatures at the interface between the condensation and the flare transition region:  $\tau = 1$ occurs at $T \approx 22,000$ K.  The maximum value of $\tau_{\rm{CC}}$ over the H$\alpha$ line profile is 1750, and there is a  decrease at larger wavelength distances from $\lambda_{\rm{cen}}$. The wavelengths over H$\alpha$ at which $\tau_{\rm{CC}} < 1$ are indicated with arrows in the emergent intensity spectrum that is shown in Figure \ref{fig:ibis_redux}(a). 
 These results suggest that the large optical depths in the low-$n_j$ Balmer line profiles probe the RHD evolution of the thermodynamics within the chromospheric condensation.  In the following analysis sections, we explain in detail how the measurements of line broadening in the model spectra quantitatively relate to the electron densities in chromospheric condensations.

\begin{rotate}

\begin{figure}
\gridline{\fig{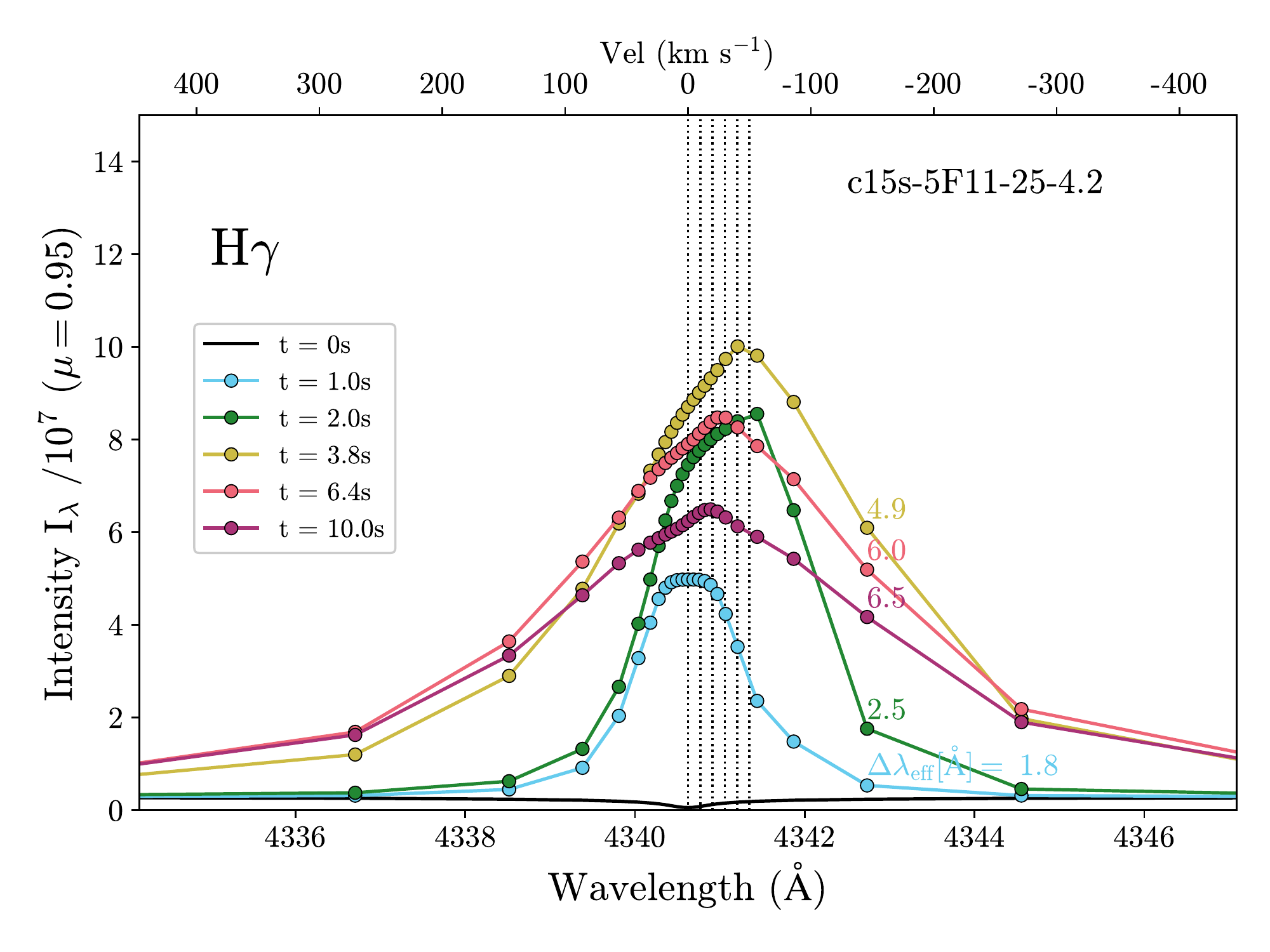}{0.45\textwidth}{(a)}
          \fig{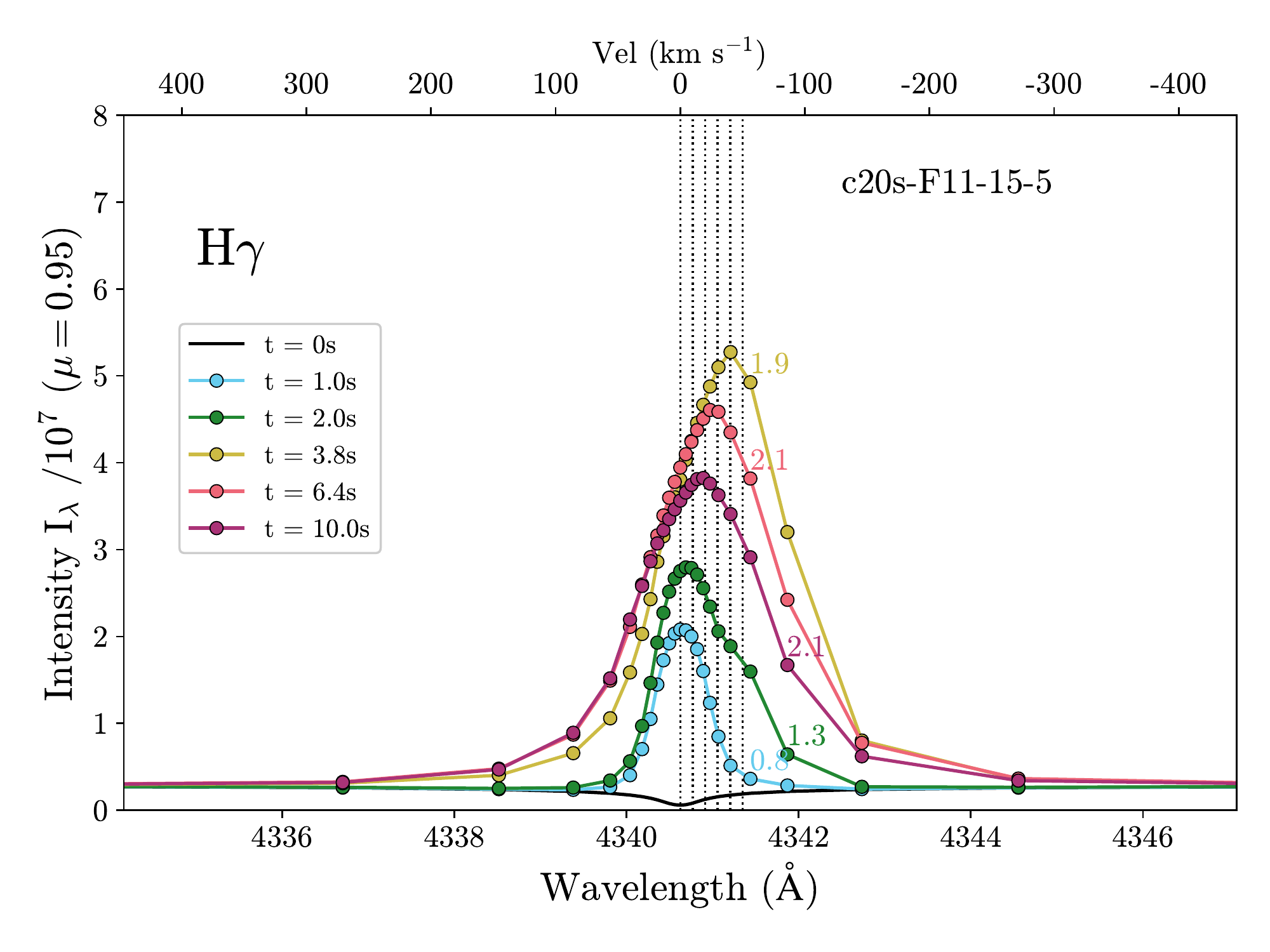}{0.45\textwidth}{(b)}
          }\caption{Time evolution of H$\gamma$ for a wavelength window ($\Delta \lambda = 12.9$ \AA) that is possible with a blue arm of the DKIST/ViSP.  The profiles are shown at the same times for the 5F11 model (a) from Paper I and the F11 model (b) from \citet{Graham2020}. Vertical dashed lines indicate velocity shifts from 0 to $-50$ km s$^{-1}$ in increments of $-10$ km s$^{-1}$, where negative velocities correspond to redshifts in the spectra and downflows in the hydrodynamics.  The spectra are labeled by their calculated effective widths (see text).  Note the different intensity scales for these two figures.    }
\label{fig:time_evol}
\end{figure}
\end{rotate}

\begin{figure}
\gridline{\fig{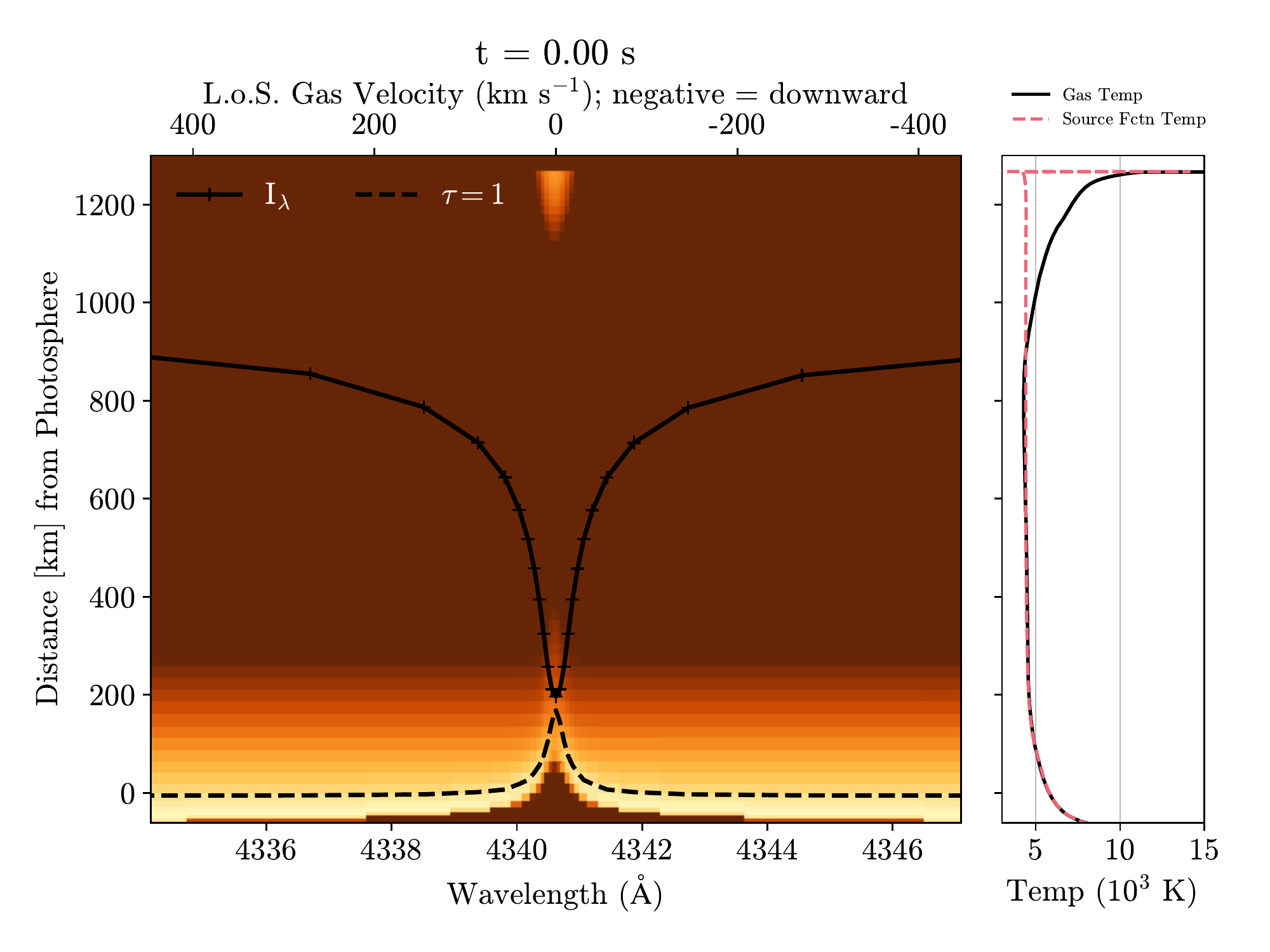}{0.7\textwidth}{(a)}}
\gridline{\fig{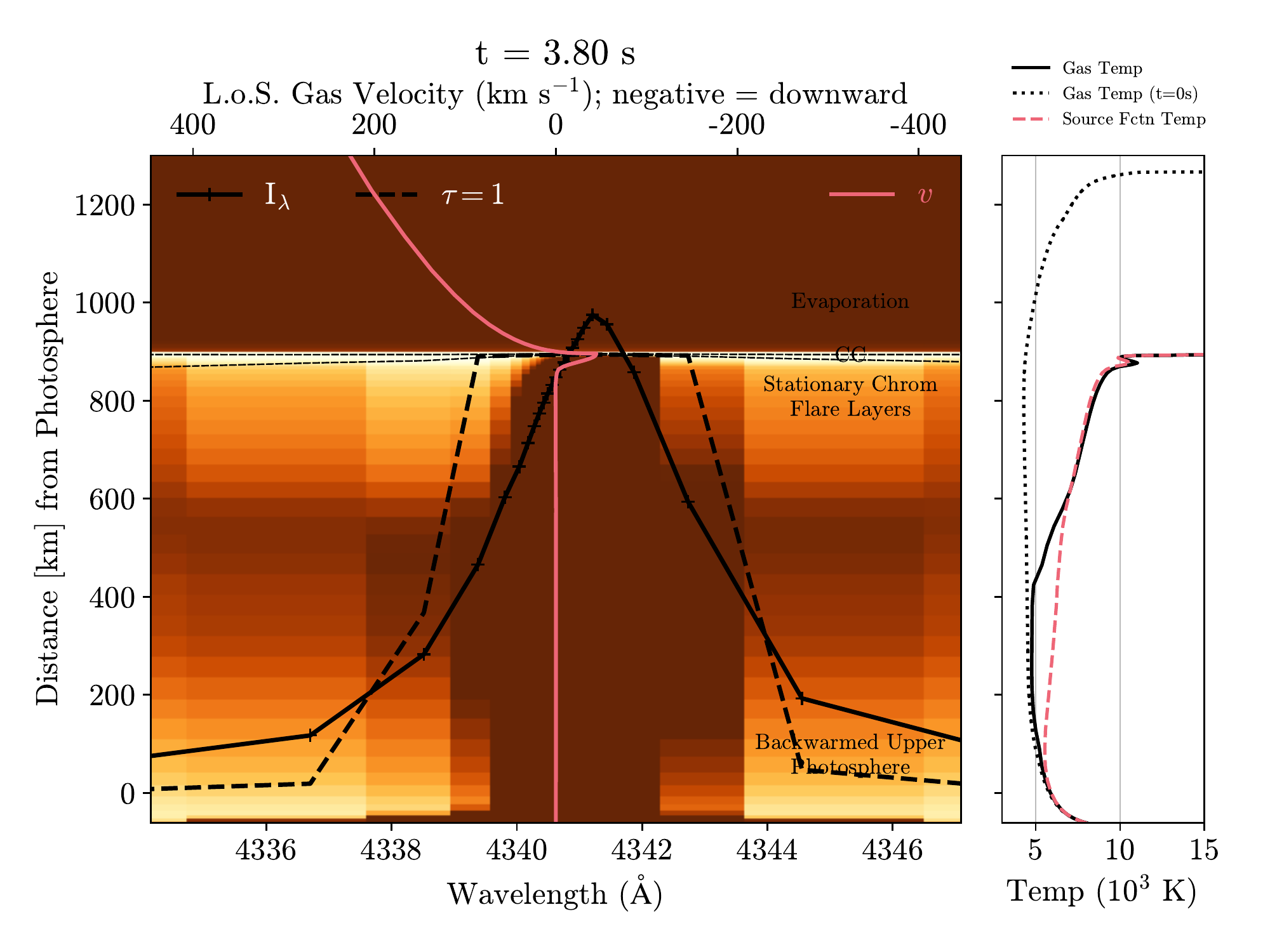}{0.7\textwidth}{(b)}}
\caption{Contribution function to the emergent intensity ($\mu=0.95$) of H$\gamma$ at (a) $t=0$~s and at (b) $t=3.8$~s in the 5F11-25-4.2 model.  The shading ranges logarithmically (log$_{10}$) from $0.001$ (red/brown) to $1$ (white/yellow) in units of erg s$^{-1}$ cm$^{-2}$ sr$^{-1}$ \AA$^{-1}$ cm$^{-1}$ in both panels.  The cumulative contribution function surfaces are shown for 10\% (upper dotted curve) and 90\% (lower dotted curve) of the emergent intensity at $t=3.8$~s, calculated using only layers above 400 km.  The line-of-sight (L.o.S.) gas velocity $v$ is defined as $\mu v_z$ where $\mu$ is related to the viewing angle $\cos$ $\theta$, and $v_z$ is the gas velocity along the loop.  The side panels in each figure on the right show the atmospheric temperature and the bound-bound source function equivalent temperature.  The departures from LTE over 3 km at the top of the condensation (at the temperature increase from 10,000 to 25,000 K) are significant but are not apparent on this scale. 
\label{fig:ci_hgamma_5F11}}
\end{figure}

\subsubsection{The Eddington-Barbier Approximation: The Origin of Decreasing Line Intensity} \label{sec:ebapprox}

 After $t=3.8$~s, the condensations decelerate and cool while accreting more material and encountering larger pressures in their paths \citep[][Figure \ref{fig:ccevol}]{Fisher1989}. The Doppler shift of $\lambda_{\rm{cen}}$ decreases from $\approx -40$ km s$^{-1}$ to $\approx -20$ km s$^{-1}$ by 10~s.  The broadest profiles are less redshifted and smaller in intensity than the profiles that are brightest and most redshifted, which occur at $t \approx 3.8$~s in both 5F11-25-4.2 and F11-15-5 models.  We first explain this decrease in emergent intensity at $\lambda_{\rm{cen}}$ using the Eddington-Barbier approximation.  We argue that since the H$\gamma$ line source function approaches LTE in the chromospheric condensation, the value of $I_{\lambda}(\lambda_{\rm{cen}})(t)$ is a probe of the temperature evolution at the top of the condensation.  
 
 We equate the H$\gamma$ NLTE source function values to a Planck function to display as an equivalent source function temperature in the right panels of Figure \ref{fig:ci_hgamma_5F11}.  In the following argument, however, the NLTE line source function and ambient Planck function values are compared in units of \ilam.    Over the heights corresponding to 10\% and 90\% of the cumulative contribution functions in Figure \ref{fig:ci_hgamma_5F11}(b), the source function is well within 10\% of the Planck function, and it is mostly  within 5\%.  At $\lambda_{\rm{cen}}$, the H$\gamma$ source function is 10\% of the Planck  function at $\tau=1$; however the departures from LTE become larger when the temperature exceeds $T \approx 15,000$ K within 3 km of the top of the condensation and into the flare transition region.  The H$\alpha$ source function is within 5\% of the Planck function where the contribution function is large in its wings.  Around $\lambda_{\rm{cen}}$, the  $\tau = 1$ surface for H$\alpha$ occurs at heights where the source function is only $0.5-0.6$ of the Planck function.  This is a much larger departure from LTE than for H$\gamma$ at this Doppler shift from $\lambda_o$, and it results in the small central reversal in the H$\alpha$ spectrum (e.g., Figure \ref{fig:ibis_redux}a).

 The source functions for all Balmer lines further approach LTE in the condensations at $t>4$~s.  For example, most of the H$\gamma$ line profile source function is within 3\% of the Planck function at $t=10$s in the 5F11.  Even H$\alpha$ is within 15\% of the Planck function at $\tau=1$ at $\lambda_{\rm{cen}}$, and its central reversal is filled in.  The condensation attains higher density over time, which results in larger radiative cooling, leading to a gradual drop in temperature below 10,000 K (Figure \ref{fig:ccevol}).  Because the source function approaches the Planck function and also decreases by a factor of $\approx 1.5$ in the region of $\lambda_{\rm{cen}}$ formation,  the emergent intensity at $\lambda_{\rm{cen}}$ decreases with this temperature drop in the condensation.  This is consistent with the Eddington-Barbier approximation for emergent intensity formation around $\lambda_{\rm{cen}}$ at the top of the condensation, where the source function decreases monotonically, and approximately linearly, as a function of $\tau$ over most of the heights where the contribution function is significant.  At earlier times, around $t=3.8$~s, the H$\alpha$ emergent intensity at $\lambda_{\rm{cen}}$ is consistent with the Eddington-Barbier approximation using the NLTE source function.

\subsubsection{Broadening due to Large Optical Depths:  The Origin of Increasing Effective Widths} \label{sec:growth}

As previously noted, the wings of H$\gamma$ in the model spectra in Figure \ref{fig:time_evol}(a)-(b) become broader as the intensity at $\lambda_{\rm{cen}}$ becomes fainter.  In the far wings, the Eddington-Barbier approximation is no longer helpful because the condensation ceases to be a ``semi-infinite'' ($\tau_{\rm{bottom}} \gg 1$)  atmosphere as the optical depth drops according to the $\tau=1$ surface in Figure \ref{fig:ci_hgamma_5F11}(b).
To quantify the wing broadening in the spectra, we calculate the effective widths of H$\gamma$ (Eq. \ref{eq:effective_width}). 
These widths vary from $\Delta \lambda_{\rm{eff}}=1.8$\AA\ at $t=1$~s, $\Delta \lambda_{\rm{eff}}=4.9$\AA\ at $t=3.8$~s, and  $\Delta \lambda_{\rm{eff}}=6.5$\AA\  at $t=10$~s in the 5F11-25-4.2 model.  In the F11-15-5 model, they vary from $\Delta \lambda_{\rm{eff}}=0.4$\AA\ to 2.1\AA.  The effective widths are far larger than expected from any electron density in the region of line formation due to optical depth effects.

The increasing bound-bound optical depths and electron densities over the evolution of the chromospheric condensation \citep[e.g.,][]{Kowalski2015} generate the variations in the broadening in the low-$n_j$ Balmer line spectra.   Figure \ref{fig:ccevol} suggests that the condensations accrue a factor of three enhancement in mass density from $t=3.8$~s to $t=10$~s while the electron density decreases by a factor of three.
Depending on the optical depth of a line, the quantities at maximum mass density in the chromospheric condensation in Figure \ref{fig:ccevol}, however, are not necessarily equal to the quantities over which most of the Balmer lines form in the condensation\footnote{For example, the maximum mass density and electron densities correspond to the same height at $t=3.8$~s, while the maximum electron density occurs higher in the condensation at later times.}.  Thus, we calculate contribution-function-weighted atmospheric parameters at each wavelength, and we subsequently compute a weighted average over the emergent intensity within two effective widths of $\lambda_{\rm{cen}}$.  The weighted mass density increases by a factor of two, and the weighted electron density increases by factors of $\approx 1.2-1.3$ from $t=3.8$~s to 10~s in the two models.   The cooling of the condensations due to increasing mass densities and diminishing amounts of beam energy reaching depths below the top of the condensation over time (Paper I) leads to decreasing ionization fractions where the Balmer lines are formed.   These combined effects generate larger bound-bound optical depths and an enhancement in broadening over that due to the ambient charged particle density alone. In the 5F11 model, the optical depth at the bottom of the condensation at $\lambda_{\rm{cen}}$ for the H$\gamma$ line remarkably increases from $\tau_{\rm{CC}} \approx 35$ at 3.8~s to $\tau_{\rm{CC}} \approx 110$ at 10~s.

The broadening of the emergent intensity of H$\gamma$ due to large optical depths was demonstrated for an isodensity, isothermal, finite slab in LTE in Figure \ref{fig:summary}, where the connection between the broadening and a curve-of-growth of the H$\gamma$ TB09$+$HM88 profile is illustrated.
To quantify the broadening in emergent spectra from a non-equilibrium, inhomogeneous atmosphere, we follow \citet{Rathore2015A} who defined an opacity broadening factor for the C II profiles in the non-flaring solar chromosphere.  Opacity broadening follows from the Eddington-Barbier approximation in a semi-infinite atmosphere with a depth-dependent source function \citep[see also][]{Wood1996, Carlsson2015, Carlsson2019}.    In the flare case, the TB09$+$HM88 profiles are dominated by Holtsmark wings rather than a Doppler core or damping/Lorentzian wings, and the opacity broadening occurs at relatively small detunings where the Eddington-Barbier relation approximately holds (Section \ref{sec:ebapprox}).  As detuning increases out to $\approx 5$ Doppler widths, $\tau_{\rm{CC}}$ is still large near 10, and the $\tau = \mu$ location occurs deeper into the condensation where there is a lower source function.  At even larger detunings from $\lambda_{\rm{cen}}$ where the emergent intensity in the wings transition to a formation over lower optical depths, a relatively low b-b wing emissivity is compensated by photons escaping over a larger physical depth range (\emph{cf.} the 90\% cumulative contribution contour in Figure \ref{fig:ci_hgamma_5F11}).  At these wavelengths, the condensation is no longer semi-infinite, and the source function exhibits a rather complex variation over the relevant heights.

To quantify all optical depth effects (opacity broadening and the relative far wing enhancement) for a line with large Holtsmark wing opacity, we define an optical depth broadening factor.   We calculate this as the effective width (Eq.\ \ref{eq:effective_width}) of the emergent intensity profile from the RADYN simulation divided by the effective width of an optically thin, bound-bound emergent intensity calculation from the same atmospheric model structure, including the gas velocities.  For the optically thin intensity calculation, we use the TB09$+$HM88 profiles and the non-equilibrium level populations from RADYN at heights above 400 km.  The optically thin calculation gives the expected broadening due solely to broadening from ambient charged particle perturbations in a dynamic, heterogeneous chromosphere.
Note that the nearby continuum\footnote{The photosphere and upper photosphere in the 5F11 experience a gradual temperature increase from $t=1$~s to $t=10$~s due to radiative backwarming, which results in enhanced optical bound-free radiation.  As discussed in Appendix A of Paper I, the emergent optical continuum intensity response in the 5F11 is a rather complicated superposition of the  H-minus emission from the backwarmed upper photosphere, stationary chromospheric flare layers, and condensation, in addition to Paschen recombination radiation from the condensation.    If we include all heights in the cumulative contribution function of H$\gamma$, 90\% of the emergent intensity originates from the condensation at wavelengths within a half effective width from $\lambda_{\rm{cen}}$.  Therefore, the photospheric response does not largely affect the emergent intensity and broadening over the Balmer lines at wavelengths that are opaque in the condensation.  For further discussion on the physics of radiative backwarming in RADYN flare models, see \citet{Allred2005} and \citet{Cheng2010}. }  is linearly interpolated and subtracted for the effective width calculation from each RADYN spectrum (Eq.\ \ref{eq:effective_width}).  In the 5F11-25-4.2 model, the optical depth broadening factors of H$\gamma$ increase from $\approx 1$ to $\approx 5$ over the time interval of $t=1-10$~s. 
 In the c20s-F11 model spectra, the effective widths are smaller than calculated from the 5F11 spectra, yet the optical depth broadening factors attain values as large as $\approx 3$.  The F11 model produces a factor of three smaller optical depth through the condensation (measured at $\lambda_{\rm{cen}}$) and a factor of three smaller contribution function-weighted electron density.

\begin{figure}
\begin{center}
\includegraphics[scale=0.7]{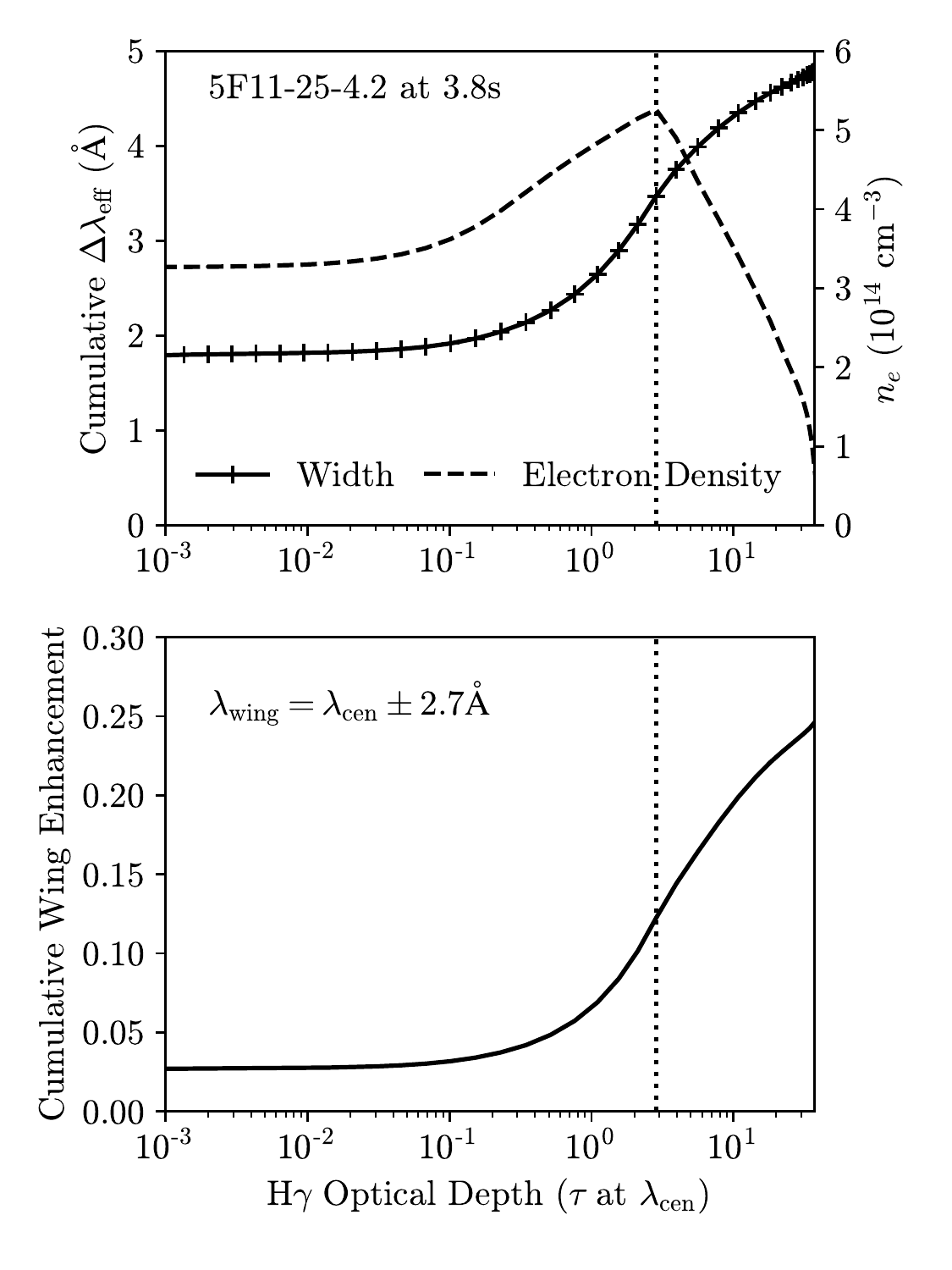}
\caption{ (Top) The curve-of-growth of the effective width of H$\gamma$ through the chromospheric condensation at $t=3.8$~s in the 5F11 model.  The physical depth range in this figure is $\Delta z \approx 30$ km.  At larger optical depths than at the maximum of $n_e$ (vertical dotted line), the widths of the line increase further until reaching values near 4.9 \AA\ in the emergent intensity spectrum.   (Bottom) The cumulative of the contribution function at $\lambda_{\rm{cen}} \pm 2.7$ \AA\ relative to the cumulative of the contribution function at $\lambda_{\rm{cen}}$.  This closely follows the cumulative variation in effective width in the top panel.
\label{fig:growth_curve_cc}}
\end{center}
\end{figure}

 To clarify the role of optical depth enhancement in the wing broadening in the chromospheric condensations, we calculate a curve-of-growth of H$\gamma$ at $t=3.8$~s in the 5F11 model.  Specifically, we calculate the effective widths of the cumulative contribution function at each height, integrating from the top of the atmosphere to height $z$.   We refer to this quantity as a cumulative effective width, which is plotted against $\tau$ at $\lambda_{\rm{cen}}$ in Figure \ref{fig:growth_curve_cc} at each height within the 5F11 chromospheric condensation.   The increase in the cumulative effective width occurs through deeper layers than the maximum of the electron density, which is indicated by a vertical dotted line in Figure \ref{fig:growth_curve_cc}.  This only be a result of further optical depth effects causing a relative enhancement of wing photons that escape.   This occurs as large values of $\tau = 5 - 35$ at $\lambda_{\rm{cen}}$ are attained (\emph{cf.} the effective width curve of growth in Figure \ref{fig:summary}b).   The variation in the cumulative contribution function at $\lambda = \lambda_{\rm{cen}} \pm 2.7$ \AA\ relative to the cumulative contribution function at $\lambda_{\rm{cen}}$ quantifies the cumulative enhancement of the wings at wavelengths where $\tau_{\rm{CC}}$ decreases below $\approx 1$.  This quantity exhibits a similar variation (Figure \ref{fig:growth_curve_cc}, bottom panel) as the cumulative effective width in the top panel.  The cumulative wing enhancement starts out at $\approx$ 0.03 at low optical depths, increases to $\approx$ 0.10 at the maximum electron density, and increases further to $\approx$ 0.25 at the bottom of the condensation.  The value attained at the bottom of the condensation is evident in the emergent intensity spectrum (e.g., Figure \ref{fig:ci_hgamma_5F11}). 
 
 The large optical depths in these models lead to another unexpected result, which is discussed in the next section: high-density chromospheric condensations produce broader H$\alpha$ spectra than H$\beta$ and H$\gamma$.

\subsection{Comparisons of Ca II K, H$\alpha$, H$\beta$, and H$\gamma$} \label{sec:dec}
Comparisons of the widths and wavelength-integrated intensities among the hydrogen series members are often employed as a  multi-line diagnostic of the flaring plasma's temperature, optical depth, and electron density \citep{Svestka1963, Svestka1965, Drake1980, JohnsKrull1997, Garcia2002, Allred2006, Crespo2006, Paulson2006, Hilton2010, Schmidt2012, Kowalski2012, KCF15}.   The TB09$+$HM88 broadening increases the wavelength-integrated emergent intensity in the Balmer lines by a factor of two \citep{Kowalski2017Broadening} (see also Figure \ref{fig:summary}), allowing for updated calculations of the Balmer decrement.  The Balmer decrement is the ratio of the wavelength-integrated intensity (or flux) in a Balmer emission line relative to another series member, usually taken to be H$\gamma$. Until relatively recently \citep{Allred2006}, most analyses \citep[e.g.,][]{KCF15} of the decrement have relied on static slab model predictions of \citet{Drake1980} for interpretation of flare decrements.
In this section, we explore the time-evolution of the decrement of the H$\alpha$, H$\beta$, and H$\gamma$ lines that are predicted in high-density chromospheric condensations in the 5F11-25-4.2 and F11-15-5 models.  The wavelength-integrated emergent intensities are inextricably related to the spectral line broadening through the effective width calculation (Eq. \ref{eq:effective_width}).

The theory of hydrogen line pressure broadening due to charged ambient particles dictates that the widths (in units of $\Delta$\AA) of hydrogen lines increase approximately linearly with increasing $n_{j}$, assuming sufficiently low densities and low $n_j$ such that non-ideal effects, as in TB09 (\emph{cf} their Figure 5), are not large.  The theory also predicts that the widths of the hydrogen series in the blue and in the Johnson $U$-band ($\lambda \approx 3200$\AA$ - 4000$\AA) should be much larger than the widths of the nearby resonance lines of Ca II H and K, which are not broadened due to energy level splittings with a linear (hydrogenic) dependence on the perturbing field. 
The impact broadening of Ca II H and K by ambient electrons have been calculated by \citet{Griem1974}, \citet{StarkB_CaIIA}, and \citet{StarkB_CaIIB}.  At the highest electron densities in the condensations, the Lorentzian FWHM values\footnote{Griem:  \url{https://griem.obspm.fr/}, Stark-B: \url{https://stark-b.obspm.fr/}}   are only $1-1.5 \times 10^{-3}$ \AA.
Large differences in broadening are often reported in echelle observations of solar \citep{Neidig1983, Neidig1984} and stellar flares \citep{HP91, Paulson2006, Fuhrmeister2020}.  The chromospheric condensation models predict dramatic differences in broadening between Ca II K and H$\gamma$, as shown in Figure \ref{fig:decrement}(a) for the 5F11 model at $t=3.8$~s.   

  In Figure \ref{fig:decrement}(b), the evolution of the effective widths from the RADYN models are shown for H$\alpha$, H$\beta$, and H$\gamma$ at $\mu=0.95$.  Widths calculated from optically thin, uniform slabs are indicated by the lines connecting square symbols, labeled by electron density.  As expected, the widths increase up the series (increasing $n_j$) for the optically thin spectra, except at the lowest electron density where thermal broadening is more important than the ambient charged particle broadening for the lowest $n_j$.  The interline trends in effective widths remain approximately constant over the evolution of the condensations in both models.  Optical depth broadening factors are shown in Figure \ref{fig:decrement}(c).  To compare the optical depth effects between the two chromospheric condensation models, we calculate the optical depth broadening factors using the ratio of effective widths of the cumulative contribution functions (Section \ref{sec:growth}) at the bottoms of the condensations. 
  Because the H$\alpha$ line transition is the most optically thick, the emergent spectra from the flare atmospheres exhibit much larger optical depth broadening factors than H$\beta$ and H$\gamma$.  In fact, the weighted electron densities over the line formation (Section \ref{sec:growth}) of H$\alpha$ are no larger than those weighted over the H$\gamma$ profile contribution function.

  The Balmer decrement (Figure \ref{fig:decrement}d) is  reversed in these models, and especially so in the 5F11-25-4.2 model. A reversed decrement results when  the H$\alpha$ wavelength-integrated intensity is less than H$\gamma$, and H$\beta$ is comparable to or less than H$\gamma$.   Though the H$\alpha$ line is broader than H$\gamma$ it is a factor of two less bright at $\lambda_{\rm{cen}}$, which is clear from the values of $I_{\lambda}(\lambda_{\rm{cen}})$ in Figure \ref{fig:ibis_redux}(a) and Figure \ref{fig:time_evol} at $t=3.8$~s.  Again, we use the Eddington-Barbier approximation to provide insight: the source function equivalent temperature for H$\alpha$ falls below the source function equivalent temperature for H$\gamma$ (Figure \ref{fig:ci_hgamma_5F11}b) in the 3 km over which $I_{\lambda}(\lambda_{\rm{cen}})$  is formed in the condensation.  Even where the source function equivalent temperatures (e.g., as in the near-wing) are nearly the same, larger emergent intensities in H$\gamma$ are expected as long as the source function equivalent temperature at $\tau = \mu$ is $ \gtrsim 10^4$ K.  By definition (Eq.\ \ref{eq:effective_width}), the effective widths in Figure \ref{fig:decrement}(b) multiplied by the maximum emergent specific intensity, $I_{\lambda}(\lambda_{\rm{max}})$, which is determined by the source function and Eddington-Barbier relation, are equal to the wavelength-integrated emergent intensities, $I$, in the decrements in Figure \ref{fig:decrement}(d).

  \subsubsection{A new line-to-continuum ratio diagnostic in optical solar flare spectra} \label{sec:ratio}
\begin{figure}
\gridline{\fig{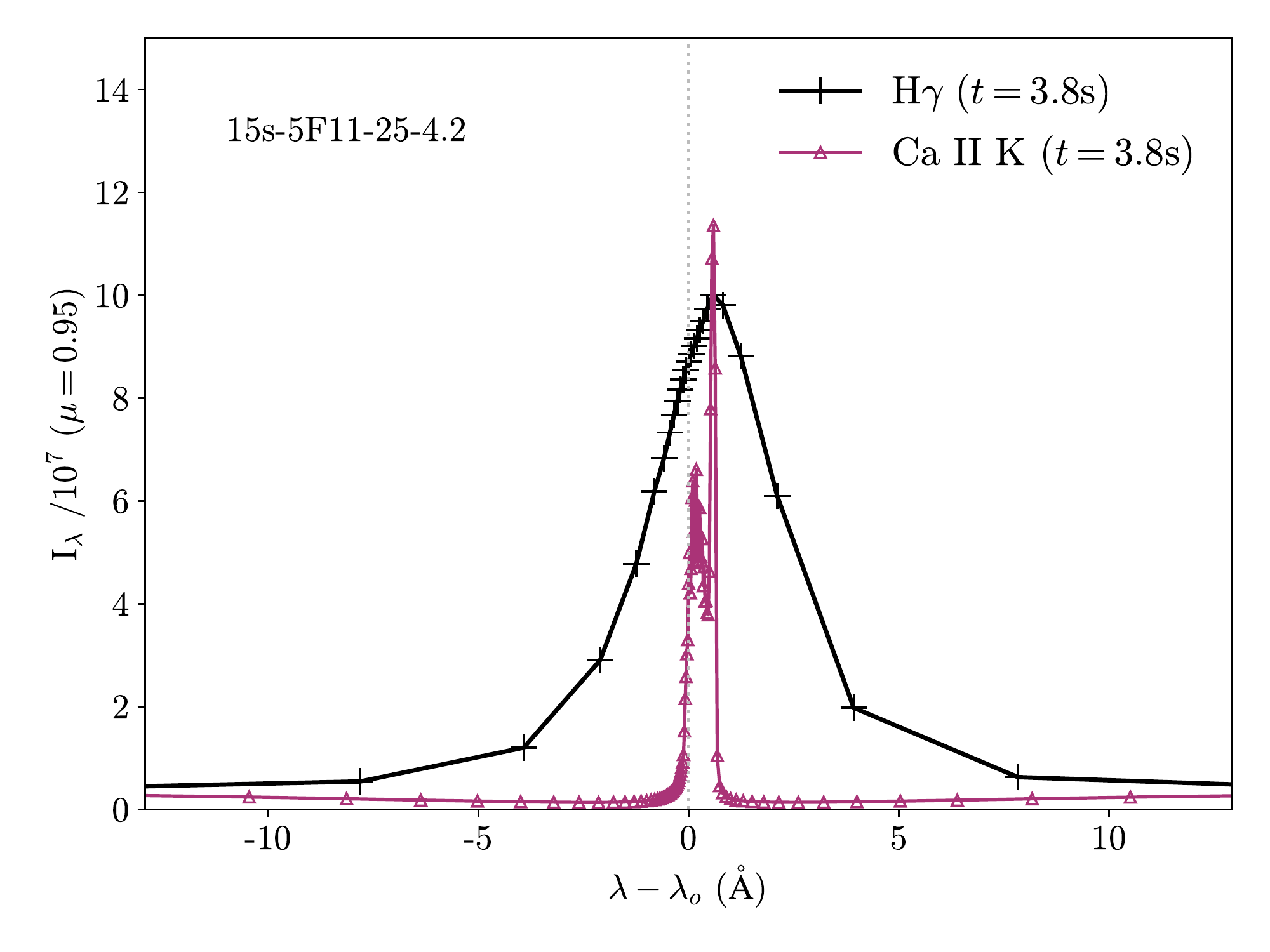}{0.45\textwidth}{(a)}
\fig{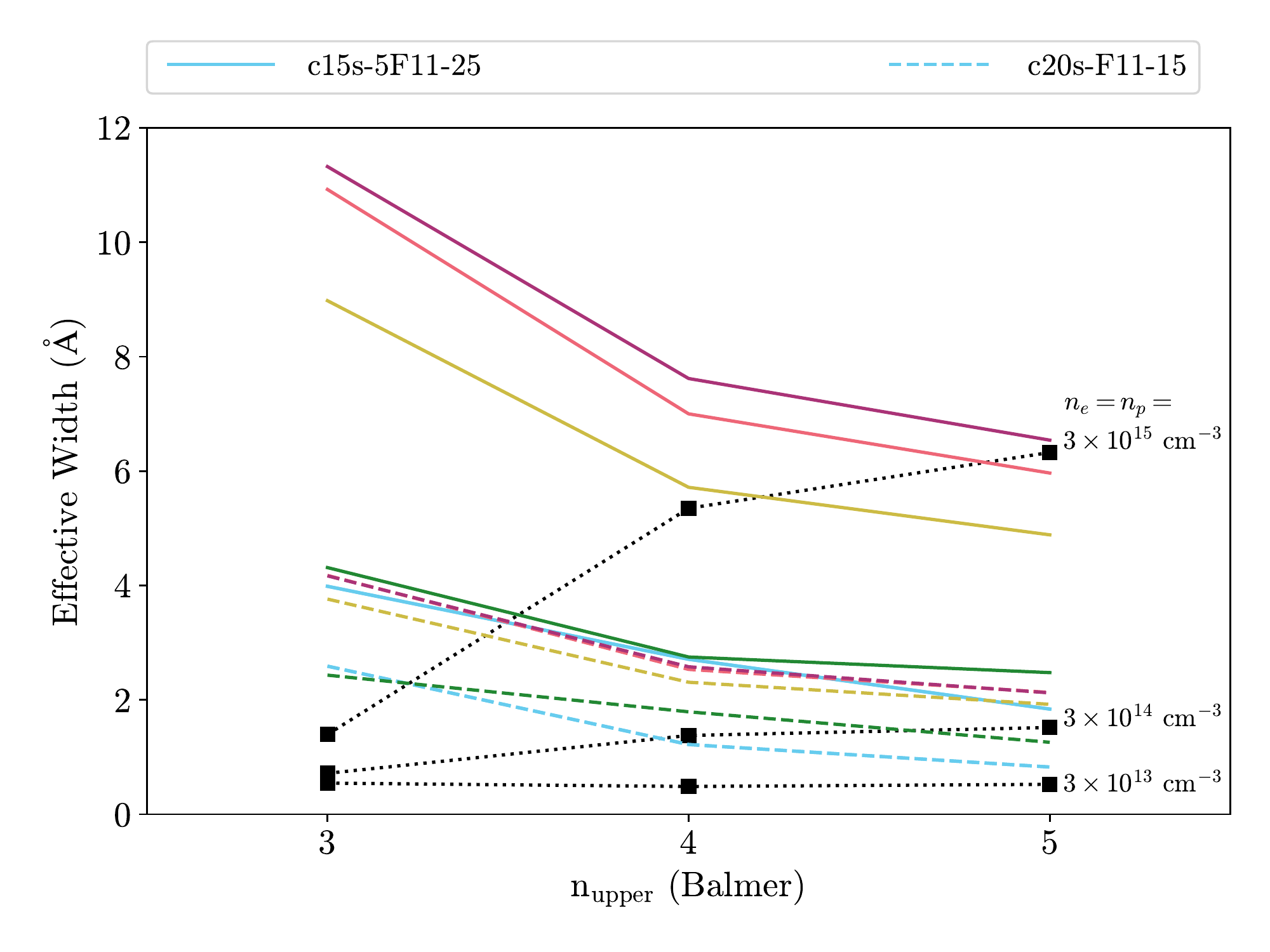}{0.45\textwidth}{(b)}}
\gridline{\fig{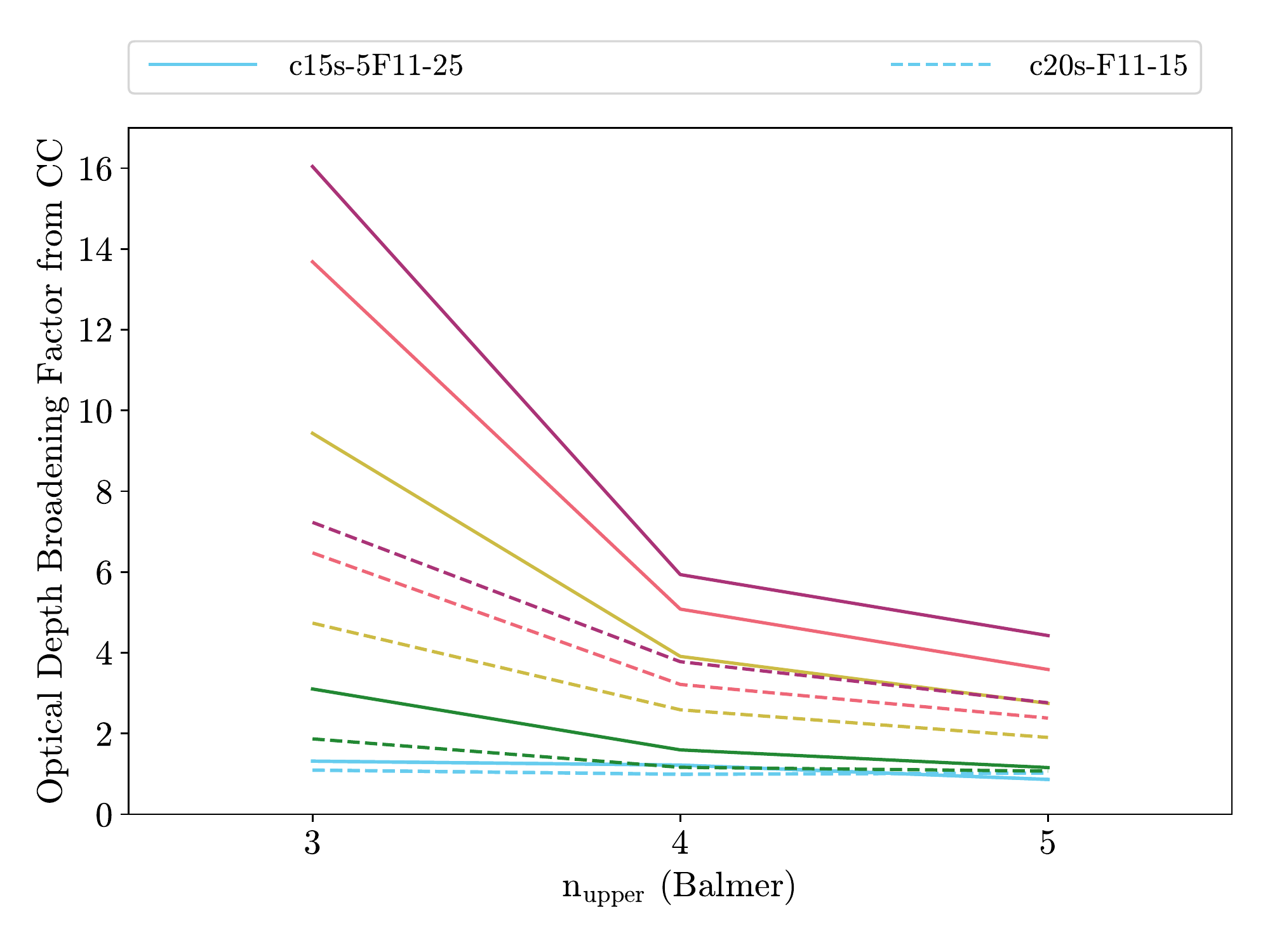}{0.45\textwidth}{(c)}
          \fig{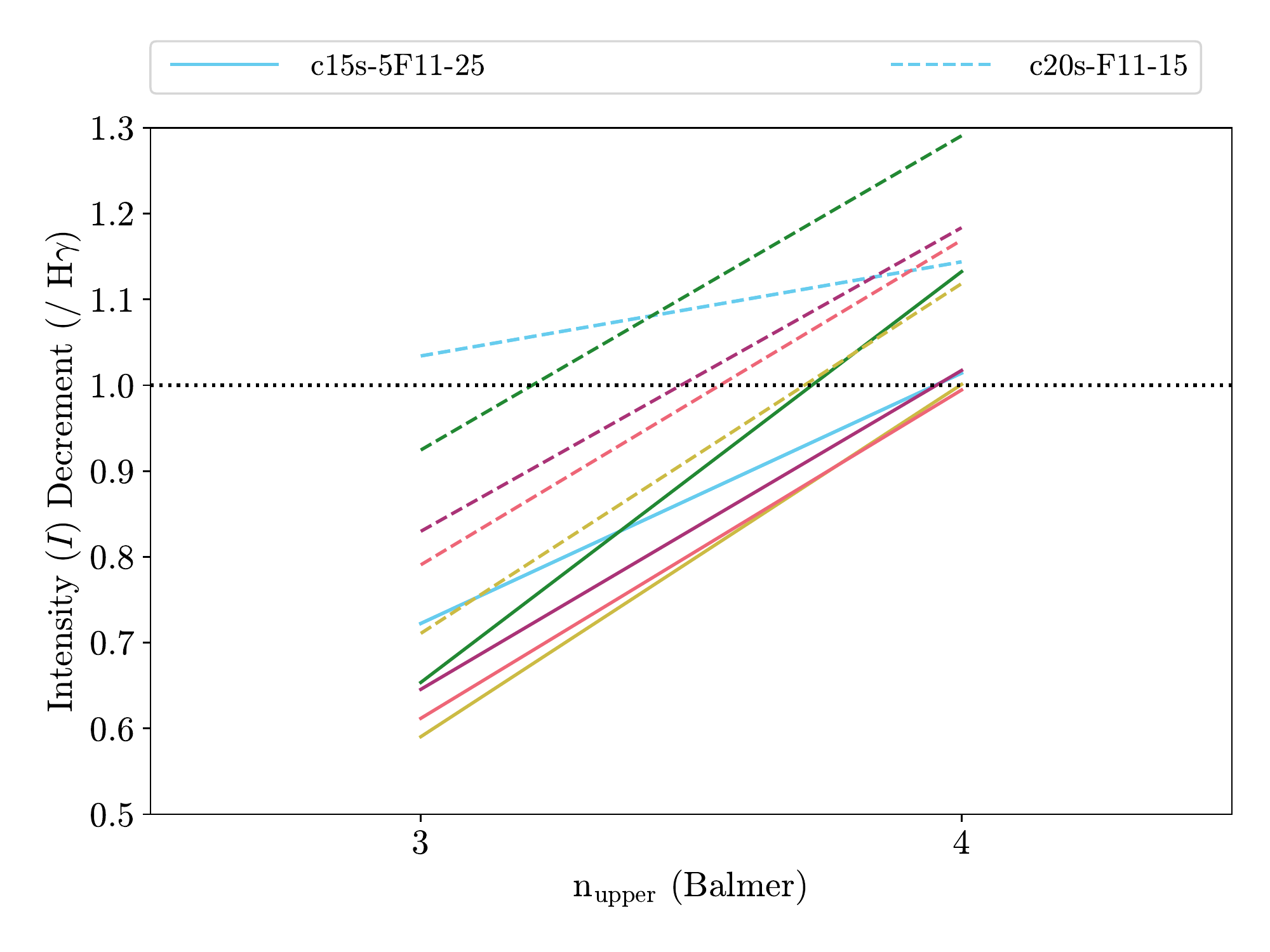}{0.45\textwidth}{(d)}}
          \caption{Comparisons of several optical emission lines in the flare models.  (a) Large differences in the broadening of H$\gamma$ and Ca II K in the 5F11-25-4.2 model agree qualitatively with the observed phenomena in solar and stellar flares \citep[e.g.,][]{Neidig1983}.  (b) Evolution of the effective widths of the H$\gamma$, H$\beta$, and H$\alpha$ lines in the 5F11-25-4.2 and F11-15-5 models.    The effective width trends of the TB09$+$HM88 profiles without optical depth effects (black dashed lines) for several representative ambient electron densities increase up the series, as expected. (c) Optical depth broadening factors through the chromospheric condensation (CC) increase from H$\gamma$ to H$\alpha$.  The effective widths in the broadening factors were calculated from the cumulative contribution function at the deepest height of the condensation at each time-step. Note that the optical depth broadening factor of H$\gamma$ at $t=1$~s in the 5F11 model is slightly less than $1$, which is due to a low but non-negligible ($\approx 0.1$) optical depth in the condensation and velocity gradients that lead to broadening in the $\tau=0$ calculation.  (d) The Balmer decrement with respect to H$\gamma$.  In panels (b), (c), and (d), the color coding according to time is the same as in Figures \ref{fig:time_evol}(a)-(b).
\label{fig:decrement} }
\end{figure}

  Reversed Balmer decrements in Figure \ref{fig:decrement}(d) provide stringent tests of the large optical depths produced in chromospheric condensation models.  However, accurate measurements of the decrements rely on intensity calibration over a broad wavelength range, 
 which is technically difficult to achieve and validate using ground-based data \citep{KCF15, Ondrej1}.
 A line-to-continuum ratio is an alternative diagnostic that can be calculated from spectra with a narrow wavelength range \citep{Silverberg2016, Kowalski2019HST}.
  The motivation is similar to the line-to-continuum ratio of Fe II$\lambda2814$ to the $\lambda \approx 2826$ \AA\ continuum radiation in IRIS/NUV solar flare spectra \citep{Kowalski2019IRIS}. Here we use the wavelength-integrated, continuum-subtracted intensity of H$\gamma$ divided by the pre-flare-subtracted continuum intensity at  $\lambda=4170$ \AA.  We denote this line-to-continuum ratio as H$\gamma$/C4170$^{\prime}$, where the prime ($^{\prime}$) symbol indicates that the pre-flare spectrum is subtracted.

 The low-$n_j$ Balmer lines are optically thick within an effective width of $\lambda_{\rm{cen}}$, and thus the wavelength-integrated intensity primarily probes the ambient charge density, temperature, and optical depth in the condensation (Section \ref{sec:visp}).  
   The nearby continuum flare intensity is much more optically thin than the Balmer lines around $\lambda_{\rm{cen}}$ in the condensation and includes larger contributions from the stationary chromospheric flare layers and the upper photospheric backwarmed layers. 
  In the solar models here, the line-to-continuum ratios  decrease from H$\gamma$/C4170$^{\prime}$ $\approx 1200$ to $\approx 800$ over $t= 1-10$s in the F11 model and from H$\gamma$/C4170$^{\prime}$ $\approx 1000$ to $\approx 500$ in the 5F11 model.  As the wavelength-integrated H$\gamma$ intensities decrease at $t \ge 6.4$s, the C4170$^\prime$ continuum intensity\footnote{ For reference, the local continuum intensity at $\lambda \approx 4170$ \AA\ is  $I_{\lambda} \approx 4.6 \times 10^6$ \ilam\ at disk center \citep{Neckel1984}.} increases to $\gtrsim 10^{6}$ \ilam\ by $t=10$~s (see Footnote 11).
  We discuss the broader empirical context of these line-to-continuum ratio values in Section \ref{sec:mdwarfs}.

\subsection{Balmer Series Broadening in the $U$-Band} \label{sec:lz}

\begin{figure}
\fig{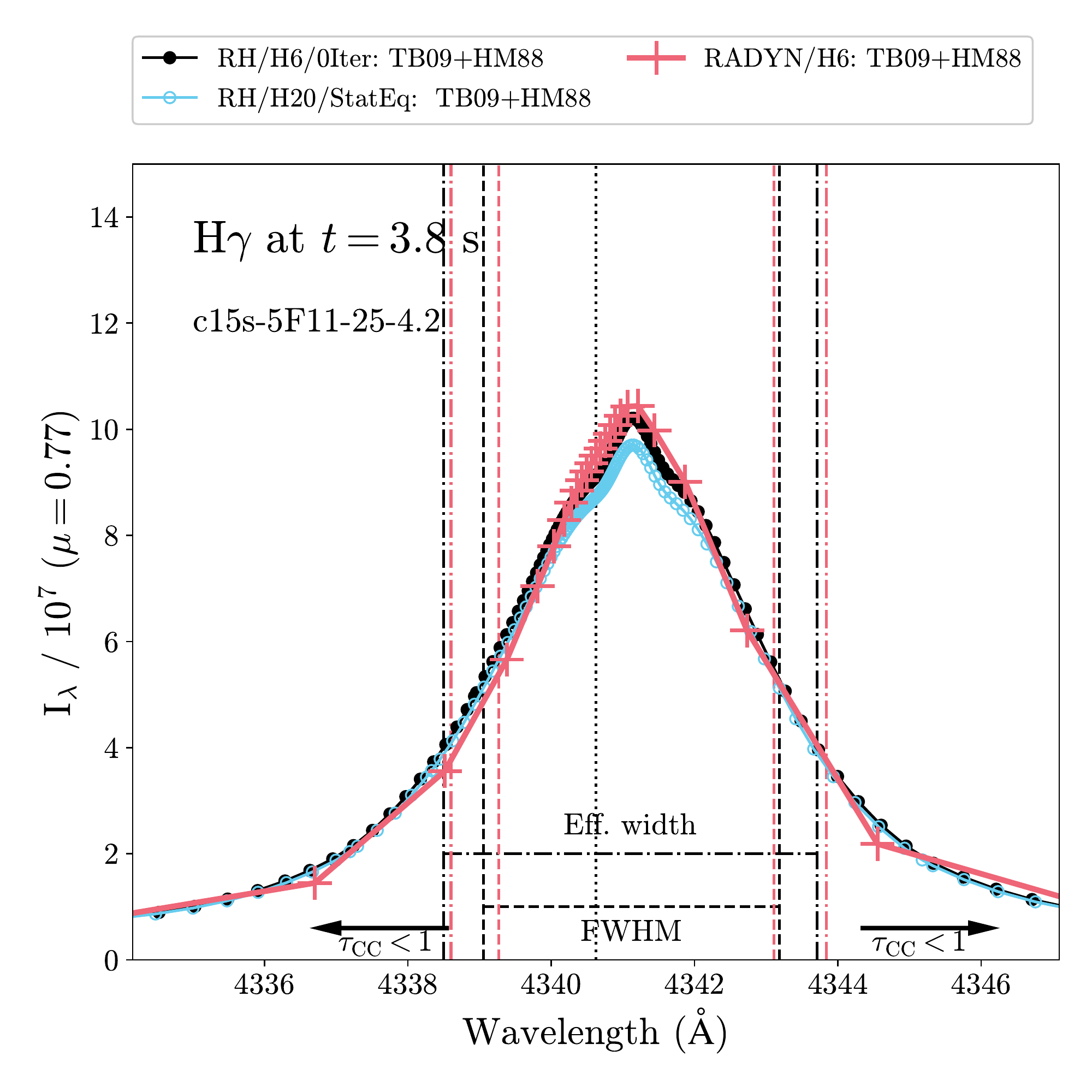}{0.45\textwidth}{}
          \caption{Calculations of the H$\gamma$ profile from the 5F11-25-4.2 model at $t=3.8$~s using a 6-level hydrogen atom in RH and RADYN.   Both codes calculate the NLTE emergent intensity with the TB09$+$HM88 profiles.  RH uses a numerical convolution with a Voigt profile for resonance, thermal, van der Waals broadening, while RADYN interpolates the thermally-convolved TB09$+$HM88 profiles.  The measurements of the full widths at half maxima and the effective widths are in agreement.  The blue-color spectrum shows the same RH calculation with a 20-level hydrogen atom.  In the 20-level calculation, the level populations are solved in statistical equilibrium (see text).  }
\label{fig:comparison}
\end{figure}

The TB09$+$HM88 profiles were incorporated into the RH code \citep{Uitenbroek2001}  in \citet{Kowalski2017Broadening} for M dwarf flare spectral modeling of the Balmer lines.  RH is a static code that employs a numerical convolution of the TB09$+$HM88 profiles with a Voigt profile, thus requiring long computation times for $1000-1500$ wavelength points across each line.  We use the RH code to compute the 5F11 model spectrum at $t=3.8$~s using the non-equilibrium electron density and the six hydrogen-level population densities from the RADYN snapshot\footnote{In RH, we set \Verb|N\_ITER=0| so that populations are fixed.}.  A comparison of the RH calculation to the RADYN spectrum is shown in Figure \ref{fig:comparison}, which demonstrates visual consistency in the two methods.   The calculated effective widths ($\Delta \lambda_{\rm{eff}} =$ 5.2 \AA) agree very well and approximately indicate the wavelength distances from $\lambda_{\rm{cen}}$ at which the wings of H$\gamma$ become optically thin in the condensation.  As expected, the effective widths are larger than the FWHM values.

We investigated the relatively small differences in the far wings and nearby continuum in Figure \ref{fig:comparison} at $\tau_{\rm{CC}} \lesssim 1$ and find that such differences are not due to the choice of line broadening calculation method, and that they are also present in similar model comparisons at $t = 0$ s.  It is possible that the treatment of background b-f opacity of several minor species (e.g., Si I, Mg I, Al I) or differences in the method of the formal solution of the radiative transfer equation (e.g., in the presence of strong velocity gradients) are sources of differences at this level.  However, further detailed investigation is required.

In addition to providing arbitrarily fine wavelength resolution, RH calculates the overlapping wings of the high$-n_j$ Balmer series self-consistently with the dissolved-level, Balmer bound-free and bound-bound opacity modifications \citep{Dappen1987}, as described in \citet{Kowalski2015} and \citet{Kowalski2017Broadening}. Dissolved-level opacity modifications are based on the occupational probability ($w_n$) formalism that was developed by \citet{Mihalas1966}, HM88, and \citet{Dappen1987}.  Following TB09, the NLTE bound-bound opacity (Eq. \ref{eq:bbopacity}) in RH is multiplied by $w_j / w_i  \approx w_j$ and the $n_i \rightarrow n_k$ bound-free opacity is multiplied by the dissolved level fraction \citep{Dappen1987}.   Occupational probability formalism is widely adopted in many\footnote{The generalized NLTE bound-bound and bound-free opacities with the occupational probability formalism  are given in \citet{Hubeny1994}.  The forms in \citet{Hubeny1994} are appropriate for code frameworks that have occupational probabilities in the collisional and radiative rates as well as the opacity calculations.}  state-of-the-art stellar, white dwarf, neutron star, and hot subdwarf atmospheric modeling codes \citep{Hubeny2014, NSatmos, Rauch2013, TLusty, Bohlin2020}.

 We extend these RH calculations of M dwarf heating models from \citet{Kowalski2017Broadening} to the 5F11-25-4.2 solar flare chromospheric condensation model.  
We calculate the $t=3.8$~s spectrum using a 20-level hydrogen atom (max $n_j=19$) and a Ca II ion as described in \citet{Kowalski2017Broadening}. It is informative to compare the RH and RADYN calculations of H$\gamma$. 
The 20-level RH calculation is shown in Figure \ref{fig:comparison} and results in a lower emergent intensity by only 5\% at $\pm 1$\AA\ from $\lambda_{\rm{cen}}$; the wings are not affected. We are thus confident that the statistical equilibrium solution in the 20-level calculation is a reasonable approximation in the large density environment that is attained in the 5F11 model flare chromosphere.  

\begin{figure*}
\gridline{\fig{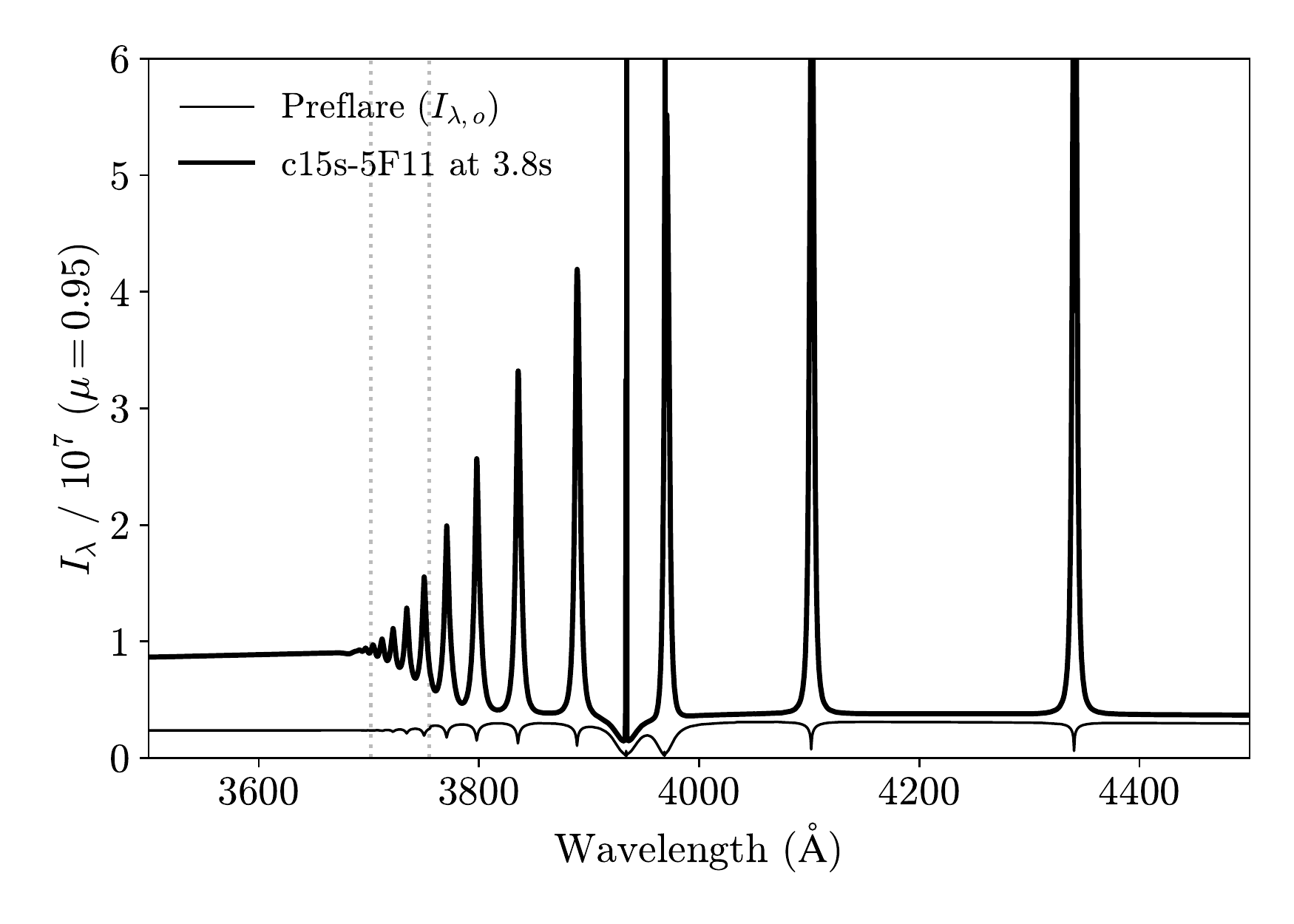}{0.5\textwidth}{(a)}
          }
\gridline{\fig{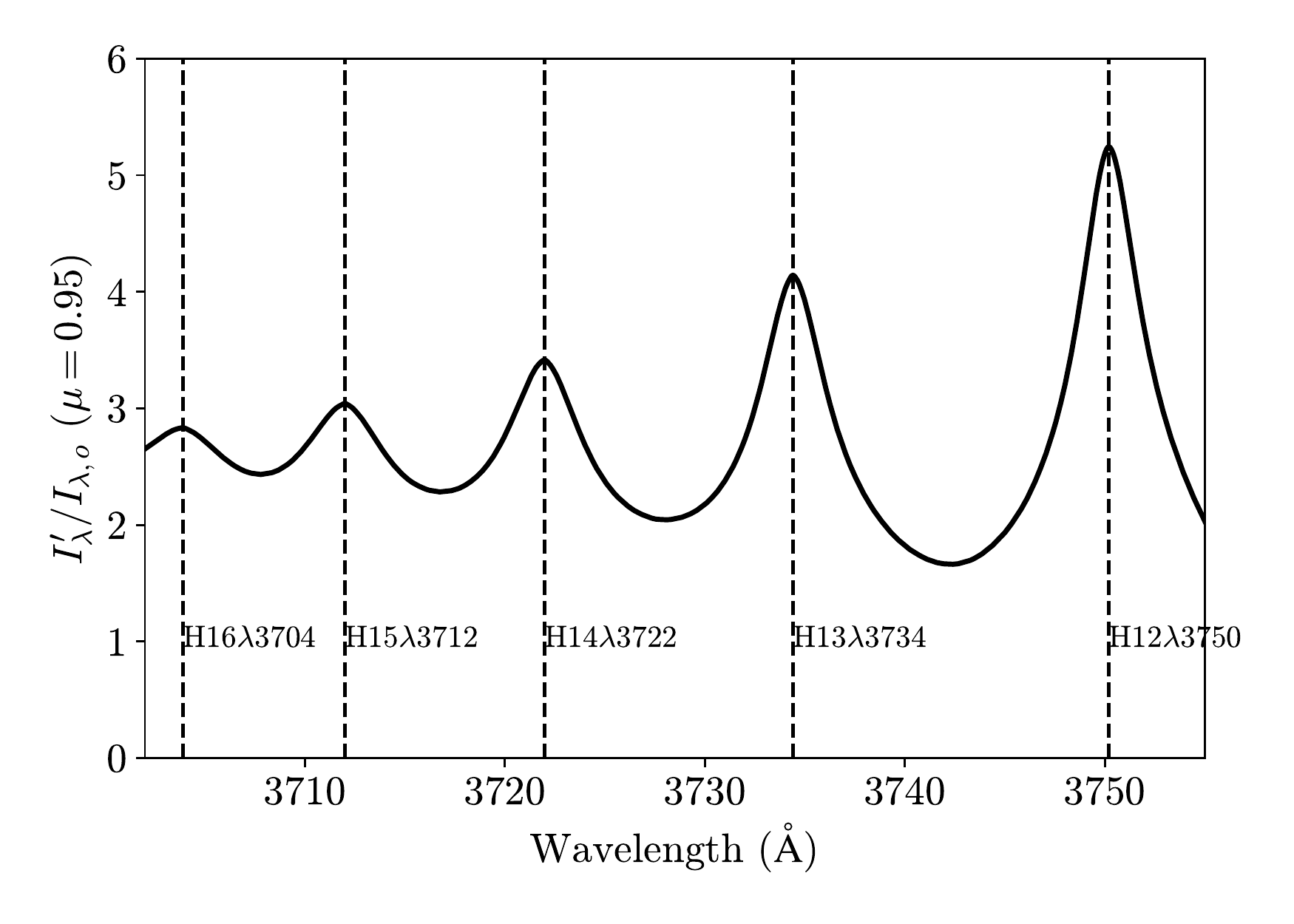}{0.5\textwidth}{(b)}
          }
          \gridline{\fig{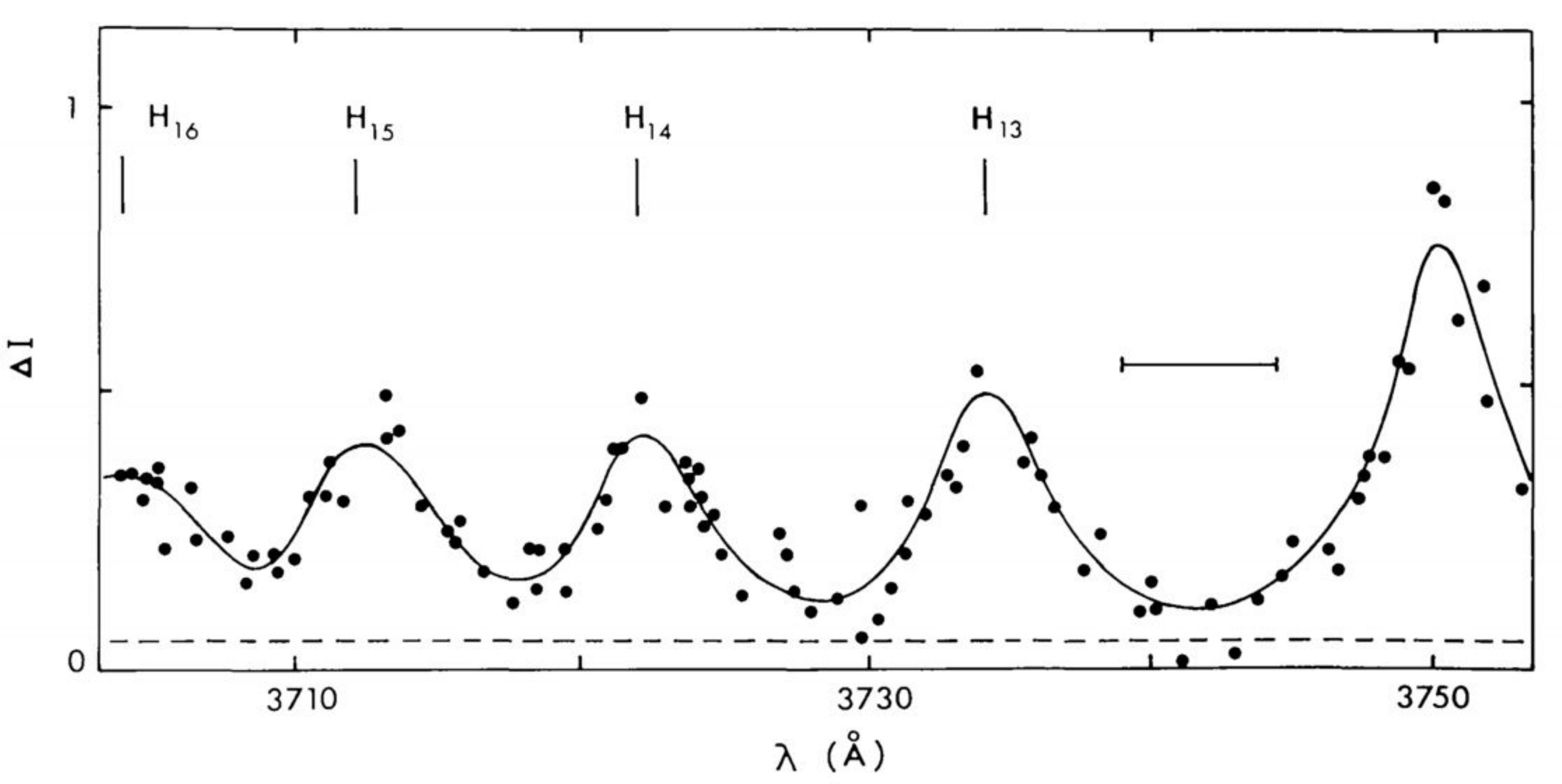}{0.46\textwidth}{(c)}
          }
\caption{ (a) Calculations of the Balmer series merging in the $U$ band using the RH code at $t=0$~s and $t=3.8$~s in the 5F11-25-4.2 model.  The calculation method is described in \citet{Kowalski2015} and \citet{Kowalski2017Broadening}.  (b) The excess flare intensity ($I_{\lambda}^{\prime} = I_{\lambda} - I_{\lambda,o}$; here, the $o$-subscript indicates the pre-flare time) contrast is shown within the wavelength range that is indicated by vertical lines in panel (a).  The Balmer lines fade into the dissolved-level Balmer continuum at $\lambda \approx 3646 - 3760$\AA.  The rest wavelengths used in the RH code are indicated.  
(c) Solar flare spectrum from Figure 6 of \citet{Neidig1983} over the same wavelength range as in panel (b).  The vertical axes of panels (b) and (c) are in units of $2.6 \times 10^6$ \ilam.  
\label{fig:balmer_limit}}
\end{figure*}

The model flare spectrum at the Balmer limit in the $U$-band is shown in Figure \ref{fig:balmer_limit}(a) compared to  the $t=0$~s model with the same RH calculation setup.
The wings of the high-$n_j$ Balmer lines broaden, fade, and merge into the dissolved-level Balmer continuum.  The main features of this spectrum are qualitatively similar to the observations of solar and stellar flare spectra \citep{Donati1985,HP91,Ondrej1}.
Figure \ref{fig:balmer_limit}(b) enlarges a wavelength region redward of the Balmer limit (3646 \AA) where the $n_j = 12$ (H12) to $n_j=16$ (H16) Balmer lines fade away due to decreasing oscillator strengths and occupational probabilities, $w_j$, of the upper levels of these transitions.

The spectral profiles of the high $n_j$ Balmer lines exhibit notable differences to the low $n_j$ Balmer line spectra in the 5F11 model.
The Doppler shifts of $\lambda_{\rm{cen}}$ for the Balmer lines in Figure \ref{fig:balmer_limit}(b) are negligible, in contrast to H$\alpha$, H$\beta$, and H$\gamma$ (Section \ref{sec:visp}).  The effective widths of the H12-H16 lines ($\Delta \lambda_{\rm{eff}} \approx 4-5$\AA, measured by linearly interpolating and subtracting the pseudo-continuum intensity in the troughs between the lines) are less than expected ($\Delta \lambda_{\rm{eff}} \gtrsim 13$ \AA) from the electron densities and optical depth broadening factors in the condensation.  Furthermore, the effective widths decrease with larger $n_j$ in Figure \ref{fig:balmer_limit}(b), which is also contrary to expectation.  To explain these effects, we investigate the contribution function to the emergent intensity in this spectral region.   The optical depth, $\tau_{\rm{CC}}$, at $\lambda_{\rm{cen}}$ for these high-$n_j$ lines and the optical depth in the troughs between the lines converge to the low optical depth values, $\tau_{\rm{CC}} \approx 0.1$, of the Balmer continuum at $\lambda < 3646$ \AA.  The relatively low optical depth through the condensation at H13 results in a large percentage, $\approx 40$\%, of the emergent intensity at $\lambda_{\rm{cen}}$ from the stationary chromospheric flare layers where the electron density is $n_e \approx 1.5- 6 \times 10^{13}$ cm$^{-3}$.  In the condensation, the occupational probabilities are much smaller (e.g., $w_{13} \approx 0.5$, $w_{16} < 0.1$) for the upper levels of these transitions because the electron density is much larger, $n_e \approx 5\times 10^{14}$ cm$^{-3}$ \citep[\emph{cf.} Fig 9 of][]{Kowalski2015}).   The condensation contributes most to the emergent intensity in the wings and in the smooth dissolved-level Balmer continuum throughout the spectral range. 

There are several effects that generate the magnitude and shape of the spectral broadening of the high $n_j$ hydrogen Balmer lines.  In the unified theory profiles, the wings at $|\lambda - \lambda_{o}| \approx 1-7$ \AA\ occur in the transition from the impact to Holtsmark 
limits in the line profile functions \citep[\emph{cf.} Figure 8 of][]{Vidal1971}.  Thus, a power-law does not accurately describe the emergent spectral intensities at these wing wavelengths.  Furthermore, the non-ideal gas effects in the TB09 recalculations cause the line profile shapes in the far wings ($|\lambda - \lambda_{\rm{cen}}| \approx 3-7$ \AA) to be steeper than a Holtsmark power-law even at relatively small electron densities \citep[\emph{cf.} Figure 1 of ][]{Kowalski2017Broadening}.  In contrast, the H$\gamma$ line profile function (Figure \ref{fig:summary}) exhibits a Holtsmark power-law at $|\lambda - \lambda_{o}| \gtrsim 1.5$ \AA.  The emergent intensity spectrum of H$\gamma$ 
from RH (Figure \ref{fig:comparison}) exhibits a Holtsmark power-law in the far wings at $|\lambda - \lambda_{\rm{cen}}| \gtrsim 4$\AA, where the line intensity decreases below 15\% of the maximum intensity (at $\lambda_{\rm{cen}}$) and where non-ideal gas effects are not important.

In summary, the measured broadening of the high$-n_j$ Balmer lines is affected by the lower electron densities in the stationary flare layers just below the condensation, the highly broadened bound-bound transitions in the condensation, and the large range of occupational probabilities (level dissolution) for a given $n_j$ in the condensation and the stationary flare layers.  In addition to the inhomogeneity of the atmosphere, the spectral line shapes are also affected by the details of the unified line broadening theory in the transition from the impact limit to the far wing limit, which is steeper than the Holtsmark limit due to non-ideal effects (level dissolution).
These issues together present a cautionary lesson for using the Inglis-Teller relation \citep{InglisTeller, Kurochka1970, Neidig1984} and generic Lorentzian or Holtsmark profiles to estimate a single electron density value from spectra formed in a heterogeneously-stratified flaring chromosphere.

Figure \ref{fig:balmer_limit}(c) displays an archival observation \citep{Neidig1983} of the large solar flare on 24 April 1981.  Despite the large differences in flare excess intensity between the model in Figure \ref{fig:balmer_limit}(b) and the observation in Figure \ref{fig:balmer_limit}(c), as already noted for the H$\alpha$ line compared to more modern observations in Section \ref{sec:mar29}, the broadening and merging of the H12$-$H16 wings are apparently satisfactory.  
The electron densities that have been inferred from Balmer line widths in this spectral region in solar flares have typically been in the range of $n_e \approx 2-6 \times 10^{13}$ cm$^{-3}$ \citep{Svestka1965,Svestka1967b, Donati1985}, which are similar to the electron densities in the stationary chromospheric flare layers just below the condensation in the 5F11 model. 
The upper levels of the H12$-$H16 transitions are very close to LTE in the chromospheric condensation and stationary flare layers in the 5F11 model, and 
a detailed parameter study of this spectral region using the approach from \citet{KowalskiAllred2018} will be presented elsewhere.

\section{Discussion \& Speculation} \label{sec:discussion}
The RH code was not originally envisioned to robustly calculate many types of atmospheres with steep velocity gradients and shocks \citep{Uitenbroek2001}.  It is thus invaluable for the accurate TB09$+$HM88 hydrogen broadening calculations to be included in the time-dependent, detailed spectra from the RADYN code as opportunities for new flare observations become available in Sunspot Cycle 25.  
The non-equilibrium ionization and bound-bound cooling rates are now self-consistently calculated with the spectral broadening in RADYN, which is especially important in the first few seconds before flare condensations accrue large densities.  RH synthesis is yet necessary for accurate treatment of overlapping transitions of higher $n_j$ hydrogen lines at the Balmer and Paschen series limits and the overlapping Mg II $h$ and $k$ lines \citep{Kerr2019A, Kerr2019B}.  The accurate theoretical treatment of the charged particle perturbations of hydrogen lines in RADYN provides a  fulcrum that anchors state-of-the-art models of flare atmospheric evolution to spectral observations.  The TB09$+$HM88 hydrogen profiles in RADYN will facilitate efficient model comparisons to future observations with the DKIST/ViSP, which will clarify if chromospheric condensations in current flare models are too dense, too optically thick, and too rapid in their dynamical evolution.

 A wide variety of H$\alpha$ broadening magnitudes, profile shapes, and Doppler shifts have been reported in previous solar flare spectra \citep{Ichimoto1984, Canfield1990, JohnsKrull1997}.
 A modern-day, comprehensive survey of the hydrogen line broadening from bright flare kernels observed at factors of $10-50$ better spatial resolution than in the past is now possible and would be enormously useful.  Spectral observations are urgently needed to test the large densities achieved in chromospheric condensations in response to high electron beam heating rates.  The DKIST/ViSP will provide spectral coverage into the wings of hydrogen lines with high spatio-temporal resolution.   One possible ViSP setup in the blue employs three spectral arms at H$\gamma$, at the flare continuum wavelengths near $\lambda \approx 4170$ \AA, and at the wavelengths around Ca II H and Balmer $\epsilon$.  These combinations mitigate relative spatial alignment uncertainties introduced by atmospheric differential refraction.  
The $\lambda \approx 4170$ \AA\  spectral region is largely free of emission lines in stellar flare observations \citep{HP91, Kowalski2016} and corresponds to an opacity minimum in the optical for a large range of atmospheric temperatures and densities.   The effective width (Eq. \ref{eq:effective_width})  is a convenient measurement that can be readily calculated from ViSP spectra.  The effective widths of the low$-n_j$ Balmer lines roughly correspond to the wavelength distances from $\lambda_{\rm{cen}}$ at which the model condensations drop below $\tau = 1$.  For symmetric profiles, the effective widths also correspond to the wavelengths at which the emergent intensity over the low-$n_j$ Balmer line spectra decreases to $e^{-1}$ of the line maximum (see Figure \ref{fig:ibis_redux} and Figure \ref{fig:comparison})  and beyond which the bound-bound wing intensity approaches a Holtsmark power-law.  Like the calculation of equivalent widths, intensity calibration is not required for interline comparisons of the effective widths.  This observational strategy facilitates comparison to and resolution of some of the longstanding problems in stellar (M dwarf) flare RHD modeling.

\subsection{Solar-Stellar Connection} \label{sec:mdwarfs}
\citet{Kowalski2017Broadening} presented F13 models of M dwarf flares with the TB09$+$HM88 profiles calculated using the RH code from a condensation model with electron density of $n_e \approx 5 \times 10^{15}$ cm$^{-3}$ and optical depth $\tau_{\rm{CC}} \approx 0.5 - 5$ in the optical and NUV continuum.  At times when the model continuum spectra reproduce the observed characteristics of M dwarf flares, the Balmer series broadening has been shown to be far too large. A linear superposition of multiple emitting regions was developed \citep{Osten2016} and was invoked to better explain both the continuum and Balmer line properties in an M dwarf ``megaflare'', including a small line-to-continuum ratio of H$\gamma$/C4170$^{\prime} \approx 40$ \citep{Kowalski2017Broadening}.  Models with extremely dense chromospheric condensations have been further explored as building blocks in multithread modeling for less energetic M dwarf flares \citep{Kowalski2019HST};  the late phase evolution of condensations  may be important in explaining the red-optical continuum with a color temperature of $T \approx 5000$ K in the gradual decay phase of impulsive-type events \citep{Kowalski2013} and in the impulsive phase of gradual-type events \citep{Kowalski2019HST}.  The beam-generated condensation models should be complemented with high-density ``post-flare'' loop broadening predictions of Balmer lines using the inferred parameters of \citet{Heinzel2018}.

It is important to consider the H$\gamma$/C4170$^{\prime}$ quantity in future observations of solar flares.  This measurement can determine if there is a major discrepancy in the models of solar flare electron beam-heated atmospheres, as in the stellar case \citep{Kowalski2013}.   The H$\gamma$/C4170$^{\prime}$ values predicted by models of solar flare chromospheric condensations are very large, H$\gamma$/C4170$^{\prime} \approx 600-1200$ (Section \ref{sec:ratio}), and F11 models without condensations also produce very large values \citep{Allred2006, Kowalski2016, Kowalski2019HST}.    The observed
H$\gamma$/C4170$^{\prime}$ line-to-continuum ratios in impulsive phase optical spectra of M dwarf flares are much smaller, $\approx 20-40$ \citep{Kowalski2013}, and ratios as large as $150-200$ are sometimes reported \citep{Kowalski2019HST}. Such small values are interpreted as evidence for the failure of low-energy electron beams in explaining the heating in M dwarf flares \citep{Kowalski2013,Kowalski2019HST}.  Even very large F13 electron beam fluxes are unable to reproduce the low-end of the distribution of H$\gamma$/C4170$^{\prime}$ values near $20$ \citep{Kowalski2017Broadening}.
Solar flare models that match the observed Balmer line broadening and intensity, while failing to reproduce the line-to-continuum ratio, would better constrain whether the model difficiencies are in the stationary flare layers where the highest-energy beam electrons heat the atmosphere \citep{Graham2020}. The problem could also lie in the physical treatment of upper photospheric heating in RADYN models, which predict that only radiative backwarming can cause a significant amount of temperature increase \citep{Allred2005, Allred2006}.  The higher-$n_j$ Balmer lines in the $U$-band or the high-$n_j$ Paschen lines, which are even more optically thin in chromospheric condensations, could provide direct measurements of the electron densities in these deeper regions.

Intriguingly, there have been several recent reports of small optical Balmer line-to-continuum ratios in solar flares \citep{KCF15, Ondrej1}, but such spectra may be representative of the so-called Type II white-light flare phenomenon \citep{Boyer1985, Fang1995, Ding1999, Ondrej1}, which is thought to result from a yet-unknown source of deep heating, possibly proton beams \citep{Ondrej2} or Alfven waves  \citep{Russell2013}.  Our models here seek instead to reproduce the fundamental RHD physics of the flaring chromosphere and the self-consistent photospheric backwarming in footpoints where high fluxes of mildly relativistic ($E=20-40$ keV) electrons are inferred from RHESSI observations \citep[e.g.,][]{Fletcher2007, Krucker2011, Kleint2016}.

Recently, \citet{Namekata2020} presented broad H$\alpha$ lines in an M dwarf flare, which was reproduced well with short-burst, F12 models using the TB09$+$HM88 profiles in RADYN.  The effective widths (which we estimate as $\approx 8-15$\AA) of their H$\alpha$ model spectra at $R \approx 2000$ are comparable to the effective widths ($\approx 12$\AA; Section \ref{sec:dec}) in the 5F11 model of solar flare chromospheric condensations.
However, the Balmer line formation is much different in the 5F11 solar model.  In the M dwarf F12 models of Namekata et al., the H$\alpha$ wing broadening is a result of deep heating from a much harder ($\delta = 3$) electron beam, combined with the self-absorption effect.  The larger flux of high-energy electrons in the $\delta=3$ F12 beam generates electron densities of $n_e \approx 10^{14}$ cm$^{-3}$ in deeper stationary chromospheric flare layers.  In the 5F11 solar flare model, only about 25\% of the far H$\alpha$ wing and nearby continuum intensity at $z>400$ km is produced in the stationary flare layers, which exhibit a lower electron density of $n_e < 10^{14}$ cm$^{-3}$ (Section \ref{sec:lz}, see also Paper I).  The remaining emergent intensity originates in the condensation region with $n_e \approx 5 \times 10^{14}$ cm$^{-3}$.   Intriguingly, chromospheric condensations do not promptly form in response to the short-duration heating employed in the Namekata et al.\ beam models.  Instead, relatively slow upflows initially develop that are similar to those within the first 10~s of the F11-25-4 solar flare model \citep[Table \ref{table:models}; \emph{cf.} Figure 10 in Namekata et al. at $z > 300$ km and Figure 8 in][]{Kuridze2015}.  For the same low-energy cutoff and power-law index as in \citet{Namekata2020}, larger beam fluxes near F13 produce dense chromospheric condensations in M dwarf models.  A larger fraction of the emergent wing and optical/NUV continuum intensity originates from the condensation due to the large optical depths that form in the compressed $T \approx 10^4$ K gas \citep{Kowalski2015, Kowalski2016, KowalskiAllred2018}.   As mentioned above, these models produce profiles that are far too broad unless several \emph{ad hoc}, spatially separated flare kernels or ribbons -- one region producing most of the blue continuum and broad Balmer wings and another producing most of the narrow emission line flux around $\lambda_{\rm{cen}}$ -- are employed \citep[see Figure 9 of][Kowalski et al. 2021, in prep]{Kowalski2017Broadening}.  High-resolution solar longslit data could directly constrain the roles of spatial inhomogeneity in stellar flare observations of Balmer wing broadening.

\subsection{The Short Timescales and Large Densities of Chromosheric Condensation Models} \label{sec:newphysics}
The equations of RHD inextricably link the timescales and densities in flare models. Throughout the atmosphere, the
time-evolution of the mass and momentum flux gradients are self-consistent with the instantaneous densities, which are constrained by accurate models of the broadening of hydrogen emission lines.
The half-lives of chromospheric condensation speeds in solar flare models are known to be $\approx 3$x too short compared to the observational development of the red-shifted satellite components in chromospheric lines \citep[see discussion related to $t_{1/2}$ in][]{Fisher1989}.  \citet{Fisher1989} speculated that this discrepancy could be due to the spatial resolution of H$\alpha$ observations, but  \citet{Graham2020} found that similar lifetimes were present  at a spatial resolution of 0.\arcsec 4 in IRIS spectra of Fe II.  In our RHD models  that produce prompt chromospheric condensations that decelerate from $-100$ km s$^{-1}$ to $-15$ km s$^{-1}$ within $\Delta t \approx 10$~s (Figure \ref{fig:ccevol}, Section \ref{sec:results}), the redshifts of $\lambda_{\rm{cen}}$ in the Balmer lines show the tail-end of this rapid decrease of condensation speed by a factor of $\approx 2$ within just $\Delta t \approx 6$~s.  The timescale problem seems to be worse in lower flux models.  \citet{Kuridze2015} discusses that the H$\alpha$ 2v and 2r variations are well-reproduced in  the F11-25-4 RADYN model here but that the modeled variations occur over a time of 20~s instead of the observed duration of 5 minutes. \citet{Kerr2021} used RADYN to reproduce the magnitude of He I 10830 \AA\ dimming that was observed at the fronts of solar flare ribbons \citep{Xu2016}, but the duration of the dimming is too short-lived in the models with beam fluxes above 5F10. The spatial resolution of the observations is discussed as a potential source of the discrepancy.  

With the ViSP, the spatial resolution along the spectrograph slit in the blue is expected to be better than 0.\arcsec 06.  Combined with high-time resolution, such new observations will clarify the inter-connected problems of larger-than-observed model intensities, faster-than-observed condensation development and deceleration, and broader-than-observed Balmer line profiles in models with large electron beam fluxes.   Evidence that the model chromospheric condensations of bright flare kernels are too dense and too optically thick could be found in observational detections of persistent emission line components around $\lambda_o$ in spectra of low-$n_j$ Balmer lines at times when the red satellite components brighten and broaden (Figure \ref{fig:time_evol}).  It will also be important to quantify the observational bias in previous spectral observations toward longer timescales of $\approx 10-30$~s:  much shorter chromospheric radiation bursts have been reported with durations of $\approx 0.3 - 4$~s  \citep[e.g.][see also Figure 3 of \citealt{Kowalski2019IRIS}]{Aschwanden1995, Radz2007, Nishizuka2009, Qiu2012, Penn2016, Knuth2020}.

Improvements to the realism of chromospheric condensation modeling in solar flares may be warranted by critical assessments of the new hydrogen wing broadening predictions.  We have identified several areas of improvement for flare RHD model calculations of the self-consistent evolution of densities, velocities, decelerations, and spectral broadening. Briefly, they are the following:

\begin{itemize}
\item  A magnetic nozzle caused by field convergence in the low atmosphere.  \citet{Emslie1992} found that downward flows are compressed/decelerated and upward flows are rarefied/accelerated when a varying cross-sectional (tapered) loop area is included in the hydrodynamic equations \citep[see also][and references therein]{Reep2020}. The magnetic field convergence in flare loops is poorly constrained but will likely be improved with the DKIST and Expanded Owens Valley Solar Array (EOVSA).

\item  Advances in conductive heat transport physics.  The nonthermal beam energy is deposited in the center of an expanding, million K temperature explosion in the middle or upper chromosphere.  Heat flux occurs from the center of the temperature explosion to the upward- and downward-moving fronts through thermal conduction.  At these fronts, there is a spike in thermal conductive energy deposition as the temperature drops to $T < 350,000$ K \citep[see the energy balance analysis of the upward-moving front in][]{Kennedy2015}.  This generates a propagating non-equilibrium, $T \approx 10^5$ K, high-density, He II$\rightarrow$III ionization front with radiative losses primarily in optically thin transitions and He II 304 \AA.  The thermal pressure spike is very narrow but is well-resolved with the adaptive grid in RADYN, and it appears to occur in all RADYN flare simulations of chromospheric condensation phenomena \citep[see also the models of][]{Gan1991}.  The hydrodynamic variables at this pressure spike are roughly consistent with the self-similar relations expected behind the front of a Sedov-Taylor blast wave \citep[e.g.,][]{Titan1994,Thorne2017}, while the cool, chromospheric condensation that is in front of the downward-propagating conduction spike resembles a ``snow-plow'' region \citep{Shu1992}.  

The thermo- and hydrodynamic properties within the spike are self-consistent with the expected relationships among the ambient macroscopic variables, such as the ideal gas equation of state and the equation of momentum conservation.  However, the conduction spike may be abated if additional microphysical processes are included in RADYN.  Though thermal heat flux saturation is employed in RADYN at steep temperature gradients \citep{Smith1980, Fisher1985V}, recent work has shown that further suppression may occur due to turbulent magnetic field fluctuations along the flare loop \citep{Bian2018, Emslie2018}; work is ongoing to include this in RADYN (Allred et al., in preparation).  Non-local conductive heat transport \citep{Karpen1987} and particle diffusion  \citep{Fontenla2002} may also ``smear'' out the conduction fronts, leading to less dramatic pressure gradients in flare simulations.

\item Limits on radiative cooling rates at flare transition region densities and temperatures \citep{LeenaartsCarlsson2012}.  As evident in Figure \ref{fig:ccevol}, the temperature in the condensation drops precipitously within the first second.  The detailed radiative cooling in the condensation at $t=0.4-1$~s originates primarily from He II 304 \AA\ and in the bound-free Lyman continuum.  However, the major source of radiative cooling occurs from minor species that are calculated in the optically thin limit using CHIANTI \citep{Dere1997}.  Thermal non-equilibrium ionization \citep[e.g.,][]{Bradshaw2003, Doyle2012} and beam collisional excitation and ionization rates \citep[e.g.,][]{Zharkova1993, Karlicky2004, Kasparova2009, Kerr2021} may also be important for minor species at early times in the beam injection.  However, the densities in the condensation (Figure \ref{fig:ccevol}) are large enough that assumptions about the opacity in transition region lines of minor species \citep{LeenaartsCarlsson2012, Kerr2019Si, Mathioudakis1999} are possibly a more important source of systematic modeling inaccuracy.

\item Time-variable injection of the beam energy.  We use a step-function for the beam heating impulse with a dwell time of $10-20$~s in the models in Table \ref{table:models}.  One may expect that in reality the injected nonthermal heating flux ramps up to a maximum, and then it ramps down.  A pulsed variation in heating \citep{Aschwanden2004} was investigated in \cite{Zhu2019} with a rise time of 9~s to a maximum beam flux of 5F11, and a pulse FWHM of 20~s.  We analyze the hydrogen lines from this simulation but do not find notable differences in the timescales of the hydrodynamics or line broadening in the spectra after the condensation starts to decelerate.

\item The horizontal expansion of plasma in  chromospheric flare footpoints, which may result from large thermal pressure gradients as described above.  If magneto-hydrodynamical forces are considered \citep[e.g.][]{Low1975, Uitenbroek2003, Abbett2007, Gudiksen2011,Rempel2017}, we speculate that a drop in density and optical depth in the model condensations may occur.  However, it is not immediately evident how currently established frameworks of solar 3D MHD models are modified  \citep{Cheung2019} to resolve geometrically thin condensation regions with high-emissivity gas/plasma in flare footpoints.  
\end{itemize}

A major improvement to RADYN flare modeling has recently been made possible with the FP code \citep{Allred2020}, allowing for return current effects and magnetic mirroring of particles to be accurately included in the simulations.  A predecessor (\verb|fp_solver.f|) to the Coulomb collision module of the FP code was used to calculate the beam heating at each time-step in the RADYN simulations in this work and in those that have been previously presented with the TB09$+$HM88 profiles.  We are working to integrate the FP code and the TB09$+$HM88 profiles into a common version of RADYN, which will be released on github at a future date.  Notably, this merged version of the RADYN source code employs 201 wavelength points across the hydrogen lines, thus allowing more detailed analysis of line profiles than with $31-51$ points over the profiles that are presented here.

\section{Summary \& Conclusions} \label{sec:conclusions}
This is the second paper in a series that explores the atmospheric response to heating from a large flux of deka-keV power-law electrons in solar flares.  In Paper I, we compared high-flux models of chromospheric condensations to observations of the IRIS/NUV Fe II lines in the 2014 Mar 29 solar flare.  In the model  spectra, these lines exhibit emission components from the condensation in the mid-upper chromosphere and from the stationary flare layers in the lower-mid chromosphere.  Here, we extend the predictions of Paper I to hydrogen Balmer lines using an updated treatment of the pressure broadening due to ambient (thermal) electrons and protons/ions.  Unlike other chromospheric flare lines, such as Fe II and Mg II, electronic and ionic pressure broadening of hydrogen lines dominates over other sources of microphysical broadening by several orders of magnitude, and its fundamental physics is well understood and extensively tested against laboratory and astrophysical plasmas.  The results from sophisticated quantum calculations are readily incorporated in stellar atmospheric modeling codes.  ``Quiet Sun'' synthesis models have started to consider the advantages of these accurate, self-consistent hydrogen broadening profiles as well \citep{Bjorgen2019, Marchenko2021}.  Here we describe the implementation of new line profile functions in the state-of-the-art RHD code RADYN, which is widely used for simulations of a wide variety of solar flare phenomena \citep{Simoes2017, Polito2019, Sadykov2020, Monson2021}, Ellerman bombs \citep{Hong2017, Reid2017}, nanoflares \citep{Testa2014, Polito2018}, and chromospheric dynamics such as spicules and shocks \citep{Carlsson2002, WedemeyerBohm2011, Guerreiro, Molnar2021}.
The updated hydrogen line profile functions are known as the TB09$+$HM88 profiles, which are extensions of the VCS unified theory calculations to higher-$n_j$ lines with self-consistent, non-ideal broadening from electron collisions and proton/ion perturbations.

We focused our analysis on H$\gamma$ spectra from two solar flare models that produce prompt chromospheric condensations with a range of ambient electron/proton densities from $\approx 1 - 6 \times 10^{14}$ cm$^{-3}$.  Previous RADYN models of chromospheric condensations show a bright redshifted component in Ca II K \citep{Abbett1999}, Mg II \citep{Zhu2019}, and Fe II (Paper I), but the small broadening that is expected from proton and electron collisions with non-hydrogenic ions is not helpful as a direct diagnostic of chromospheric flare electron densities.  
To quantify the broadening in the emergent model spectra and characterize its relationship to electron density in flare atmospheres, we calculate effective widths, which are more dependent on the wing broadening than the traditional FWHM measure.  
Large optical depths in the condensations create line-core saturation in the emergent intensity that is given by the Eddington-Barbier approximation.  The optical depths generate an opacity broadening in the emergent intensity around $\lambda_{\rm{cen}}$ where the Eddington-Barbier approximation holds and the source function decreases.  Additionally, a  broadening enhancement occurs through the transition to the optically thin wings where photons escape over a much larger fraction of the high-density condensation. The ``optical depth broadening factors'' of the emergent intensity spectra of H$\gamma$ range from a factor of two to five  above the optically thin spectral broadening calculated from the heterogeneous distribution of charged particle densities in the dynamic flare chromosphere.  It has long been known that column hydrogen density in $n=2$ (optical depth) and charge particle density must be fit simultaneously to infer electron densities from solar flare spectra of Balmer lines \citep{Svestka1965}.  The TB09$+$HM88 profiles facilitate self-consistent inferences of these parameters from non-uniform, non-equilibrium flare models.

Unlike emission line flare models of Fe II in the NUV \citep[Paper I and][]{Graham2020},
the entire line profiles of H$\alpha$, H$\beta$, and H$\gamma$ shift by the maximum downward velocity in the condensation and are symmetric about this shift \citep[see also the Mg II spectra in][]{Zhu2019} by $t=4$~s.   This is due to the very large optical depths of the low-$n_j$ Balmer lines in the chromospheric condensation at $T\approx10,000 - 20,000$ K.   None of the emergent intensity at $\lambda_{\rm{cen}}$  originates from the stationary flare layers with lower electron density below the condensation.   An optical line-to-continuum ratio (H$\gamma$/C4170$^{\prime}$) is thus useful in determining  missing physics in the deeper regions of the atmosphere.  The condensation models exhibit much broader H$\gamma$ lines than Ca II H and K, which constrain the contributions from nonthermal unresolved mass motions (e.g., micro- and macro-turbulence) in the flare chromosphere.  The models also predict that the H$\alpha$ line is broadened more than the H$\gamma$ line due to much larger optical depths in H$\alpha$ through the condensations.  However, the intensity at $\lambda_{\rm{cen}}$ for H$\alpha$ is less than H$\gamma$, which we attribute to differences in the respective source functions over the heights where $\lambda_{\rm{cen}}$ is formed.  The use of the Eddington-Barbier approximation is justified by the approximately linear decrease in source function with $\tau$ as $\tau \gg 1$ within the upper layers of the condensation.  The wavelength-integrated intensity is a product of the intensity at $\lambda_{\rm{cen}} \approx \lambda_{\rm{max}}$ and the effective width (through Eq. \ref{eq:effective_width}), which explains the reversed Balmer decrement values in the model spectra.  

In each spectrum of the low$-n_j$ hydrogen Balmer lines, the broadening changes rapidly over the first 4~s of beam heating for a range of beam fluxes.  
The condensations in our models begin in a state that is far from equilibrium for Balmer line formation but evolve toward LTE after only several seconds as collisional rates for Balmer lines increase in the condensation.  As the condensation continues to plow through the lower atmosphere and accrue mass after 4~s, the Balmer line source functions further approach LTE.   Around the centroid ($\lambda_{\rm{cen}}$) of the low-$n_j$ Balmer lines, the emergent intensity is closely related to the decreasing temperature evolution in the upper $\approx 3$ km of the chromospheric condensation.
 The optically thick wavelengths over the profiles thus become fainter over time as the condensations cool.   Within half of the effective widths of the low$-n_j$ Balmer line centroids, the spectra broaden over time due to increasing bound-bound optical depths and electron/proton densities in the region of line formation in the chromospheric condensation.   In the far wings, the optical depth broadening and large charged particle densities cause the large widths, and there are minor amounts of additional broadening in the emergent intensity spectra due to increased bound-free radiation that escapes from the backwarmed photosphere.

The high flux (5F11) beam heating model results in broadening of the emergent spectra of high-$n_j$ Balmer series members (H12-H16) that is in reasonable agreement with a unique archival observation of a solar flare that was observed in the early 1980s.   The high-$n_j$ Balmer lines at $\lambda = 3700-3760$\AA\ are much more optically thin than H$\gamma$, H$\beta$, and H$\alpha$, and their upper levels are largely dissolved in the chromospheric condensation ($n_e \approx 5 \times 10^{14}$ cm$^{-3}$).  These two effects create a translucent opacity window into the deeper layers of the flare atmosphere and explain why the bright 5F11 model spectrum is consistent with  previous inferences of much lower electron densities using this spectral range.  The widths of H12 - H16 lines are formed by lower charged particle densities  ($n_e \approx 1.5 - 6 \times 10^{13}$ cm$^{-3}$) and small curve-of-growth broadening enhancements in the stationary flare layers just below the condensation.  Thus, the Inglis-Teller relationship cannot be accurately applied to spectra from heterogeneous flare atmosphere models; detailed analyses of contribution functions and occupational probabilities are crucial.   The higher-$n_j$ lines from H$\gamma$ through H16 are much closer to LTE than H$\alpha$.  This offers a promising path toward modeling beam-heated chromospheric condensations late into their evolution using simplified approaches \citep[e.g.,][]{KowalskiAllred2018}.

The extreme broadening and large redshifts of H$\alpha$ in our high-flux model are ostensibly consistent with recent IBIS observations \citep{Rubio2016} of flat profiles over a narrow spectral range ($\lambda_o \pm 1.5$ \AA).   A different modeling approach in \citet{Druett2018} reached a similar conclusion from their comparisons to these observations, but a factor of 100 smaller densities produce the Balmer line broadening in their chromospheric condensation models.  
 To our knowledge, the electron/proton densities that broaden the low-$n_j$ Balmer line profiles in our chromospheric condensation models are higher than any previously inferred values from broadening analyses of solar flare hydrogen lines, which are typically in the range of $n_e \approx 2-6 \times 10^{13}$ cm$^{-3}$ \citep{Suemoto1959, Svestka1962, Svestka1963, Svestka1965, Svestka1967, Svestka1967b, Machado1974, Neidig1983, Donati1985, JohnsKrull1997}.  However, lower electron densities have been inferred as well \citep{Hiei1982, Neidig1984, KCF15}. 
 
 The new pressure-broadened, redshifted hydrogen profiles in RHD models need to be verified against future observations of solar flares.
 Spectral observations from the ViSP on the DKIST (or the Horizontal Spectrograph on the Dunn Solar Telescope) into the wings of hydrogen lines will readily determine if the large densities, optical depths, and the rapid development and deceleration of chromospheric condensations as predicted by state-of-the-art RADYN flare models are realistic.  Notable differences can provide specific guidance for concerted efforts to improve the physics in models of solar and stellar flares.  Agreement between hydrogen spectral observations and models may be crucial in resolving some of the most widely debated issues from Sunspot Cycle 24. For example,  \citet{Kennedy2015} discusses how chromospheric  condensations in RADYN models descend (e.g., middle panel of Figure \ref{fig:ccevol}) toward the temperature minimum region and may explain the surprisingly low heights of the SDO/HMI and hard X-ray sources that were reported in \citet{Martinez2012} \citep[but see also the height inferences in the same flare reported in ][]{Battaglia2011}.  Our new RADYN chromospheric condensation models predict very broad, symmetric spectra of H$\alpha$, H$\beta$, and H$\gamma$ that are redshifted by less than $15$ km s$^{-1}$ at late times in the atmospheric response.   Until high-resolution observations of the wings of the Balmer lines are available, we suggest a closer inspection of the vast amount of IBIS and CRISP data of solar flares  from Sunspot Cycle 24. Many unpublished spectra from IBIS exhibit flat profiles with slopes increasing to the red side (A. Tritschler, 2019, priv. communication; L. Kleint 2018, priv. communication).   

All spectral profiles of the Balmer lines from the models in this work are readily available as FITS files through Zenodo.  The atmospheric variables and contribution functions are available upon request.

\acknowledgments
We thank an anonymous referee for their careful reading and constructive feedback, which helped improve the presentation of the ideas in this paper.  A.F.K. thanks Yingjie Zhu for helpful conversations on collisional broadening theory.  A.F.K. is grateful for funding support from NSF Award 1916511 and NASA ADAP 80NSSC21K0632.

\bibliography{main_final}{}

\begin{thebibliography}{}
\expandafter\ifx\csname natexlab\endcsname\relax\def\natexlab#1{#1}\fi
\providecommand{\url}[1]{\href{#1}{#1}}
\providecommand{\dodoi}[1]{doi:~\href{http://doi.org/#1}{\nolinkurl{#1}}}
\providecommand{\doeprint}[1]{\href{http://ascl.net/#1}{\nolinkurl{http://ascl.net/#1}}}
\providecommand{\doarXiv}[1]{\href{https://arxiv.org/abs/#1}{\nolinkurl{https://arxiv.org/abs/#1}}}

\bibitem[{{Abbett}(1998)}]{Abbett1998}
{Abbett}, W.~P. 1998, PhD thesis, MICHIGAN STATE UNIVERSITY

\bibitem[{{Abbett}(2007)}]{Abbett2007}
---. 2007, \apj, 665, 1469, \dodoi{10.1086/519788}

\bibitem[{{Abbett} \& {Hawley}(1999)}]{Abbett1999}
{Abbett}, W.~P., \& {Hawley}, S.~L. 1999, \apj, 521, 906,
  \dodoi{10.1086/307576}

\bibitem[{{Alaoui} \& {Holman}(2017)}]{Alaoui2017}
{Alaoui}, M., \& {Holman}, G.~D. 2017, \apj, 851, 78,
  \dodoi{10.3847/1538-4357/aa98de}

\bibitem[{{Allred} {et~al.}(2020){Allred}, {Alaoui}, {Kowalski}, \&
  {Kerr}}]{Allred2020}
{Allred}, J.~C., {Alaoui}, M., {Kowalski}, A.~F., \& {Kerr}, G.~S. 2020, \apj,
  902, 16, \dodoi{10.3847/1538-4357/abb239}

\bibitem[{{Allred} {et~al.}(2005){Allred}, {Hawley}, {Abbett}, \&
  {Carlsson}}]{Allred2005}
{Allred}, J.~C., {Hawley}, S.~L., {Abbett}, W.~P., \& {Carlsson}, M. 2005,
  \apj, 630, 573, \dodoi{10.1086/431751}

\bibitem[{{Allred} {et~al.}(2006){Allred}, {Hawley}, {Abbett}, \&
  {Carlsson}}]{Allred2006}
---. 2006, \apj, 644, 484, \dodoi{10.1086/503314}

\bibitem[{{Allred} {et~al.}(2015){Allred}, {Kowalski}, \&
  {Carlsson}}]{Allred2015}
{Allred}, J.~C., {Kowalski}, A.~F., \& {Carlsson}, M. 2015, \apj, 809, 104,
  \dodoi{10.1088/0004-637X/809/1/104}

\bibitem[{{Arnold} {et~al.}(2021){Arnold}, {Drake}, {Swisdak}, {Guo}, {Dahlin},
  {Chen}, {Fleishman}, {Glesener}, {Kontar}, {Phan}, \& {Shen}}]{Arnold2021}
{Arnold}, H., {Drake}, J.~F., {Swisdak}, M., {et~al.} 2021, \prl, 126, 135101,
  \dodoi{10.1103/PhysRevLett.126.135101}

\bibitem[{{Aschwanden}(2004)}]{Aschwanden2004}
{Aschwanden}, M.~J. 2004, \apj, 608, 554, \dodoi{10.1086/392494}

\bibitem[{{Aschwanden} {et~al.}(1995){Aschwanden}, {Schwartz}, \&
  {Alt}}]{Aschwanden1995}
{Aschwanden}, M.~J., {Schwartz}, R.~A., \& {Alt}, D.~M. 1995, \apj, 447, 923,
  \dodoi{10.1086/175930}

\bibitem[{{Ashfield} \& {Longcope}(2021)}]{Ashfield2021}
{Ashfield}, W.~H., \& {Longcope}, D.~W. 2021, \apj, 912, 25,
  \dodoi{10.3847/1538-4357/abedb4}

\bibitem[{{Baranger}(1958{\natexlab{a}})}]{Baranger1958}
{Baranger}, M. 1958{\natexlab{a}}, Physical Review, 112, 855,
  \dodoi{10.1103/PhysRev.112.855}

\bibitem[{{Baranger}(1958{\natexlab{b}})}]{Baranger1958b}
---. 1958{\natexlab{b}}, Physical Review, 111, 494,
  \dodoi{10.1103/PhysRev.111.494}

\bibitem[{{Baranger}(1962)}]{Baranger1962}
---. 1962, Pure and Applied Physics, 13, 493,
  \dodoi{10.1016/B978-0-12-081450-3.50017-5}

\bibitem[{{Barklem}(2016)}]{Barklem2016}
{Barklem}, P.~S. 2016, \aapr, 24, 9, \dodoi{10.1007/s00159-016-0095-9}

\bibitem[{{Barklem} {et~al.}(2000){Barklem}, {Piskunov}, \&
  {O'Mara}}]{Barklem2000}
{Barklem}, P.~S., {Piskunov}, N., \& {O'Mara}, B.~J. 2000, \aap, 363, 1091.
\newblock \doarXiv{astro-ph/0010022}

\bibitem[{{Battaglia} \& {Kontar}(2011)}]{Battaglia2011}
{Battaglia}, M., \& {Kontar}, E.~P. 2011, \aap, 533, L2,
  \dodoi{10.1051/0004-6361/201117605}

\bibitem[{{Bian} {et~al.}(2018){Bian}, {Emslie}, {Horne}, \&
  {Kontar}}]{Bian2018}
{Bian}, N., {Emslie}, A.~G., {Horne}, D., \& {Kontar}, E.~P. 2018, \apj, 852,
  127, \dodoi{10.3847/1538-4357/aa9f29}

\bibitem[{{Bj{\o}rgen} {et~al.}(2019){Bj{\o}rgen}, {Leenaarts}, {Rempel},
  {Cheung}, {Danilovic}, {de la Cruz Rodr{\'\i}guez}, \&
  {Sukhorukov}}]{Bjorgen2019}
{Bj{\o}rgen}, J.~P., {Leenaarts}, J., {Rempel}, M., {et~al.} 2019, \aap, 631,
  A33, \dodoi{10.1051/0004-6361/201834919}

\bibitem[{{Bohlin} {et~al.}(2014){Bohlin}, {Gordon}, \&
  {Tremblay}}]{Bohlin2014}
{Bohlin}, R.~C., {Gordon}, K.~D., \& {Tremblay}, P.-E. 2014, \pasp, 126, 711,
  \dodoi{10.1086/677655}

\bibitem[{{Bohlin} {et~al.}(2020){Bohlin}, {Hubeny}, \& {Rauch}}]{Bohlin2020}
{Bohlin}, R.~C., {Hubeny}, I., \& {Rauch}, T. 2020, \aj, 160, 21,
  \dodoi{10.3847/1538-3881/ab94b4}

\bibitem[{{B{\"o}hm-Vitense}(1989)}]{BohmVitense1989}
{B{\"o}hm-Vitense}, E. 1989, {Introduction to stellar astrophysics. Vol. 2.
  Stellar atmospheres.}, Vol.~2

\bibitem[{{Boyer} {et~al.}(1985){Boyer}, {Sotirovsky}, {Machado}, \&
  {Rust}}]{Boyer1985}
{Boyer}, R., {Sotirovsky}, P., {Machado}, M.~E., \& {Rust}, D.~M. 1985,
  \solphys, 98, 255, \dodoi{10.1007/BF00152459}

\bibitem[{{Bradshaw} \& {Mason}(2003)}]{Bradshaw2003}
{Bradshaw}, S.~J., \& {Mason}, H.~E. 2003, \aap, 401, 699,
  \dodoi{10.1051/0004-6361:20030089}

\bibitem[{Bransden \& Joachain(2000)}]{Bransden}
Bransden, B.~H., \& Joachain, C.~J. 2000, Quantum Mechanics (Harlow, England ::
  Pearson, Prentice Hall)

\bibitem[{{Brissaud} \& {Frisch}(1971)}]{MMM1}
{Brissaud}, A., \& {Frisch}, U. 1971, \jqsrt, 11, 1767,
  \dodoi{10.1016/0022-4073(71)90021-5}

\bibitem[{{Canfield} \& {Gayley}(1987)}]{Canfield1987}
{Canfield}, R.~C., \& {Gayley}, K.~G. 1987, \apj, 322, 999,
  \dodoi{10.1086/165795}

\bibitem[{{Canfield} {et~al.}(1984){Canfield}, {Gunkler}, \&
  {Ricchiazzi}}]{Canfield1984}
{Canfield}, R.~C., {Gunkler}, T.~A., \& {Ricchiazzi}, P.~J. 1984, \apj, 282,
  296, \dodoi{10.1086/162203}

\bibitem[{{Canfield} {et~al.}(1990){Canfield}, {Penn}, {Wulser}, \&
  {Kiplinger}}]{Canfield1990}
{Canfield}, R.~C., {Penn}, M.~J., {Wulser}, J.-P., \& {Kiplinger}, A.~L. 1990,
  \apj, 363, 318, \dodoi{10.1086/169345}

\bibitem[{{Carlsson}(1986)}]{MULTI1}
{Carlsson}, M. 1986, Uppsala Astronomical Observatory Reports, 33

\bibitem[{{Carlsson}(1992)}]{MULTI2}
{Carlsson}, M. 1992, in Astronomical Society of the Pacific Conference Series,
  Vol.~26, Cool Stars, Stellar Systems, and the Sun, ed. M.~S. {Giampapa} \&
  J.~A. {Bookbinder}, 499

\bibitem[{{Carlsson} {et~al.}(2019){Carlsson}, {De Pontieu}, \&
  {Hansteen}}]{Carlsson2019}
{Carlsson}, M., {De Pontieu}, B., \& {Hansteen}, V.~H. 2019, \araa, 57, 189,
  \dodoi{10.1146/annurev-astro-081817-052044}

\bibitem[{{Carlsson} \& {Leenaarts}(2012)}]{LeenaartsCarlsson2012}
{Carlsson}, M., \& {Leenaarts}, J. 2012, \aap, 539, A39,
  \dodoi{10.1051/0004-6361/201118366}

\bibitem[{{Carlsson} {et~al.}(2015){Carlsson}, {Leenaarts}, \& {De
  Pontieu}}]{Carlsson2015}
{Carlsson}, M., {Leenaarts}, J., \& {De Pontieu}, B. 2015, \apjl, 809, L30,
  \dodoi{10.1088/2041-8205/809/2/L30}

\bibitem[{{Carlsson} \& {Rutten}(1992)}]{Carlsson1992}
{Carlsson}, M., \& {Rutten}, R.~J. 1992, \aap, 259, L53

\bibitem[{{Carlsson} \& {Stein}(1992)}]{Carlsson1992B}
{Carlsson}, M., \& {Stein}, R.~F. 1992, \apjl, 397, L59, \dodoi{10.1086/186544}

\bibitem[{{Carlsson} \& {Stein}(1995)}]{Carlsson1995}
---. 1995, \apjl, 440, L29, \dodoi{10.1086/187753}

\bibitem[{{Carlsson} \& {Stein}(1997)}]{Carlsson1997}
---. 1997, \apj, 481, 500

\bibitem[{{Carlsson} \& {Stein}(2002)}]{Carlsson2002}
---. 2002, \apj, 572, 626, \dodoi{10.1086/340293}

\bibitem[{{Cheng} {et~al.}(2010){Cheng}, {Ding}, \& {Carlsson}}]{Cheng2010}
{Cheng}, J.~X., {Ding}, M.~D., \& {Carlsson}, M. 2010, \apj, 711, 185,
  \dodoi{10.1088/0004-637X/711/1/185}

\bibitem[{{Cheung} {et~al.}(2019){Cheung}, {Rempel}, {Chintzoglou}, {Chen},
  {Testa}, {Mart{\'\i}nez-Sykora}, {Sainz Dalda}, {DeRosa}, {Malanushenko},
  {Hansteen}, {De Pontieu}, {Carlsson}, {Gudiksen}, \& {McIntosh}}]{Cheung2019}
{Cheung}, M.~C.~M., {Rempel}, M., {Chintzoglou}, G., {et~al.} 2019, Nature
  Astronomy, 3, 160, \dodoi{10.1038/s41550-018-0629-3}

\bibitem[{{Condon} \& {Shortley}(1963)}]{Condon1963}
{Condon}, E.~U., \& {Shortley}, G.~H. 1963, {The theory of atomic spectra}

\bibitem[{{Cooper}(1966)}]{Cooper1966}
{Cooper}, J. 1966, Reports on Progress in Physics, 29, 35,
  \dodoi{10.1088/0034-4885/29/1/302}

\bibitem[{{Cooper} {et~al.}(1989){Cooper}, {Ballagh}, \& {Hubeny}}]{Cooper1989}
{Cooper}, J., {Ballagh}, R.~J., \& {Hubeny}, I. 1989, \apj, 344, 949,
  \dodoi{10.1086/167863}

\bibitem[{{Cram} \& {Woods}(1982)}]{Cram1982}
{Cram}, L.~E., \& {Woods}, D.~T. 1982, \apj, 257, 269, \dodoi{10.1086/159985}

\bibitem[{{Crespo-Chac{\'o}n} {et~al.}(2006){Crespo-Chac{\'o}n}, {Montes},
  {Garc{\'\i}a-Alvarez}, {Fern{\'a}ndez-Figueroa}, {L{\'o}pez-Santiago}, \&
  {Foing}}]{Crespo2006}
{Crespo-Chac{\'o}n}, I., {Montes}, D., {Garc{\'\i}a-Alvarez}, D., {et~al.}
  2006, \aap, 452, 987, \dodoi{10.1051/0004-6361:20053615}

\bibitem[{{Dappen} {et~al.}(1987){Dappen}, {Anderson}, \&
  {Mihalas}}]{Dappen1987}
{Dappen}, W., {Anderson}, L., \& {Mihalas}, D. 1987, \apj, 319, 195,
  \dodoi{10.1086/165446}

\bibitem[{{de Feiter}(1964)}]{deFeiter1964}
{de Feiter}, L.~D. 1964, {On the Stark Broadening of High Balmer Lines Emitted
  by Solar Flares}, Vol.~50, 81

\bibitem[{{De Pontieu} {et~al.}(2021){De Pontieu}, {Polito}, {Hansteen},
  {Testa}, {Reeves}, {Antolin}, {N{\'o}brega-Siverio}, {Kowalski},
  {Martinez-Sykora}, {Carlsson}, {McIntosh}, {Liu}, {Daw}, \&
  {Kankelborg}}]{DePontieu2021}
{De Pontieu}, B., {Polito}, V., {Hansteen}, V., {et~al.} 2021, \solphys, 296,
  84, \dodoi{10.1007/s11207-021-01826-0}

\bibitem[{{de Wijn} {et~al.}(2012){de Wijn}, {Casini}, {Nelson}, \&
  {Huang}}]{dewijn2012}
{de Wijn}, A.~G., {Casini}, R., {Nelson}, P.~G., \& {Huang}, P. 2012, in
  Society of Photo-Optical Instrumentation Engineers (SPIE) Conference Series,
  Vol. 8446, Ground-based and Airborne Instrumentation for Astronomy IV, ed.
  I.~S. {McLean}, S.~K. {Ramsay}, \& H.~{Takami}, 84466X,
  \dodoi{10.1117/12.926497}

\bibitem[{{Dere} {et~al.}(1997){Dere}, {Landi}, {Mason}, {Monsignori Fossi}, \&
  {Young}}]{Dere1997}
{Dere}, K.~P., {Landi}, E., {Mason}, H.~E., {Monsignori Fossi}, B.~C., \&
  {Young}, P.~R. 1997, \aaps, 125, 149, \dodoi{10.1051/aas:1997368}

\bibitem[{{Dimitrijevi{\'c}} \& {Sahal-Br{\'e}chot}(1992)}]{StarkB_CaIIA}
{Dimitrijevi{\'c}}, M.~S., \& {Sahal-Br{\'e}chot}, S. 1992, Bulletin
  Astronomique de Belgrade, 145, 83

\bibitem[{{Dimitrijevic} \& {Sahal-Brechot}(1993)}]{StarkB_CaIIB}
{Dimitrijevic}, M.~S., \& {Sahal-Brechot}, S. 1993, \jqsrt, 49, 157,
  \dodoi{10.1016/0022-4073(93)90056-N}

\bibitem[{{Ding} {et~al.}(1999){Ding}, {Fang}, \& {Yun}}]{Ding1999}
{Ding}, M.~D., {Fang}, C., \& {Yun}, H.~S. 1999, \apj, 512, 454,
  \dodoi{10.1086/306776}

\bibitem[{{Donati-Falchi} {et~al.}(1985){Donati-Falchi}, {Falciani}, \&
  {Smaldone}}]{Donati1985}
{Donati-Falchi}, A., {Falciani}, R., \& {Smaldone}, L.~A. 1985, \aap, 152, 165

\bibitem[{{Doyle} {et~al.}(2012){Doyle}, {Giunta}, {Singh}, {Madjarska},
  {Summers}, {Kellett}, \& {O'Mullane}}]{Doyle2012}
{Doyle}, J.~G., {Giunta}, A., {Singh}, A., {et~al.} 2012, \solphys, 280, 111,
  \dodoi{10.1007/s11207-012-0025-6}

\bibitem[{{Drake} \& {Ulrich}(1980)}]{Drake1980}
{Drake}, S.~A., \& {Ulrich}, R.~K. 1980, \apjs, 42, 351, \dodoi{10.1086/190654}

\bibitem[{{Druett} \& {Zharkova}(2018)}]{Druett2018}
{Druett}, M.~K., \& {Zharkova}, V.~V. 2018, \aap, 610, A68,
  \dodoi{10.1051/0004-6361/201731053}

\bibitem[{{Emslie} \& {Bian}(2018)}]{Emslie2018}
{Emslie}, A.~G., \& {Bian}, N.~H. 2018, \apj, 865, 67,
  \dodoi{10.3847/1538-4357/aad961}

\bibitem[{{Emslie} {et~al.}(1992){Emslie}, {Li}, \& {Mariska}}]{Emslie1992}
{Emslie}, A.~G., {Li}, P., \& {Mariska}, J.~T. 1992, \apj, 399, 714,
  \dodoi{10.1086/171964}

\bibitem[{{Falchi} {et~al.}(1997){Falchi}, {Qiu}, \& {Cauzzi}}]{Falchi1997}
{Falchi}, A., {Qiu}, J., \& {Cauzzi}, G. 1997, \aap, 328, 371

\bibitem[{{Falcon} {et~al.}(2015){Falcon}, {Rochau}, {Bailey}, {Gomez},
  {Montgomery}, {Winget}, \& {Nagayama}}]{Falcon2015}
{Falcon}, R.~E., {Rochau}, G.~A., {Bailey}, J.~E., {et~al.} 2015, \apj, 806,
  214, \dodoi{10.1088/0004-637X/806/2/214}

\bibitem[{{Fang} \& {Ding}(1995)}]{Fang1995}
{Fang}, C., \& {Ding}, M.~D. 1995, \aaps, 110, 99

\bibitem[{{Fisher}(1989)}]{Fisher1989}
{Fisher}, G.~H. 1989, \apj, 346, 1019, \dodoi{10.1086/168084}

\bibitem[{{Fisher} {et~al.}(1985){Fisher}, {Canfield}, \&
  {McClymont}}]{Fisher1985V}
{Fisher}, G.~H., {Canfield}, R.~C., \& {McClymont}, A.~N. 1985, \apj, 289, 414,
  \dodoi{10.1086/162901}

\bibitem[{{Fletcher} {et~al.}(2007){Fletcher}, {Hannah}, {Hudson}, \&
  {Metcalf}}]{Fletcher2007}
{Fletcher}, L., {Hannah}, I.~G., {Hudson}, H.~S., \& {Metcalf}, T.~R. 2007,
  \apj, 656, 1187, \dodoi{10.1086/510446}

\bibitem[{{Foley}(1946)}]{Foley1946}
{Foley}, H.~M. 1946, Physical Review, 69, 616, \dodoi{10.1103/PhysRev.69.616}

\bibitem[{{Fontenla} {et~al.}(2002){Fontenla}, {Avrett}, \&
  {Loeser}}]{Fontenla2002}
{Fontenla}, J.~M., {Avrett}, E.~H., \& {Loeser}, R. 2002, \apj, 572, 636,
  \dodoi{10.1086/340227}

\bibitem[{{Frisch} \& {Brissaud}(1971)}]{MMM2}
{Frisch}, U., \& {Brissaud}, A. 1971, \jqsrt, 11, 1753,
  \dodoi{10.1016/0022-4073(71)90020-3}

\bibitem[{{Fritzov{\'a}-{\v{S}}vestkov{\'a}} \&
  {{\v{S}}vestka}(1967)}]{Svestka1967b}
{Fritzov{\'a}-{\v{S}}vestkov{\'a}}, L., \& {{\v{S}}vestka}, Z. 1967, \solphys,
  2, 87, \dodoi{10.1007/BF00155894}

\bibitem[{{Fuhrmeister} {et~al.}(2020){Fuhrmeister}, {Czesla}, {Hildebrandt},
  {Nagel}, {Schmitt}, {Jeffers}, {Caballero}, {Hintz}, {Johnson},
  {Sch{\"o}fer}, {Zechmeister}, {Reiners}, {Ribas}, {Amado}, {Quirrenbach},
  {Nortmann}, {Bauer}, {B{\'e}jar}, {Cort{\'e}s-Contreras}, {Dreizler},
  {Galad{\'\i}-Enr{\'\i}quez}, {Hatzes}, {Kaminski}, {K{\"u}rster}, {Lafarga},
  \& {Montes}}]{Fuhrmeister2020}
{Fuhrmeister}, B., {Czesla}, S., {Hildebrandt}, L., {et~al.} 2020, \aap, 640,
  A52, \dodoi{10.1051/0004-6361/202038279}

\bibitem[{{Gallagher}(2006)}]{Gallagher2006}
{Gallagher}, T. 2006, {Rydberg Atoms}, 235,
  \dodoi{10.1007/978-0-387-26308-3\_14}

\bibitem[{{Gan} \& {Mauas}(1994)}]{Gan1994}
{Gan}, W.~Q., \& {Mauas}, P.~J.~D. 1994, \apj, 430, 891, \dodoi{10.1086/174459}

\bibitem[{{Gan} {et~al.}(1993){Gan}, {Rieger}, {Fang}, \& {Zhang}}]{Gan1993}
{Gan}, W.~Q., {Rieger}, E., {Fang}, C., \& {Zhang}, H.~Q. 1993, \solphys, 143,
  141, \dodoi{10.1007/BF00619101}

\bibitem[{{Gan} {et~al.}(1991){Gan}, {Zhang}, \& {Fang}}]{Gan1991}
{Gan}, W.~Q., {Zhang}, H.~Q., \& {Fang}, C. 1991, \aap, 241, 618

\bibitem[{{Garc{\'{\i}}a-Alvarez} {et~al.}(2002){Garc{\'{\i}}a-Alvarez},
  {Jevremovi{\'c}}, {Doyle}, \& {Butler}}]{Garcia2002}
{Garc{\'{\i}}a-Alvarez}, D., {Jevremovi{\'c}}, D., {Doyle}, J.~G., \& {Butler},
  C.~J. 2002, \aap, 383, 548, \dodoi{10.1051/0004-6361:20011743}

\bibitem[{{Gayley} \& {Canfield}(1991)}]{Gayley1991}
{Gayley}, K.~G., \& {Canfield}, R.~C. 1991, \apj, 380, 660,
  \dodoi{10.1086/170621}

\bibitem[{{Gehmeyr} \& {Mihalas}(1994)}]{Titan1994}
{Gehmeyr}, M., \& {Mihalas}, D. 1994, Physica D Nonlinear Phenomena, 77, 320,
  \dodoi{10.1016/0167-2789(94)90143-0}

\bibitem[{{Gigosos}(2014)}]{Gigosos2014}
{Gigosos}, M.~A. 2014, Journal of Physics D Applied Physics, 47, 343001,
  \dodoi{10.1088/0022-3727/47/34/343001}

\bibitem[{{Goldman} \& {Cassar}(2006)}]{Goldman2006}
{Goldman}, S., \& {Cassar}, M. 2006, {Atoms in Strong Fields}, 227,
  \dodoi{10.1007/978-0-387-26308-3\_13}

\bibitem[{{Gomez} {et~al.}(2016){Gomez}, {Nagayama}, {Kilcrease}, {Montgomery},
  \& {Winget}}]{Gomez2016}
{Gomez}, T.~A., {Nagayama}, T., {Kilcrease}, D.~P., {Montgomery}, M.~H., \&
  {Winget}, D.~E. 2016, \pra, 94, 022501, \dodoi{10.1103/PhysRevA.94.022501}

\bibitem[{{Graham} {et~al.}(2020){Graham}, {Cauzzi}, {Zangrilli}, {Kowalski},
  {Sim{\~o}es}, \& {Allred}}]{Graham2020}
{Graham}, D.~R., {Cauzzi}, G., {Zangrilli}, L., {et~al.} 2020, \apj, 895, 6,
  \dodoi{10.3847/1538-4357/ab88ad}

\bibitem[{{Griem}(1960)}]{Griem1960}
{Griem}, H.~R. 1960, \apj, 132, 883, \dodoi{10.1086/146987}

\bibitem[{{Griem}(1962)}]{Griem1962}
---. 1962, \apj, 136, 422, \dodoi{10.1086/147394}

\bibitem[{{Griem}(1974)}]{Griem1974}
---. 1974, {Spectral line broadening by plasmas}

\bibitem[{{Griem} {et~al.}(1959){Griem}, {Kolb}, \& {Shen}}]{Griem1959}
{Griem}, H.~R., {Kolb}, A.~C., \& {Shen}, K.~Y. 1959, Physical Review, 116, 4,
  \dodoi{10.1103/PhysRev.116.4}

\bibitem[{{Gudiksen} {et~al.}(2011){Gudiksen}, {Carlsson}, {Hansteen}, {Hayek},
  {Leenaarts}, \& {Mart{\'\i}nez-Sykora}}]{Gudiksen2011}
{Gudiksen}, B.~V., {Carlsson}, M., {Hansteen}, V.~H., {et~al.} 2011, \aap, 531,
  A154, \dodoi{10.1051/0004-6361/201116520}

\bibitem[{{Guerreiro} {et~al.}(2013){Guerreiro}, {Carlsson}, \&
  {Hansteen}}]{Guerreiro}
{Guerreiro}, N., {Carlsson}, M., \& {Hansteen}, V. 2013, \apj, 766, 128,
  \dodoi{10.1088/0004-637X/766/2/128}

\bibitem[{{Hawley} \& {Fisher}(1994)}]{HF94}
{Hawley}, S.~L., \& {Fisher}, G.~H. 1994, \apj, 426, 387,
  \dodoi{10.1086/174075}

\bibitem[{{Hawley} \& {Pettersen}(1991)}]{HP91}
{Hawley}, S.~L., \& {Pettersen}, B.~R. 1991, \apj, 378, 725,
  \dodoi{10.1086/170474}

\bibitem[{{Hawley} {et~al.}(2007){Hawley}, {Walkowicz}, {Allred}, \&
  {Valenti}}]{Hawley2007}
{Hawley}, S.~L., {Walkowicz}, L.~M., {Allred}, J.~C., \& {Valenti}, J.~A. 2007,
  \pasp, 119, 67, \dodoi{10.1086/510561}

\bibitem[{{Heinzel} \& {Shibata}(2018)}]{Heinzel2018}
{Heinzel}, P., \& {Shibata}, K. 2018, \apj, 859, 143,
  \dodoi{10.3847/1538-4357/aabe78}

\bibitem[{{Hiei}(1982)}]{Hiei1982}
{Hiei}, E. 1982, \solphys, 80, 113, \dodoi{10.1007/BF00153427}

\bibitem[{{Hilton} {et~al.}(2010){Hilton}, {West}, {Hawley}, \&
  {Kowalski}}]{Hilton2010}
{Hilton}, E.~J., {West}, A.~A., {Hawley}, S.~L., \& {Kowalski}, A.~F. 2010,
  \aj, 140, 1402, \dodoi{10.1088/0004-6256/140/5/1402}

\bibitem[{{Hong} {et~al.}(2017){Hong}, {Carlsson}, \& {Ding}}]{Hong2017}
{Hong}, J., {Carlsson}, M., \& {Ding}, M.~D. 2017, \apj, 845, 144,
  \dodoi{10.3847/1538-4357/aa80e3}

\bibitem[{{Hubeny} {et~al.}(1994){Hubeny}, {Hummer}, \& {Lanz}}]{Hubeny1994}
{Hubeny}, I., {Hummer}, D.~G., \& {Lanz}, T. 1994, \aap, 282, 151

\bibitem[{{Hubeny} \& {Lanz}(2017)}]{TLusty}
{Hubeny}, I., \& {Lanz}, T. 2017, arXiv e-prints, arXiv:1706.01935.
\newblock \doarXiv{1706.01935}

\bibitem[{{Hubeny} \& {Mihalas}(2014)}]{Hubeny2014}
{Hubeny}, I., \& {Mihalas}, D. 2014, {Theory of Stellar Atmospheres}

\bibitem[{{Hummer} \& {Mihalas}(1988)}]{HM88}
{Hummer}, D.~G., \& {Mihalas}, D. 1988, \apj, 331, 794, \dodoi{10.1086/166600}

\bibitem[{{Ichimoto} \& {Kurokawa}(1984)}]{Ichimoto1984}
{Ichimoto}, K., \& {Kurokawa}, H. 1984, \solphys, 93, 105,
  \dodoi{10.1007/BF00156656}

\bibitem[{{Inglis} \& {Teller}(1939)}]{InglisTeller}
{Inglis}, D.~R., \& {Teller}, E. 1939, \apj, 90, 439, \dodoi{10.1086/144118}

\bibitem[{{Jefferies}(1968)}]{Jefferies1968}
{Jefferies}, J.~T. 1968, {Spectral line formation}

\bibitem[{{Johns-Krull} {et~al.}(1997){Johns-Krull}, {Hawley}, {Basri}, \&
  {Valenti}}]{JohnsKrull1997}
{Johns-Krull}, C.~M., {Hawley}, S.~L., {Basri}, G., \& {Valenti}, J.~A. 1997,
  \apjs, 112, 221, \dodoi{10.1086/313030}

\bibitem[{{Karlick{\'y}} {et~al.}(2004){Karlick{\'y}}, {Ka{\v{s}}parov{\'a}},
  \& {Heinzel}}]{Karlicky2004}
{Karlick{\'y}}, M., {Ka{\v{s}}parov{\'a}}, J., \& {Heinzel}, P. 2004, \aap,
  416, L13, \dodoi{10.1051/0004-6361:20040034}

\bibitem[{{Karpen} \& {DeVore}(1987)}]{Karpen1987}
{Karpen}, J.~T., \& {DeVore}, C.~R. 1987, \apj, 320, 904,
  \dodoi{10.1086/165608}

\bibitem[{{Ka{\v{s}}parov{\'a}} {et~al.}(2009){Ka{\v{s}}parov{\'a}}, {Varady},
  {Heinzel}, {Karlick{\'y}}, \& {Moravec}}]{Kasparova2009}
{Ka{\v{s}}parov{\'a}}, J., {Varady}, M., {Heinzel}, P., {Karlick{\'y}}, M., \&
  {Moravec}, Z. 2009, \aap, 499, 923, \dodoi{10.1051/0004-6361/200811559}

\bibitem[{{Kennedy} {et~al.}(2015){Kennedy}, {Milligan}, {Allred},
  {Mathioudakis}, \& {Keenan}}]{Kennedy2015}
{Kennedy}, M.~B., {Milligan}, R.~O., {Allred}, J.~C., {Mathioudakis}, M., \&
  {Keenan}, F.~P. 2015, \aap, 578, A72, \dodoi{10.1051/0004-6361/201425144}

\bibitem[{{Kerr} {et~al.}(2019{\natexlab{a}}){Kerr}, {Allred}, \&
  {Carlsson}}]{Kerr2019A}
{Kerr}, G.~S., {Allred}, J.~C., \& {Carlsson}, M. 2019{\natexlab{a}}, \apj,
  883, 57, \dodoi{10.3847/1538-4357/ab3c24}

\bibitem[{{Kerr} {et~al.}(2019{\natexlab{b}}){Kerr}, {Carlsson}, \&
  {Allred}}]{Kerr2019B}
{Kerr}, G.~S., {Carlsson}, M., \& {Allred}, J.~C. 2019{\natexlab{b}}, \apj,
  885, 119, \dodoi{10.3847/1538-4357/ab48ea}

\bibitem[{{Kerr} {et~al.}(2019{\natexlab{c}}){Kerr}, {Carlsson}, {Allred},
  {Young}, \& {Daw}}]{Kerr2019Si}
{Kerr}, G.~S., {Carlsson}, M., {Allred}, J.~C., {Young}, P.~R., \& {Daw}, A.~N.
  2019{\natexlab{c}}, \apj, 871, 23, \dodoi{10.3847/1538-4357/aaf46e}

\bibitem[{{Kerr} {et~al.}(2021){Kerr}, {Xu}, {Allred}, {Polito}, {Sadykov},
  {Huang}, \& {Wang}}]{Kerr2021}
{Kerr}, G.~S., {Xu}, Y., {Allred}, J.~C., {et~al.} 2021, \apj, 912, 153,
  \dodoi{10.3847/1538-4357/abf42d}

\bibitem[{{Kerr} {et~al.}(2020{\natexlab{a}}){Kerr}, {Alaoui}, {Allred},
  {Bian}, {Dennis}, {Emslie}, {Fletcher}, {Guidoni}, {Hayes}, {Holman},
  {Hudson}, {Karpen}, {Kowalski}, {Milligan}, {Polito}, {Qiu}, \&
  {Ryan}}]{Kerr2020A}
{Kerr}, G.~S., {Alaoui}, M., {Allred}, J.~C., {et~al.} 2020{\natexlab{a}},
  arXiv e-prints, arXiv:2009.08400.
\newblock \doarXiv{2009.08400}

\bibitem[{{Kerr} {et~al.}(2020{\natexlab{b}}){Kerr}, {Alaoui}, {Allred},
  {Bian}, {Dennis}, {Emslie}, {Fletcher}, {Guidoni}, {Hayes}, {Holman},
  {Hudson}, {Karpen}, {Kowalski}, {Milligan}, {Polito}, {Qiu}, \&
  {Ryan}}]{Kerr2020B}
---. 2020{\natexlab{b}}, arXiv e-prints, arXiv:2009.08407.
\newblock \doarXiv{2009.08407}

\bibitem[{{Kleint} {et~al.}(2015){Kleint}, {Battaglia}, {Reardon}, {Sainz
  Dalda}, {Young}, \& {Krucker}}]{Kleint2015}
{Kleint}, L., {Battaglia}, M., {Reardon}, K., {et~al.} 2015, \apj, 806, 9,
  \dodoi{10.1088/0004-637X/806/1/9}

\bibitem[{{Kleint} {et~al.}(2016){Kleint}, {Heinzel}, {Judge}, \&
  {Krucker}}]{Kleint2016}
{Kleint}, L., {Heinzel}, P., {Judge}, P., \& {Krucker}, S. 2016, \apj, 816, 88,
  \dodoi{10.3847/0004-637X/816/2/88}

\bibitem[{{Kleppner} {et~al.}(1981){Kleppner}, {Littman}, \&
  {Zimmerman}}]{Kleppner1981}
{Kleppner}, D., {Littman}, M.~G., \& {Zimmerman}, M.~L. 1981, Scientific
  American, 244, 130, \dodoi{10.1038/scientificamerican0581-130}

\bibitem[{{Knuth} \& {Glesener}(2020)}]{Knuth2020}
{Knuth}, T., \& {Glesener}, L. 2020, \apj, 903, 63,
  \dodoi{10.3847/1538-4357/abb779}

\bibitem[{{Kolb} \& {Griem}(1958)}]{Kolb1958}
{Kolb}, A.~C., \& {Griem}, H. 1958, Physical Review, 111, 514,
  \dodoi{10.1103/PhysRev.111.514}

\bibitem[{{Kowalski}(2012)}]{Kowalski2012}
{Kowalski}, A.~F. 2012, PhD thesis, University of Washington

\bibitem[{{Kowalski} \& {Allred}(2018)}]{KowalskiAllred2018}
{Kowalski}, A.~F., \& {Allred}, J.~C. 2018, \apj, 852, 61,
  \dodoi{10.3847/1538-4357/aa9d91}

\bibitem[{{Kowalski} {et~al.}(2017{\natexlab{a}}){Kowalski}, {Allred}, {Daw},
  {Cauzzi}, \& {Carlsson}}]{Kowalski2017Mar29}
{Kowalski}, A.~F., {Allred}, J.~C., {Daw}, A., {Cauzzi}, G., \& {Carlsson}, M.
  2017{\natexlab{a}}, \apj, 836, 12, \dodoi{10.3847/1538-4357/836/1/12}

\bibitem[{{Kowalski} {et~al.}(2019{\natexlab{a}}){Kowalski}, {Butler}, {Daw},
  {Fletcher}, {Allred}, {De Pontieu}, {Kerr}, \& {Cauzzi}}]{Kowalski2019IRIS}
{Kowalski}, A.~F., {Butler}, E., {Daw}, A.~N., {et~al.} 2019{\natexlab{a}},
  \apj, 878, 135, \dodoi{10.3847/1538-4357/ab1f8b}

\bibitem[{{Kowalski} {et~al.}(2015{\natexlab{a}}){Kowalski}, {Cauzzi}, \&
  {Fletcher}}]{KCF15}
{Kowalski}, A.~F., {Cauzzi}, G., \& {Fletcher}, L. 2015{\natexlab{a}}, \apj,
  798, 107, \dodoi{10.1088/0004-637X/798/2/107}

\bibitem[{{Kowalski} {et~al.}(2015{\natexlab{b}}){Kowalski}, {Hawley},
  {Carlsson}, {Allred}, {Uitenbroek}, {Osten}, \& {Holman}}]{Kowalski2015}
{Kowalski}, A.~F., {Hawley}, S.~L., {Carlsson}, M., {et~al.}
  2015{\natexlab{b}}, \solphys, 290, 3487, \dodoi{10.1007/s11207-015-0708-x}

\bibitem[{{Kowalski} {et~al.}(2013){Kowalski}, {Hawley}, {Wisniewski}, {Osten},
  {Hilton}, {Holtzman}, {Schmidt}, \& {Davenport}}]{Kowalski2013}
{Kowalski}, A.~F., {Hawley}, S.~L., {Wisniewski}, J.~P., {et~al.} 2013, \apjs,
  207, 15, \dodoi{10.1088/0067-0049/207/1/15}

\bibitem[{{Kowalski} {et~al.}(2016){Kowalski}, {Mathioudakis}, {Hawley},
  {Wisniewski}, {Dhillon}, {Marsh}, {Hilton}, \& {Brown}}]{Kowalski2016}
{Kowalski}, A.~F., {Mathioudakis}, M., {Hawley}, S.~L., {et~al.} 2016, \apj,
  820, 95, \dodoi{10.3847/0004-637X/820/2/95}

\bibitem[{{Kowalski} {et~al.}(2017{\natexlab{b}}){Kowalski}, {Allred},
  {Uitenbroek}, {Tremblay}, {Brown}, {Carlsson}, {Osten}, {Wisniewski}, \&
  {Hawley}}]{Kowalski2017Broadening}
{Kowalski}, A.~F., {Allred}, J.~C., {Uitenbroek}, H., {et~al.}
  2017{\natexlab{b}}, \apj, 837, 125, \dodoi{10.3847/1538-4357/aa603e}

\bibitem[{{Kowalski} {et~al.}(2019{\natexlab{b}}){Kowalski}, {Wisniewski},
  {Hawley}, {Osten}, {Brown}, {Fari{\~n}a}, {Valenti}, {Brown}, {Xilouris},
  {Schmidt}, \& {Johns-Krull}}]{Kowalski2019HST}
{Kowalski}, A.~F., {Wisniewski}, J.~P., {Hawley}, S.~L., {et~al.}
  2019{\natexlab{b}}, \apj, 871, 167, \dodoi{10.3847/1538-4357/aaf058}

\bibitem[{{Krucker} {et~al.}(2011){Krucker}, {Hudson}, {Jeffrey}, {Battaglia},
  {Kontar}, {Benz}, {Csillaghy}, \& {Lin}}]{Krucker2011}
{Krucker}, S., {Hudson}, H.~S., {Jeffrey}, N.~L.~S., {et~al.} 2011, \apj, 739,
  96, \dodoi{10.1088/0004-637X/739/2/96}

\bibitem[{{Kunze}(2009)}]{Kunze2009}
{Kunze}, H.-J. 2009, {Introduction to Plasma Spectroscopy}, Vol.~56,
  \dodoi{10.1007/978-3-642-02233-3}

\bibitem[{{Kuridze} {et~al.}(2020){Kuridze}, {Mathioudakis}, {Heinzel}, {Koza},
  {Morgan}, {Oliver}, {Kowalski}, \& {Allred}}]{Kuridze2020}
{Kuridze}, D., {Mathioudakis}, M., {Heinzel}, P., {et~al.} 2020, \apj, 896,
  120, \dodoi{10.3847/1538-4357/ab9603}

\bibitem[{{Kuridze} {et~al.}(2015){Kuridze}, {Mathioudakis}, {Sim{\~o}es},
  {Rouppe van der Voort}, {Carlsson}, {Jafarzadeh}, {Allred}, {Kowalski},
  {Kennedy}, {Fletcher}, {Graham}, \& {Keenan}}]{Kuridze2015}
{Kuridze}, D., {Mathioudakis}, M., {Sim{\~o}es}, P.~J.~A., {et~al.} 2015, \apj,
  813, 125, \dodoi{10.1088/0004-637X/813/2/125}

\bibitem[{{Kuridze} {et~al.}(2016){Kuridze}, {Mathioudakis}, {Christian},
  {Kowalski}, {Jess}, {Grant}, {Kawate}, {Sim{\~o}es}, {Allred}, \&
  {Keenan}}]{Kuridze2016}
{Kuridze}, D., {Mathioudakis}, M., {Christian}, D.~J., {et~al.} 2016, \apj,
  832, 147, \dodoi{10.3847/0004-637X/832/2/147}

\bibitem[{{Kurochka} \& {Maslennikova}(1970)}]{Kurochka1970}
{Kurochka}, L.~N., \& {Maslennikova}, L.~B. 1970, \solphys, 11, 33,
  \dodoi{10.1007/BF00156548}

\bibitem[{{Kurucz}(1979)}]{Kurucz1979}
{Kurucz}, R.~L. 1979, \apjs, 40, 1, \dodoi{10.1086/190589}

\bibitem[{{Lacatus} {et~al.}(2017){Lacatus}, {Judge}, \& {Donea}}]{Lacatus2017}
{Lacatus}, D.~A., {Judge}, P.~G., \& {Donea}, A. 2017, \apj, 842, 15,
  \dodoi{10.3847/1538-4357/aa725d}

\bibitem[{{Leenaarts} \& {Carlsson}(2009)}]{MULTI3D}
{Leenaarts}, J., \& {Carlsson}, M. 2009, in Astronomical Society of the Pacific
  Conference Series, Vol. 415, The Second Hinode Science Meeting: Beyond
  Discovery-Toward Understanding, ed. B.~{Lites}, M.~{Cheung}, T.~{Magara},
  J.~{Mariska}, \& K.~{Reeves}, 87

\bibitem[{{Leone} {et~al.}(2004){Leone}, {Paoletti}, \& {Robotti}}]{Leone2004}
{Leone}, M., {Paoletti}, A., \& {Robotti}, N. 2004, Physics in Perspective, 6,
  271, \dodoi{10.1007/s00016-003-0170-2}

\bibitem[{{Libbrecht} {et~al.}(2019){Libbrecht}, {de la Cruz Rodr{\'\i}guez},
  {Danilovic}, {Leenaarts}, \& {Pazira}}]{Libbrecht2019}
{Libbrecht}, T., {de la Cruz Rodr{\'\i}guez}, J., {Danilovic}, S., {Leenaarts},
  J., \& {Pazira}, H. 2019, \aap, 621, A35, \dodoi{10.1051/0004-6361/201833610}

\bibitem[{{Liu} {et~al.}(2009){Liu}, {Petrosian}, \& {Mariska}}]{Liu2009}
{Liu}, W., {Petrosian}, V., \& {Mariska}, J.~T. 2009, \apj, 702, 1553,
  \dodoi{10.1088/0004-637X/702/2/1553}

\bibitem[{{Livshits} {et~al.}(1981){Livshits}, {Badalian}, {Kosovichev}, \&
  {Katsova}}]{Livshits1981}
{Livshits}, M.~A., {Badalian}, O.~G., {Kosovichev}, A.~G., \& {Katsova}, M.~M.
  1981, \solphys, 73, 269, \dodoi{10.1007/BF00151682}

\bibitem[{{Longair}(2013)}]{Longair2013}
{Longair}, M. 2013, {Quantum Concepts in Physics}

\bibitem[{{Low}(1975)}]{Low1975}
{Low}, B.~C. 1975, \apj, 197, 251, \dodoi{10.1086/153508}

\bibitem[{{Machado} \& {Rust}(1974)}]{Machado1974}
{Machado}, M.~E., \& {Rust}, D.~M. 1974, \solphys, 38, 499,
  \dodoi{10.1007/BF00155084}

\bibitem[{{Marchenko} {et~al.}(2021){Marchenko}, {Criscuoli}, {DeLand},
  {Choudhary}, \& {Kopp}}]{Marchenko2021}
{Marchenko}, S., {Criscuoli}, S., {DeLand}, M.~T., {Choudhary}, D.~P., \&
  {Kopp}, G. 2021, \aap, 646, A81, \dodoi{10.1051/0004-6361/202037767}

\bibitem[{{Mart{\'\i}nez Oliveros} {et~al.}(2012){Mart{\'\i}nez Oliveros},
  {Hudson}, {Hurford}, {Krucker}, {Lin}, {Lindsey}, {Couvidat}, {Schou}, \&
  {Thompson}}]{Martinez2012}
{Mart{\'\i}nez Oliveros}, J.-C., {Hudson}, H.~S., {Hurford}, G.~J., {et~al.}
  2012, \apjl, 753, L26, \dodoi{10.1088/2041-8205/753/2/L26}

\bibitem[{{Mathioudakis} {et~al.}(1999){Mathioudakis}, {McKenny}, {Keenan},
  {Williams}, \& {Phillips}}]{Mathioudakis1999}
{Mathioudakis}, M., {McKenny}, J., {Keenan}, F.~P., {Williams}, D.~R., \&
  {Phillips}, K.~J.~H. 1999, \aap, 351, L23

\bibitem[{{McIntosh} {et~al.}(2020){McIntosh}, {Chapman}, {Leamon}, {Egeland},
  \& {Watkins}}]{McIntosh2020}
{McIntosh}, S.~W., {Chapman}, S., {Leamon}, R.~J., {Egeland}, R., \& {Watkins},
  N.~W. 2020, \solphys, 295, 163, \dodoi{10.1007/s11207-020-01723-y}

\bibitem[{{Melrose}(2018)}]{Melrose2018}
{Melrose}, D.~B. 2018, Plasma Science and Technology, 20, 074003,
  \dodoi{10.1088/2058-6272/aab966}

\bibitem[{{Metcalf} {et~al.}(1990){Metcalf}, {Canfield}, \&
  {Saba}}]{Metcalf1990}
{Metcalf}, T.~R., {Canfield}, R.~C., \& {Saba}, J.~L.~R. 1990, \apj, 365, 391,
  \dodoi{10.1086/169494}

\bibitem[{{Mihalas}(1966)}]{Mihalas1966}
{Mihalas}, D. 1966, \apjs, 13, 1, \dodoi{10.1086/190135}

\bibitem[{{Mihalas}(1978)}]{Mihalas1978}
---. 1978, {Stellar atmospheres /2nd edition/}

\bibitem[{{Milligan} {et~al.}(2014){Milligan}, {Kerr}, {Dennis}, {Hudson},
  {Fletcher}, {Allred}, {Chamberlin}, {Ireland}, {Mathioudakis}, \&
  {Keenan}}]{Milligan2014}
{Milligan}, R.~O., {Kerr}, G.~S., {Dennis}, B.~R., {et~al.} 2014, \apj, 793,
  70, \dodoi{10.1088/0004-637X/793/2/70}

\bibitem[{{Molnar} {et~al.}(2021){Molnar}, {Reardon}, {Cranmer}, {Kowalski},
  {Chai}, \& {Gary}}]{Molnar2021}
{Molnar}, M.~E., {Reardon}, K.~P., {Cranmer}, S.~R., {et~al.} 2021, arXiv
  e-prints, arXiv:2107.08952.
\newblock \doarXiv{2107.08952}

\bibitem[{{Monson} {et~al.}(2021){Monson}, {Mathioudakis}, {Reid}, {Milligan},
  \& {Kuridze}}]{Monson2021}
{Monson}, A.~J., {Mathioudakis}, M., {Reid}, A., {Milligan}, R., \& {Kuridze},
  D. 2021, \apj, 915, 16, \dodoi{10.3847/1538-4357/abfda8}

\bibitem[{{Namekata} {et~al.}(2020){Namekata}, {Maehara}, {Sasaki}, {Kawai},
  {Notsu}, {Kowalski}, {Allred}, {Iwakiri}, {Tsuboi}, {Murata}, {Niwano},
  {Shiraishi}, {Adachi}, {Iida}, {Oeda}, {Honda}, {Tozuka}, {Katoh}, {Onozato},
  {Okamoto}, {Isogai}, {Kimura}, {Kojiguchi}, {Wakamatsu}, {Tampo}, {Nogami},
  \& {Shibata}}]{Namekata2020}
{Namekata}, K., {Maehara}, H., {Sasaki}, R., {et~al.} 2020, \pasj,
  \dodoi{10.1093/pasj/psaa051}

\bibitem[{{Nayfonov} {et~al.}(1999){Nayfonov}, {D{\"a}ppen}, {Hummer}, \&
  {Mihalas}}]{Nayfonov1999}
{Nayfonov}, A., {D{\"a}ppen}, W., {Hummer}, D.~G., \& {Mihalas}, D. 1999, \apj,
  526, 451, \dodoi{10.1086/307972}

\bibitem[{{Neckel} \& {Labs}(1984)}]{Neckel1984}
{Neckel}, H., \& {Labs}, D. 1984, \solphys, 90, 205, \dodoi{10.1007/BF00173953}

\bibitem[{{Neidig}(1983)}]{Neidig1983}
{Neidig}, D.~F. 1983, \solphys, 85, 285, \dodoi{10.1007/BF00148655}

\bibitem[{{Neidig} {et~al.}(1994){Neidig}, {Grosser}, \& {Hrovat}}]{Neidig1994}
{Neidig}, D.~F., {Grosser}, H., \& {Hrovat}, M. 1994, \solphys, 155, 199,
  \dodoi{10.1007/BF00670740}

\bibitem[{{Neidig} {et~al.}(1993){Neidig}, {Kiplinger}, {Cohl}, \&
  {Wiborg}}]{Neidig1993}
{Neidig}, D.~F., {Kiplinger}, A.~L., {Cohl}, H.~S., \& {Wiborg}, P.~H. 1993,
  \apj, 406, 306, \dodoi{10.1086/172442}

\bibitem[{{Neidig} \& {Wiborg}(1984)}]{Neidig1984}
{Neidig}, D.~F., \& {Wiborg}, P.~H., J. 1984, \solphys, 92, 217,
  \dodoi{10.1007/BF00157247}

\bibitem[{{Nelson} {et~al.}(2010){Nelson}, {Casini}, {de Wijn}, \&
  {Knoelker}}]{Nelson2010}
{Nelson}, P.~G., {Casini}, R., {de Wijn}, A.~G., \& {Knoelker}, M. 2010, in
  Society of Photo-Optical Instrumentation Engineers (SPIE) Conference Series,
  Vol. 7735, Ground-based and Airborne Instrumentation for Astronomy III, ed.
  I.~S. {McLean}, S.~K. {Ramsay}, \& H.~{Takami}, 77358C,
  \dodoi{10.1117/12.857610}

\bibitem[{{Nishizuka} {et~al.}(2009){Nishizuka}, {Asai}, {Takasaki},
  {Kurokawa}, \& {Shibata}}]{Nishizuka2009}
{Nishizuka}, N., {Asai}, A., {Takasaki}, H., {Kurokawa}, H., \& {Shibata}, K.
  2009, \apjl, 694, L74, \dodoi{10.1088/0004-637X/694/1/L74}

\bibitem[{{Osten} {et~al.}(2016){Osten}, {Kowalski}, {Drake}, {Krimm}, {Page},
  {Gazeas}, {Kennea}, {Oates}, {Page}, {de Miguel}, {Nov{\'a}k}, {Apeltauer},
  \& {Gehrels}}]{Osten2016}
{Osten}, R.~A., {Kowalski}, A., {Drake}, S.~A., {et~al.} 2016, \apj, 832, 174,
  \dodoi{10.3847/0004-637X/832/2/174}

\bibitem[{{Paulson} {et~al.}(2006){Paulson}, {Allred}, {Anderson}, {Hawley},
  {Cochran}, \& {Yelda}}]{Paulson2006}
{Paulson}, D.~B., {Allred}, J.~C., {Anderson}, R.~B., {et~al.} 2006, \pasp,
  118, 227, \dodoi{10.1086/499497}

\bibitem[{{Penn} {et~al.}(2016){Penn}, {Krucker}, {Hudson}, {Jhabvala},
  {Jennings}, {Lunsford}, \& {Kaufmann}}]{Penn2016}
{Penn}, M., {Krucker}, S., {Hudson}, H., {et~al.} 2016, \apjl, 819, L30,
  \dodoi{10.3847/2041-8205/819/2/L30}

\bibitem[{{Polito} {et~al.}(2018){Polito}, {Testa}, {Allred}, {De Pontieu},
  {Carlsson}, {Pereira}, {Go{\v{s}}i{\'c}}, \& {Reale}}]{Polito2018}
{Polito}, V., {Testa}, P., {Allred}, J., {et~al.} 2018, \apj, 856, 178,
  \dodoi{10.3847/1538-4357/aab49e}

\bibitem[{{Polito} {et~al.}(2019){Polito}, {Testa}, \& {De
  Pontieu}}]{Polito2019}
{Polito}, V., {Testa}, P., \& {De Pontieu}, B. 2019, \apjl, 879, L17,
  \dodoi{10.3847/2041-8213/ab290b}

\bibitem[{{Proch{\'a}zka} {et~al.}(2017){Proch{\'a}zka}, {Milligan}, {Allred},
  {Kowalski}, {Kotr{\v c}}, \& {Mathioudakis}}]{Ondrej1}
{Proch{\'a}zka}, O., {Milligan}, R.~O., {Allred}, J.~C., {et~al.} 2017, \apj,
  837, 46, \dodoi{10.3847/1538-4357/aa5da8}

\bibitem[{{Proch{\'a}zka} {et~al.}(2018){Proch{\'a}zka}, {Reid}, {Milligan},
  {Sim{\~o}es}, {Allred}, \& {Mathioudakis}}]{Ondrej2}
{Proch{\'a}zka}, O., {Reid}, A., {Milligan}, R.~O., {et~al.} 2018, \apj, 862,
  76, \dodoi{10.3847/1538-4357/aaca37}

\bibitem[{{Qiu} {et~al.}(2012){Qiu}, {Cheng}, {Hurford}, {Xu}, \&
  {Wang}}]{Qiu2012}
{Qiu}, J., {Cheng}, J.~X., {Hurford}, G.~J., {Xu}, Y., \& {Wang}, H. 2012,
  \aap, 547, A72, \dodoi{10.1051/0004-6361/201118609}

\bibitem[{{Radziszewski} {et~al.}(2007){Radziszewski}, {Rudawy}, \&
  {Phillips}}]{Radz2007}
{Radziszewski}, K., {Rudawy}, P., \& {Phillips}, K.~J.~H. 2007, \aap, 461, 303,
  \dodoi{10.1051/0004-6361:20053460}

\bibitem[{{Rast} {et~al.}(2021){Rast}, {Bello Gonz{\'a}lez}, {Bellot Rubio},
  {Cao}, {Cauzzi}, {Deluca}, {de Pontieu}, {Fletcher}, {Gibson}, {Judge},
  {Katsukawa}, {Kazachenko}, {Khomenko}, {Landi}, {Mart{\'\i}nez Pillet},
  {Petrie}, {Qiu}, {Rachmeler}, {Rempel}, {Schmidt}, {Scullion}, {Sun},
  {Welsch}, {Andretta}, {Antolin}, {Ayres}, {Balasubramaniam}, {Ballai},
  {Berger}, {Bradshaw}, {Campbell}, {Carlsson}, {Casini}, {Centeno}, {Cranmer},
  {Criscuoli}, {Deforest}, {Deng}, {Erd{\'e}lyi}, {Fedun}, {Fischer},
  {Gonz{\'a}lez Manrique}, {Hahn}, {Harra}, {Henriques}, {Hurlburt}, {Jaeggli},
  {Jafarzadeh}, {Jain}, {Jefferies}, {Keys}, {Kowalski}, {Kuckein}, {Kuhn},
  {Kuridze}, {Liu}, {Liu}, {Longcope}, {Mathioudakis}, {McAteer}, {McIntosh},
  {McKenzie}, {Miralles}, {Morton}, {Muglach}, {Nelson}, {Panesar}, {Parenti},
  {Parnell}, {Poduval}, {Reardon}, {Reep}, {Schad}, {Schmit}, {Sharma},
  {Socas-Navarro}, {Srivastava}, {Sterling}, {Suematsu}, {Tarr}, {Tiwari},
  {Tritschler}, {Verth}, {Vourlidas}, {Wang}, {Wang}, {NSO and DKIST Project},
  {DKIST Instrument Scientists}, {DKIST Science Working Group}, \& {DKIST
  Critical Science Plan Community}}]{Rast2021}
{Rast}, M.~P., {Bello Gonz{\'a}lez}, N., {Bellot Rubio}, L., {et~al.} 2021,
  \solphys, 296, 70, \dodoi{10.1007/s11207-021-01789-2}

\bibitem[{{Rathore} \& {Carlsson}(2015)}]{Rathore2015A}
{Rathore}, B., \& {Carlsson}, M. 2015, \apj, 811, 80,
  \dodoi{10.1088/0004-637X/811/2/80}

\bibitem[{{Rauch} {et~al.}(2013){Rauch}, {Werner}, {Bohlin}, \&
  {Kruk}}]{Rauch2013}
{Rauch}, T., {Werner}, K., {Bohlin}, R., \& {Kruk}, J.~W. 2013, \aap, 560,
  A106, \dodoi{10.1051/0004-6361/201322336}

\bibitem[{{Redman} \& {Suemoto}(1954)}]{Redman1954}
{Redman}, R.~O., \& {Suemoto}, Z. 1954, \mnras, 114, 524,
  \dodoi{10.1093/mnras/114.5.524}

\bibitem[{{Reep} {et~al.}(2020){Reep}, {Antolin}, \& {Bradshaw}}]{Reep2020}
{Reep}, J.~W., {Antolin}, P., \& {Bradshaw}, S.~J. 2020, \apj, 890, 100,
  \dodoi{10.3847/1538-4357/ab6bdc}

\bibitem[{{Reid} {et~al.}(2017){Reid}, {Mathioudakis}, {Kowalski}, {Doyle}, \&
  {Allred}}]{Reid2017}
{Reid}, A., {Mathioudakis}, M., {Kowalski}, A., {Doyle}, J.~G., \& {Allred},
  J.~C. 2017, \apjl, 835, L37, \dodoi{10.3847/2041-8213/835/2/L37}

\bibitem[{{Rempel}(2017)}]{Rempel2017}
{Rempel}, M. 2017, \apj, 834, 10, \dodoi{10.3847/1538-4357/834/1/10}

\bibitem[{{Rimmele} {et~al.}(2020){Rimmele}, {Warner}, {Keil}, {Goode},
  {Kn{\"o}lker}, {Kuhn}, {Rosner}, {McMullin}, {Casini}, {Lin}, {W{\"o}ger},
  {von der L{\"u}he}, {Tritschler}, {Davey}, {de Wijn}, {Elmore}, {Fehlmann},
  {Harrington}, {Jaeggli}, {Rast}, {Schad}, {Schmidt}, {Mathioudakis},
  {Mickey}, {Anan}, {Beck}, {Marshall}, {Jeffers}, {Oschmann}, {Beard},
  {Berst}, {Cowan}, {Craig}, {Cross}, {Cummings}, {Donnelly}, {de Vanssay},
  {Eigenbrot}, {Ferayorni}, {Foster}, {Galapon}, {Gedrites}, {Gonzales},
  {Goodrich}, {Gregory}, {Guzman}, {Guzzo}, {Hegwer}, {Hubbard}, {Hubbard},
  {Johansson}, {Johnson}, {Liang}, {Liang}, {McQuillen}, {Mayer}, {Newman},
  {Onodera}, {Phelps}, {Puentes}, {Richards}, {Rimmele}, {Sekulic}, {Shimko},
  {Simison}, {Smith}, {Starman}, {Sueoka}, {Summers}, {Szabo}, {Szabo},
  {Wampler}, {Williams}, \& {White}}]{Rimmele2020}
{Rimmele}, T.~R., {Warner}, M., {Keil}, S.~L., {et~al.} 2020, \solphys, 295,
  172, \dodoi{10.1007/s11207-020-01736-7}

\bibitem[{{Rubio da Costa} \& {Kleint}(2017)}]{Rubio2017}
{Rubio da Costa}, F., \& {Kleint}, L. 2017, \apj, 842, 82,
  \dodoi{10.3847/1538-4357/aa6eaf}

\bibitem[{{Rubio da Costa} {et~al.}(2016){Rubio da Costa}, {Kleint},
  {Petrosian}, {Liu}, \& {Allred}}]{Rubio2016}
{Rubio da Costa}, F., {Kleint}, L., {Petrosian}, V., {Liu}, W., \& {Allred},
  J.~C. 2016, \apj, 827, 38, \dodoi{10.3847/0004-637X/827/1/38}

\bibitem[{{Rubio da Costa} {et~al.}(2015){Rubio da Costa}, {Liu}, {Petrosian},
  \& {Carlsson}}]{Rubio2015}
{Rubio da Costa}, F., {Liu}, W., {Petrosian}, V., \& {Carlsson}, M. 2015, \apj,
  813, 133, \dodoi{10.1088/0004-637X/813/2/133}

\bibitem[{{Russell} \& {Fletcher}(2013)}]{Russell2013}
{Russell}, A.~J.~B., \& {Fletcher}, L. 2013, \apj, 765, 81,
  \dodoi{10.1088/0004-637X/765/2/81}

\bibitem[{{Rutten}(2003)}]{Rutten2003}
{Rutten}, R.~J. 2003, {Radiative Transfer in Stellar Atmospheres}

\bibitem[{{Sadykov} {et~al.}(2020){Sadykov}, {Kosovichev}, {Kitiashvili}, \&
  {Kerr}}]{Sadykov2020}
{Sadykov}, V.~M., {Kosovichev}, A.~G., {Kitiashvili}, I.~N., \& {Kerr}, G.~S.
  2020, \apj, 893, 24, \dodoi{10.3847/1538-4357/ab7b6a}

\bibitem[{{Schmidt} {et~al.}(2012){Schmidt}, {Kowalski}, {Hawley}, {Hilton},
  {Wisniewski}, \& {Tofflemire}}]{Schmidt2012}
{Schmidt}, S.~J., {Kowalski}, A.~F., {Hawley}, S.~L., {et~al.} 2012, \apj, 745,
  14, \dodoi{10.1088/0004-637X/745/1/14}

\bibitem[{{Seaton}(1990)}]{Seaton1990}
{Seaton}, M.~J. 1990, Journal of Physics B Atomic Molecular Physics, 23, 3255,
  \dodoi{10.1088/0953-4075/23/19/012}

\bibitem[{{Seidel}(1977)}]{Seidel1977}
{Seidel}, J. 1977, Zeitschrift Naturforschung Teil A, 32, 1195,
  \dodoi{10.1515/zna-1977-1101}

\bibitem[{{Shu}(1992)}]{Shu1992}
{Shu}, F.~H. 1992, {Physics of Astrophysics, Vol. II}

\bibitem[{{Silverberg} {et~al.}(2016){Silverberg}, {Kowalski}, {Davenport},
  {Wisniewski}, {Hawley}, \& {Hilton}}]{Silverberg2016}
{Silverberg}, S.~M., {Kowalski}, A.~F., {Davenport}, J.~R.~A., {et~al.} 2016,
  \apj, 829, 129, \dodoi{10.3847/0004-637X/829/2/129}

\bibitem[{{Sim{\~o}es} {et~al.}(2017){Sim{\~o}es}, {Kerr}, {Fletcher},
  {Hudson}, {Gim{\'e}nez de Castro}, \& {Penn}}]{Simoes2017}
{Sim{\~o}es}, P. J.~A., {Kerr}, G.~S., {Fletcher}, L., {et~al.} 2017, \aap,
  605, A125, \dodoi{10.1051/0004-6361/201730856}

\bibitem[{{Smith} \& {Auer}(1980)}]{Smith1980}
{Smith}, D.~F., \& {Auer}, L.~H. 1980, \apj, 238, 1126, \dodoi{10.1086/158078}

\bibitem[{{Smith} {et~al.}(1969){Smith}, {Vidal}, \& {Cooper}}]{Smith1969}
{Smith}, E.~W., {Vidal}, C.~R., \& {Cooper}, J. 1969, J Res Natl Bur Stand A
  Phys Chem., 73A, 389, \dodoi{10.6028/jres.073A.030}

\bibitem[{{Stehle}(1994)}]{Stehle1994}
{Stehle}, C. 1994, \aaps, 104, 509

\bibitem[{{Stehl{\'e}} \& {Feautrier}(1984)}]{Stehle1984}
{Stehl{\'e}}, C., \& {Feautrier}, N. 1984, Annales de Physique, 9, 697,
  \dodoi{10.1051/anphys:0198400904069700}

\bibitem[{{Stehl{\'e}} \& {Hutcheon}(1999)}]{Stehle1999}
{Stehl{\'e}}, C., \& {Hutcheon}, R. 1999, \aaps, 140, 93,
  \dodoi{10.1051/aas:1999118}

\bibitem[{{Stehle} \& {Jacquemot}(1993)}]{Stehle1993}
{Stehle}, C., \& {Jacquemot}, S. 1993, \aap, 271, 348

\bibitem[{{Stehle} {et~al.}(1983){Stehle}, {Mazure}, {Nollez}, \&
  {Feautrier}}]{Stehle1983}
{Stehle}, C., {Mazure}, A., {Nollez}, G., \& {Feautrier}, N. 1983, \aap, 127,
  263

\bibitem[{{Stehle} {et~al.}(1988){Stehle}, {Mazure}, {Nollez}, \&
  {Feautrier}}]{Stehle1988}
---. 1988, \aap, 205, 368

\bibitem[{{Suemoto} \& {Hiei}(1959)}]{Suemoto1959}
{Suemoto}, Z., \& {Hiei}, E. 1959, \pasj, 11, 185

\bibitem[{{Suleimanov} {et~al.}(2014){Suleimanov}, {Klochkov}, {Pavlov}, \&
  {Werner}}]{NSatmos}
{Suleimanov}, V.~F., {Klochkov}, D., {Pavlov}, G.~G., \& {Werner}, K. 2014,
  \apjs, 210, 13, \dodoi{10.1088/0067-0049/210/1/13}

\bibitem[{{Sutton}(1978)}]{Sutton1978}
{Sutton}, K. 1978, \jqsrt, 20, 333, \dodoi{10.1016/0022-4073(78)90102-4}

\bibitem[{{Svestka}(1962)}]{Svestka1962}
{Svestka}, Z. 1962, Bulletin of the Astronomical Institutes of Czechoslovakia,
  13, 236

\bibitem[{{Svestka}(1963)}]{Svestka1963}
---. 1963, Bulletin of the Astronomical Institutes of Czechoslovakia, 14, 234

\bibitem[{{Testa} {et~al.}(2014){Testa}, {De Pontieu}, {Allred}, {Carlsson},
  {Reale}, {Daw}, {Hansteen}, {Martinez-Sykora}, {Liu}, {DeLuca}, {Golub},
  {McKillop}, {Reeves}, {Saar}, {Tian}, {Lemen}, {Title}, {Boerner},
  {Hurlburt}, {Tarbell}, {Wuelser}, {Kleint}, {Kankelborg}, \&
  {Jaeggli}}]{Testa2014}
{Testa}, P., {De Pontieu}, B., {Allred}, J., {et~al.} 2014, Science, 346,
  1255724, \dodoi{10.1126/science.1255724}

\bibitem[{{Thorne} \& {Blandford}(2017)}]{Thorne2017}
{Thorne}, K.~S., \& {Blandford}, R.~D. 2017, {Modern Classical Physics: Optics,
  Fluids, Plasmas, Elasticity, Relativity, and Statistical Physics}

\bibitem[{{Tremblay} \& {Bergeron}(2009)}]{Tremblay2009}
{Tremblay}, P.-E., \& {Bergeron}, P. 2009, \apj, 696, 1755,
  \dodoi{10.1088/0004-637X/696/2/1755}

\bibitem[{{Tremblay} {et~al.}(2011){Tremblay}, {Bergeron}, \&
  {Gianninas}}]{Tremblay2011}
{Tremblay}, P.~E., {Bergeron}, P., \& {Gianninas}, A. 2011, \apj, 730, 128,
  \dodoi{10.1088/0004-637X/730/2/128}

\bibitem[{{Uitenbroek}(2001)}]{Uitenbroek2001}
{Uitenbroek}, H. 2001, \apj, 557, 389, \dodoi{10.1086/321659}

\bibitem[{{Uitenbroek}(2003)}]{Uitenbroek2003}
{Uitenbroek}, H. 2003, in Astronomical Society of the Pacific Conference
  Series, Vol. 286, Current Theoretical Models and Future High Resolution Solar
  Observations: Preparing for ATST, ed. A.~A. {Pevtsov} \& H.~{Uitenbroek}, 403

\bibitem[{{van Dien}(1949)}]{vanDien1949}
{van Dien}, E. 1949, \apj, 109, 452, \dodoi{10.1086/145150}

\bibitem[{{Vidal} {et~al.}(1970){Vidal}, {Cooper}, \& {Smith}}]{Vidal1970}
{Vidal}, C.~R., {Cooper}, J., \& {Smith}, E.~W. 1970, \jqsrt, 10, 1011,
  \dodoi{10.1016/0022-4073(70)90121-4}

\bibitem[{{Vidal} {et~al.}(1971){Vidal}, {Cooper}, \& {Smith}}]{Vidal1971}
---. 1971, \jqsrt, 11, 263, \dodoi{10.1016/0022-4073(71)90013-6}

\bibitem[{{Vidal} {et~al.}(1973){Vidal}, {Cooper}, \& {Smith}}]{Vidal1973}
---. 1973, \apjs, 25, 37, \dodoi{10.1086/190264}

\bibitem[{{{\v{S}}vestka}(1965)}]{Svestka1965}
{{\v{S}}vestka}, Z. 1965, Advances in Astronomy and Astrophysics, 3, 119,
  \dodoi{10.1016/B978-1-4831-9921-4.50010-7}

\bibitem[{{{\v{S}}vestka} \&
  {Fritzov{\'a}-{\v{S}}vestkov{\'a}}(1967)}]{Svestka1967}
{{\v{S}}vestka}, Z., \& {Fritzov{\'a}-{\v{S}}vestkov{\'a}}, L. 1967, \solphys,
  2, 75, \dodoi{10.1007/BF00155893}

\bibitem[{{Wedemeyer-B{\"o}hm} \& {Carlsson}(2011)}]{WedemeyerBohm2011}
{Wedemeyer-B{\"o}hm}, S., \& {Carlsson}, M. 2011, \aap, 528, A1,
  \dodoi{10.1051/0004-6361/201016186}

\bibitem[{{Wood} {et~al.}(1996){Wood}, {Harper}, {Linsky}, \&
  {Dempsey}}]{Wood1996}
{Wood}, B.~E., {Harper}, G.~M., {Linsky}, J.~L., \& {Dempsey}, R.~C. 1996,
  \apj, 458, 761, \dodoi{10.1086/176857}

\bibitem[{{Worden} {et~al.}(1984){Worden}, {Schneeberger}, {Giampapa},
  {Deluca}, \& {Cram}}]{Worden1984}
{Worden}, S.~P., {Schneeberger}, T.~J., {Giampapa}, M.~S., {Deluca}, E.~E., \&
  {Cram}, L.~E. 1984, \apj, 276, 270, \dodoi{10.1086/161611}

\bibitem[{{Xu} {et~al.}(2016){Xu}, {Cao}, {Ding}, {Kleint}, {Su}, {Liu}, {Ji},
  {Chae}, {Jing}, {Cho}, {Cho}, {Gary}, \& {Wang}}]{Xu2016}
{Xu}, Y., {Cao}, W., {Ding}, M., {et~al.} 2016, \apj, 819, 89,
  \dodoi{10.3847/0004-637X/819/2/89}

\bibitem[{{Zharkova} \& {Kobylinskii}(1993)}]{Zharkova1993}
{Zharkova}, V.~V., \& {Kobylinskii}, V.~A. 1993, \solphys, 143, 259,
  \dodoi{10.1007/BF00646487}

\bibitem[{{Zhu} {et~al.}(2019){Zhu}, {Kowalski}, {Tian}, {Uitenbroek},
  {Carlsson}, \& {Allred}}]{Zhu2019}
{Zhu}, Y., {Kowalski}, A.~F., {Tian}, H., {et~al.} 2019, \apj, 879, 19,
  \dodoi{10.3847/1538-4357/ab2238}

\end{thebibliography}
\bibliographystyle{aasjournal}

\end{document}